\documentclass[aps,pre,twocolumn,nofootinbib,tightenlines,superscriptaddress,nopacs,amsmath,amssymb,final,letterpaper]{revtex4-1}

\usepackage[utf8]{inputenc}
\usepackage{calc}
\usepackage{graphicx}
\usepackage{amsmath,amssymb,amsthm}
\usepackage{float}
\makeatletter
\let\newfloat\newfloat@ltx
\makeatother
\usepackage{algorithm}
\usepackage{algpseudocode}
\usepackage{bbold}
\usepackage{bm}
\usepackage{color}
\usepackage[dvipsnames,xcdraw,table]{xcolor}
\usepackage{enumitem}
\usepackage{tikz}
\usepackage[percent]{overpic}
\usepackage{tabularx}
\usepackage{multirow}

\DeclareMathOperator*{\argmax}{arg\,max}
\DeclareMathOperator*{\argmin}{arg\,min}

\newcommand{\pder}[2]{\frac{\partial #1}{\partial #2}}

\newcommand{\abs}[1]{\vert #1\vert}

\newcommand{\expec}[1]{\langle #1\rangle}

\def\multiset#1#2{\ensuremath{\left(\kern-.3em\left(\genfrac{}{}{0pt}{}{#1}{#2}\right)\kern-.3em\right)}}

\newtheorem{proposition}{Proposition}

\usepackage[colorlinks=true,linkcolor=blue,urlcolor=blue,citecolor=blue,anchorcolor=blue]{hyperref}

\begin{document}
\title{Estimation of partial rankings from sparse, noisy comparisons}

\author{Sebastian \surname{Morel-Balbi}}
\email{sebastian.morel@gmail.com}
\affiliation{Institute of Data Science, University of Hong Kong, Hong Kong}

\author{Alec \surname{Kirkley}}
\email{alec.w.kirkley@gmail.com}
\affiliation{Institute of Data Science, University of Hong Kong, Hong Kong}
\affiliation{Department of Urban Planning and Design, University of Hong Kong, Hong Kong}
\affiliation{Urban Systems Institute, University of Hong Kong, Hong Kong}

\begin{abstract}

Ranking items based on pairwise comparisons is common, from using match outcomes to rank sports teams to using purchase or survey data to rank consumer products. Statistical inference-based methods such as the Bradley-Terry model, which extract rankings based on an underlying generative model, have emerged as flexible and powerful tools to tackle ranking in empirical data. In situations with limited and/or noisy comparisons, it is often challenging to confidently distinguish performance of different items based on evidence available in the data. However, most inference-based ranking methods choose to assign each item to a unique rank or score, suggesting a meaningful distinction when there is none. Here, we develop a principled nonparametric Bayesian method, adaptable to any statistical ranking method, for learning partial rankings (rankings with ties) that distinguishes among the ranks of different items only when there is sufficient evidence available in the data. We develop a fast agglomerative algorithm to perform Maximum A Posteriori (MAP) inference of partial rankings under our framework and examine the performance of our method on a variety of real and synthetic network datasets, finding that it frequently gives a more parsimonious summary of the data than traditional ranking, particularly when observations are sparse.

\end{abstract}


\maketitle

\section{Introduction}
\label{sec:intro}

In a broad range of applications it can be useful to rank a set of items or players according to some predetermined notion of importance or strength. For example, rankings of players or teams, such as FIFA rankings in soccer or Elo ratings in chess, are used to determine match pairings and tournament seedings. Search engines rank web pages so as to deliver the most relevant results to users. In market analytics, products are often ranked based on sales, reviews, and customer satisfaction. In academia, research funding is often allocated by ranking grant applications, while in finance, credit scoring ranks individuals by creditworthiness, significantly influencing their access to loans, mortgages, and other financial services.

Given the prominence of rankings across domains, it has long been of interest to develop algorithms that automatically rank items based on some set of data about the items and their associations~\cite{bradley1952rank, mallows1957non, luce1959individual, plackett1975analysis, langville2012s}. An extensively studied subset of ranking problems is that of ranking from \emph{pairwise comparisons}~\cite{whelan2017prior, davidson1973bayesian, langville2012s, david1963method, bradley1952rank, jerdee2024luck}, where rankings are inferred from comparisons among pairs of entities. In this approach, the outcome of a single interaction between two items---for example, a win or loss in sports or one product being preferred over another---serves as the basis for the ranking. This approach allows one to infer rankings in situations where underlying ratings of the entities are unavailable, unobserved, or difficult to compute. For example, instead of trying to assess the absolute skill of individual tennis players, we can infer their relative rankings from their head-to-head match results, assuming that matches are more often won by the better player.

Since pairwise comparisons inherently involve some degree of uncertainty due to
randomness, noise, or incomplete information in the observed outcomes, pairwise
ranking models are generally formulated as probabilistic models whose
parameters can then be estimated, for example, via Maximum Likelihood
Estimation (MLE) or Maximum A Posteriori (MAP) inference~\cite{bradley1952rank,
elo1978rating, herbrich2006trueskill, de2018physical, peixoto2022ordered,
cantwell2022belief}. It is also useful to model pairwise comparison data as a network in which items are nodes and each comparison is an edge between its participating nodes, pointing from the winner of the corresponding match/preferred item to the loser of the match/non-preferred item. Multiple edges may exist between a pair of nodes if multiple comparisons were performed, each edge providing evidence for the relative ranking of the two items on its endpoints. Therefore, we will use the network-centric terminology ``nodes'' interchangeably with the entities being ranked (e.g. players, teams, etc) and ``edges'' interchangeably with the pairwise comparisons used for ranking (e.g. matches, hiring flows, etc).

In the classic Bradley-Terry (BT) model \cite{bradley1952rank}, the outcome of a comparison between two nodes is modeled as a Bernoulli random variable with a probability that depends on the latent scores associated with the items. After inference with the model, one obtains a set of continuous scores that can be ordered to obtain a final ranking. The BT model has inspired numerous extensions \cite{agresti2012categorical,jerdee2024luck} to more flexibly model variations in match outcomes, including the possibility of ties in the match outcomes \cite{rao1967ties,davidson1970extending,yan2016ranking,baker2021modifying} (rather than ties in the final \emph{rankings} as we study here). Another popular inference-based method for ranking from pairwise comparisons is SpringRank \cite{de2018physical}, which ranks nodes by finding the ground state of a physical system which models the comparison network as a set of directed springs and has also inspired numerous generalizations \cite{iacovissi2022interplay,della2024model}. These methods can parsimoniously model heterogeneity and noise in match outcomes but do not allow for the possibility of ties in the inferred rankings and, as such, are unable to determine when there is enough structure in the comparison data to justify distinguishing the rankings of different items. (In principle, two nodes can have identical scores under such methods, but this is rarely ever achieved in practice numerically.) Such methods can therefore infer rankings that overfit the data when limited observations are available, resulting in rankings that are highly sensitive to small changes in outcomes. Several methods have been proposed to reduce this overfitting and promote ties in the inferred rankings (see Sec.~\ref{subsec:partial-ranking}). However, these models often require extensive parameter tuning, a priori knowledge about the number and size of the groups being inferred, or incorporate information other than the outcomes of the pairwise comparisons, making them difficult to adapt to a broad range of existing ranking methods.

In this work, we introduce a statistical inference-based ranking framework that intrinsically incorporates partial rankings---rankings where multiple nodes can have the same rank---into its underlying generative model.
\footnote{In the context of ranking algorithms, the term \emph{partial ranking} can generally refer either to rankings where only a subset of the items are fully ordered~\cite{pearce2024bayesian, vanichbuncha2017modelling}, or where some of the items are tied in ranking~\cite{fagin2004comparing}. In this work, we explicitly refer to partial rankings as rankings with ties.}
This is accomplished by introducing a hierarchy of uniform Bayesian priors for the underlying rankings of the nodes, which encourages parsimony in the final scores or ranks inferred using the model. This allows us to account for the uncertainty introduced by sparsity in the data by grouping players into the same rank when the data does not provide enough evidence to separate them. Furthermore, our method can be extended to accommodate any inference-based ranking method that models comparison outcomes as a function of the underlying scores or ranks of the nodes involved ~\cite{bradley1952rank,de2018physical,jerdee2024luck}. We provide a fast nonparametric agglomerative algorithm to infer the partial rankings according to our method, and by fitting our model to a wide range of synthetic and real-world datasets, we find that it often provides a more parsimonious description of the data in the regime of sparse observations. As a case study, we apply our method to a faculty hiring network among Computer Science departments at U.S. universities \cite{clauset2015systematic}. The picture that emerges is that of a well-separated hierarchy dominated by a small group of elite universities whose rankings are not meaningfully different statistically, with very little upward mobility across the ranking groups.

The paper is organized as follows. In Sec.~\ref{subsec:standard-ranking} and Sec.~\ref{subsec:partial-ranking}, we introduce the BT model and review literature relating to partial ranking inference. In Sec.~\ref{subsec:model} and Sec.~\ref{subsec:optimization} and we introduce our partial rankings model and the agglomerative algorithm we use to perform MAP inference with our model. In Sec.~\ref{sec:synthetic} and Sec.~\ref{sec:real}, we apply our method to a wide range of synthetically generated datasets as well as a corpus of empirical networks, finding that partial rankings can provide a more parsimonious description of the data in cases where the networks of outcomes are not sufficiently dense. In Sec.~\ref{sec:casestudy}, we focus on a network of faculty hiring among computer science departments in the U.S. and show that our algorithm can be used to extract ambiguities in the rankings of these departments. We finalize in Sec.~\ref{sec:conclusions} with our conclusions.

\section{Methods}
\label{sec:methods}

\subsection{Bayesian Ranking and the Bradley-Terry Model}
\label{subsec:standard-ranking}

For illustrative purposes, it will be convenient to discuss ranking in the context of matches among individuals---for example, tennis matches among tennis players. But all upcoming discussion can be equivalently framed in the context of comparing two items, in which case each match is a comparison (e.g. from a respondent in a marketing survey or a consumer purchasing one of multiple products), and the winner of the match is the preferred item. All such entities can be captured by the nodes in a network of the pairwise comparisons.

Let $\bm{W}$ be an $N\times N$ matrix of match outcomes for a set of $N$ competing individuals (nodes) such that $w_{ij}$ is the number of matches in which node $i$ beat node $j$. For simplicity, we will assume that ties in the matches are not allowed. The problem we want to solve is that of inferring a ranking $\bm{r}=[r_1,...,r_N]$ of the $N$ individuals such that $r_i<r_j$ indicates a higher probability of $i$ beating $j$ than $j$ beating $i$. Given the inherently probabilistic nature of match outcomes, a common approach is to assign a latent real-valued score $s_i\in \mathbb{R}$ to each player such that the probability $p_{ij}$ that player $i$ beats player $j$ is given by some function of the difference of their scores: $p_{ij} = f(s_i - s_j)$. For $f(s)$ to be a suitable scoring function, it must satisfy a set of constraints~\cite{jerdee2024luck}: (1) It must be a monotonically increasing function of $s$, ensuring that stronger players are assigned a higher probability of winning; (2) It must be bounded within $[0, 1]$ as it must represent a valid probability; (3) It must be antisymmetric around zero, satisfying $f(-s) = 1 - f(s)$ to ensure that the probability of losing is one minus the probability of winning. While these requirements do not uniquely determine $f(s)$, a commonly adopted choice is the logistic function, $f(s) = 1/(1+e^{-s})$, which gives
\begin{equation} \label{eq:bt_s}
    p_{ij} = \frac{e^{s_i}}{e^{s_i} + e^{s_j}}.
\end{equation}
For convenience, one generally introduces the (non-negative) quantities $\pi_i = e^{s_i}$, which, keeping with previous literature, we shall call the player \emph{strengths}~\cite{zermelo1929berechnung, newman2023efficient}. Eq.~\eqref{eq:bt_s} can then be written as
\begin{equation} \label{eq:bt_pi}
    p_{ij} = \frac{\pi_i}{\pi_i + \pi_j}.
\end{equation}
This win probability is the basis of the well known Bradley-Terry (BT) model, first introduced by Zermelo~\cite{zermelo1929berechnung} and then, independently, by Bradley and Terry~\cite{bradley1952rank}.

Given the matrix $\bm{W}$ and assuming all match outcomes are independent of each other, the likelihood of observing $\bm{W}$ is given by
\begin{equation} \label{eq:bt_likelihood}
    P(\bm{W}\vert \bm{\pi}) = \prod_{ij}\left(\frac{\pi_i}{\pi_i+\pi_j}\right)^{w_{ij}}.   
\end{equation}
Differentiating the log-likelihood (i.e. the logarithm of Eq.~\eqref{eq:bt_likelihood}) with respect to the $\pi_i$ and equating the result to zero, we obtain the following set of self-consistent equations:
\begin{equation}
    \pi_i = \frac{\sum_{j=1}^N w_{ij}}{\sum_{j=1}^N(w_{ij} + w_{ji})/(\pi_i + \pi_j)},
\end{equation}
which can be solved by iteration to obtain the MLE estimates of the player scores. Then, these scores can be converted to an implied ranking $\bm{r}$ by simply taking the ordering of the nodes with respect to the inferred strengths $\bm{\pi}$~\cite{zermelo1929berechnung, bradley1952rank, ford1957solution, newman2023efficient}. This iterative procedure is known as Zermelo's algorithm~\cite{zermelo1929berechnung}. More generally, Zermelo's algorithm has been shown to belong to a larger family of iterative algorithms of the form
\begin{equation}  \label{eq:alpha-zermelo}
    \pi_i = \frac{\sum_{j=1}^N w_{ij}(\alpha \pi_i + \pi_j) / (\pi_i + \pi_j)}{\sum_{j=1}^N(\alpha w_{ij} + w_{ji}) / (\pi_i + \pi_j)},
\end{equation}
which have all been proven to converge to the same MLE solution (if one exists) for $0 \leq \alpha \leq 1$. Furthermore, it has been shown that the convergence of this family of algorithms becomes monotonically slower with increasing $\alpha$, with $\alpha = 0$ converging fastest and Zermelo's algorithm (corresponding to the $\alpha = 1$ case) being the slowest to converge. Eq.~\eqref{eq:alpha-zermelo} was first derived in~\cite{newman2023efficient}, and we use its $\alpha=0$ realization throughout this work.

The MLE approach described above presents a series of drawbacks. First, we note that Eq.~\eqref{eq:bt_pi}, and therefore Eq.~\eqref{eq:bt_likelihood}, is invariant under a constant rescaling of all player strengths, meaning that player scores are not uniquely identifiable. To resolve this ambiguity, one generally imposes a suitable normalization. A common choice is to fix the average player score to zero
\begin{equation}
    \expec{s} = \frac{1}{N}\sum_{i=1}^N s_i = 0.
\end{equation}
This choice centers the log-strengths around zero, making them interpretable as relative to an average player. In turn, this ensures that the strength of the average player is given by $\expec{\pi} = 1$, so that a player with $\pi_i > 1$ is stronger than average, and one with $\pi_i <1$ is weaker than average. This normalisation choice also has the advantage of providing an intuitive interpretation of the player's strengths. Following~\cite{newman2023efficient}, let $p_1$ be the probability that a player with strength $\pi$ has of beating the average player of strength one. Then, according to the BT model, $p_1$ is given by $p_1 = \pi / (\pi + 1)$, which in turn implies that $\pi = p_1 / (1- p_1)$. A player's strength is then simply the odds that the player has of beating the average player.

More problematic is the fact that Eq.~\eqref{eq:alpha-zermelo}---and all MLE methods for that matter---have been shown to converge only under the stringent condition that the underlying network of matches is strongly connected, i.e. there exists a directed path between any pair of nodes in the network \cite{zermelo1929berechnung, ford1957solution, hunter2004mm}. If this is not the case, then no maximum of the likelihood function exists, and the scores in Eq.~\eqref{eq:alpha-zermelo} diverge.

Both of these issues can be addressed by adopting a Bayesian perspective and imposing a prior $P(\bm{\pi})$ on the strengths $\bm{\pi}$ so that a full posterior distribution over $\bm{\pi}$ can be inferred~\cite{whelan2017prior, newman2023efficient, davidson1973bayesian, caron2012efficient}. A common choice for this prior is to assume a uniform distribution over the probability $p_1$ that a player with strength $\pi$ defeats the average player, $P(p_1) = 1$. By probability density transformation rules, this is equivalent to setting a logistic prior on the scores $s_i=\log \pi_i$. Indeed,
\begin{equation}
    P(s) = P(p_1) \cdot \frac{dp_1}{ds} = \frac{dp_1}{d\pi}\frac{d\pi}{ds} = \frac{\pi}{(\pi + 1)^2} = \frac{e^{s}}{(1 + e^{s})^2},
\end{equation}
which is a logistic distribution with mean zero and scale one. If we assume the scores of all players to be independently distributed, we have the following prior on the player strengths
\begin{equation}
    P(\bm{\pi}) = \prod_{i=1}^N\frac{\pi_i}{(\pi_i+1)^2}.
\end{equation}
The resulting posterior probability for the scores is then given (up to a multiplicative constant) by
\begin{equation} \label{eq:BT_posterior}
    P(\bm{\pi} | \bm{W}) \propto \prod_{i,j = 1}^N \left( \frac{\pi_i}{\pi_i + \pi_j} \right)^{w_{ij}} \prod_{i=1}^N \frac{\pi_i}{(\pi_i + 1)^2}.
\end{equation}
Eq.~\eqref{eq:BT_posterior} removes the invariance under a rescaling of the scores. Furthermore, it also prevents the scores from diverging. As pointed out in~\cite{whelan2017prior}, the prior for an individual $i$ can be written as
\begin{equation}
    \frac{\pi_i}{(\pi_i + 1)^2} = \frac{\pi_i}{\pi_i + 1} \cdot \frac{1}{\pi_i + 1},
\end{equation}
which corresponds to the probability that player $i$ plays two matches against the average player, winning one and losing the other. Eq.~\eqref{eq:BT_posterior} can then be seen as the likelihood of a BT model where two pseudo-games have been added for each player, one won and one lost. The resulting network of interactions is now strongly connected, ensuring that the maximum of the posterior always exists, irrespective of whether the original interaction network was fully connected or not.

For this Bayesian model, Maximum A Posteriori (MAP) estimates for $\bm{\pi}$ can be inferred by solving the following optimisation problem:
\begin{equation}
 \bm{\hat \pi} = \argmax_{\bm{\pi}} P(\bm{\pi}\vert \bm{W})
 = \argmax_{\bm{\pi}} P(\bm{W}\vert \bm{\pi})P(\bm{\pi}).
\end{equation}
Again, differentiating with respect to the $\pi_i$ and equating the result to zero, we obtain the following set of self-consistent equations for the MAP estimates of the strengths:
\begin{equation} \label{eq:MAP_update}
    \hat{\pi}_i = \frac{1 + \sum_{j \neq i} w_{ij}\pi_j / (\pi_i + \pi_j)}{2/(\pi_i + 1) + \sum_{j \neq i} w_{ji}/(\pi_i + \pi_j)},
\end{equation}
which is the generalization of Eq.~\eqref{eq:alpha-zermelo} (with $\alpha=0$) to the MAP setting and can be solved in the same way as for the MLEs. The primary advantage of Eq.~\eqref{eq:MAP_update} over the MLE updates is that the iterative updates described by Eq.~\eqref{eq:MAP_update} are guaranteed to converge, and do so quite efficiently in practice~\cite{newman2023efficient}. The inferred strengths are then converted into an implied ranking $\bm{r}$ by ordering the nodes with respect to the inferred strengths $\bm{\hat \pi}$.

\subsection{Partial Rankings}
\label{subsec:partial-ranking}

As discussed in Sec.~\ref{sec:intro}, a limitation of standard MAP estimation for the BT model is that it lacks a mechanism to handle partial rankings. In principle, one could apply a heuristic to assign the same rank to nodes $i$ and $j$ when their scores satisfy $\abs{s_i-s_j}<\epsilon$ for some small $\epsilon$. However, this approach does not offer any statistical justification for grouping $i$ and $j$ together (i.e., whether the observed score difference is simply due to statistical noise). Alternatively, one could apply 1D numerical clustering to the final score values, but this approach would not account for the significance of score differences in the context of the model's likelihood or prior. More principled approaches have been developed for 1D numerical clustering of the scores by applying $L1$ regularization to the adjacent score differences~\cite{masarotto2012ranking}, analogous to the fused lasso \cite{tibshirani2005sparsity}. This \emph{ranking lasso} has been extended in a number of different ways to increase its flexibility for different applications \cite{fan2001variable, zou2006adaptive,masarotto2012ranking, tutz2015extended, vana2016computing, varin2016statistical, tutz2015extended, jeon2018sparse, tutz2015extended, hermes2024joint}.

Despite their widespread application, regularized ranking models suffer from significant drawbacks. The most obvious is the need to tune the lasso penalty parameter, which requires us to fix the level of sparsity a-priori, forcing the groupings in a predefined way and complicating the interpretation of the inferred parameters. Furthermore, as mentioned previously, maximum likelihood estimates for the Bradley-Terry model (and general Bradley-Terry-Luce family of models) without a score prior can be difficult and computationally expensive to perform, although considerable work has been dedicated to developing efficient algorithms~\cite{hunter2004mm, maystre2015fast, agarwal2018accelerated, nguyen2023efficient, newman2023efficient}. Also, as pointed out in~\cite{pearce2024bayesian}, uncertainty quantification can be hard to perform in lasso-based methods~\cite{fan2001variable}, and prior knowledge on the number or size of the rank groups cannot be incorporated.

Many of these issues can be resolved by resorting to Bayesian approaches. However, these approaches generally require Monte Carlo estimation, which can be slow for estimation~\cite{davidson1973bayesian, caron2012efficient}. In~\cite{pearce2024bayesian}, Pearce and Erosheva propose the Rank-Clustered Bradley-Terry-Luce (RC-BTL) model, a Bayesian method based on the use of a spike-and-slab prior~\cite{mitchell1988bayesian, george1997approaches, ishwaran2005spike} to induce parameter fusion. Unlike other methodologies, Pearce and Erosheva's model requires neither the specification of the number or sizes of the rank cluster nor the tuning of lasso-type penalties, while exploiting conjugacy relationships for estimation of the full posterior. 
However, this model still requires the selection and tuning of a series of hyperparameters, among which is setting the expected number of rank clusters. Although this choice can be advantageous, as it allows the user to tune the desired level of rank clustering, poor choices of this parameter could potentially introduce unwanted biases in the inferred results. 


An alternative approach is to formulate fully nonparametric models, which do not require any hyperparameter tuning. In the ordered Stochastic Block Model (OSBM)~\cite{peixoto2022ordered}, Peixoto proposes a nonparametric model based on a modification of the Stochastic Block Model~\cite{holland1983stochastic, peixoto2019bayesian} in which nodes are grouped according to both the mixing structure of the network and hierarchies among the nodes as evidenced by edge directionality. Although the OSBM does allow nodes to be grouped in part by the edge hierarchy, it is heavily influenced by the mixing structure of the nodes, which largely depends on the positions of the edges rather than their directionality (the latter aspect being of primary interest for pairwise ranking). In contrast, the method we propose here focuses only on the directionality of the observed edges and so is adaptable to a broad class of ranking methods, including most popular generalizations of the BT model \cite{jerdee2024luck}.   

\subsection{Bayesian Nonparametric Model}
\label{subsec:model}

A principled approach to handling partial rankings is to extend the Bayesian hierarchy for the prior on the strengths $P(\bm{\pi})$ by making the strength of each node a function of its underlying ranking $\bm{r}$. This results in the following new MAP objective:
\begin{align} \label{eq:pr_map_objective}
 \bm{\hat r},\bm{\hat \pi} &= \argmax_{\bm{r},\bm{\pi}} P(\bm{r},\bm{\pi}\vert \bm{W}) \notag \\ 
 &= \argmax_{\bm{r},\bm{\pi}} P(\bm{W}\vert \bm{\pi})P(\bm{\pi}\vert \bm{r})P(\bm{r}) ,  
\end{align}
where $\bm{\hat r}$ represents the optimal partial ranking to be estimated from the observed data and $\bm{\hat\pi}$ is the optimal set of scores, as defined previously.

Given Eq.~\eqref{eq:pr_map_objective}, we now need to define the functional forms of the likelihood and priors. The easiest choice is to use the BT likelihood of Eq.~\eqref{eq:bt_likelihood}, although any choice of likelihood is possible to incorporate into this MAP estimation objective so long as it is a function of latent strengths---i.e., $P(\bm{W}\vert \bm{\pi})$---or ranks---i.e., $P(\bm{W}\vert \bm{r})$ (the latter allowing us to remove the prior $P(\bm{\pi}\vert \bm{r})$). This flexibility allows for the inclusion of a wide range of ranking objectives, including extensions of BT models \cite{jerdee2024luck} and SpringRank \cite{de2018physical}. For our experiments in Sec.~\ref{sec:results}, we will proceed with using the BT likelihood of Eq.~\eqref{eq:bt_likelihood} for subsequent analysis, which facilitates an efficient optimization scheme and a simple comparison with the original BT model. 

For the prior on $\bm{r}$, we choose to be agnostic with respect to both the number and sizes of the underlying rankings. This results in a hierarchy of uniform priors, as described by the following generative model:
\begin{enumerate}
    \item Draw the number of unique ranks $R$ uniformly at random from the range $[1,N]$. As there are $N$ maximum possible unique ranks, the probability of observing any value of $R \in [1, N]$ is given by
    \begin{equation}
        P(R) = \frac{1}{N}.
    \end{equation}

    \item Draw the histogram $\bm{n}=[n_1,...,n_R]$ of the number of nodes of each rank uniformly from the set of all histograms compatible with the number of ranks $R$. This amounts to selecting a list of $R$ positive integer values that sum to $N$. As there are $N-1\choose R-1$ such lists, each draw occurs with probability
    \begin{equation}
        P(\bm{n}\vert R) = \frac{1}{{N-1\choose R-1}}.
    \end{equation}

    \item Draw the ranks $\bm{r}$ uniformly over the set of rankings compatible with the rank counts $\bm{n}$. Since there are $N\choose n_{1},...,n_{R}$ possible partitions of $N$ objects into $R$ groups such that each group $r$ has size $n_r$, the total probability of this step is given by
    \begin{equation}
        P(\bm{r}\vert\bm{n}) = \frac{1}{{N\choose n_{1},...,n_{R}}}.
    \end{equation}
\end{enumerate}
The resulting prior is then given by
\begin{align}  \label{eq:partition_prior}
    P(\bm{r}) &= P(\bm{r}\vert \bm{n})P(\bm{n}\vert R)P(R) \notag \\
    &= \frac{1}{{N\choose n_{1},...,n_{R}}}\times\frac{1}{{N-1\choose R-1}}\times  \frac{1}{N}.
\end{align}

A reasonable choice for the prior $P(\bm{\pi}\vert \bm{r})$ is a hierarchical prior consisting of two steps: (1) Draw the set of (non-negative) unique strengths $\bm{\sigma} = \{\sigma_1,\cdots,\sigma_R\}$ based on the number of unique ranks $R$; (2) Draw the individual strengths $\bm{\pi}$ using the deterministic prior $P(\bm{\pi}\vert \bm{\sigma})=\prod_{i=1}^{N}\delta(\pi_i,\sigma_{r_i})$. Combining these gives the prior
\begin{align} \label{eq:strength_prior}
 P(\bm{\pi}\vert \bm{r}) &= P(\bm{\pi}\vert \bm{\sigma})P(\bm{\sigma}\vert \bm{r}) \notag \\
 &=  \prod_{i=1}^{N}\delta(\pi_i,\sigma_{r_i})\prod_{r=1}^{R}P(\sigma_r)\prod_{r=1}^{R-1}\Theta(\sigma_{r}-\sigma_{r+1}), 
\end{align}
where $\Theta$ is the Heaviside step function (which ensures that the unique strengths are ordered) and $P(\sigma_r)$ is given by
\begin{equation}
    P(\sigma_r) = \frac{\sigma_r}{(\sigma_r + 1)^2}
\end{equation}
as in the BT model.

The MAP estimation objective can now be written as the following minimization problem in terms of the (negative) log posterior
\begin{align}
 \bm{\hat r},\bm{\hat \sigma}
 &= \argmin_{\bm{r},\bm{\sigma}} \left[-\log P(\bm{r},\bm{\pi}(\bm{\sigma})\vert \bm{W})\right] \\
 &= \argmin_{\bm{r},\bm{\sigma}}  \{\mathcal{L}(\bm{r},\bm{\sigma}) \},
\end{align}
where
\begin{align} \label{eq:pr_dl}
 \mathcal{L}(\bm{r},\bm{\sigma}) &= \log N + \log {N-1\choose R-1}+\log {N\choose n_1,...,n_R} \notag \\
 &+\sum_{r=1}^{R}\log \left[\frac{(\sigma_r+1)^2}{\sigma_r}\right]
 +\sum_{r,r'=1}^{R}\omega_{rr'}\log \left[\frac{\sigma_r+\sigma_{r'}}{\sigma_r}\right],
\end{align}
with
\begin{align}
  \omega_{rr'}\equiv \sum_{ij}w_{ij}\delta_{r_i,r}\delta_{r_j,r'}    
\end{align}
denoting the number of edges going from nodes of rank $r$ to nodes of rank $r'$ (i.e. the number of times nodes of rank $r$ beat nodes of rank $r'$). 

The generative process described above, and the associated objective, are similar to those of the RC-BTL model introduced by Pearce and Erosheva~\cite{pearce2024bayesian}, which places a Gamma prior on the unique cluster strengths. The main difference between our model and RC-BTL lies in the prior over partitions. In our Partial Ranking (PR) formulation, players are assigned to groups via a hierarchy of uniform priors over the number, sizes, and compositions of the clusters. By contrast, RC-BTL employs a Poisson prior on the number of clusters and assigns equal probability to all partitions with the same $R$, regardless of cluster sizes. As a result, PR and RC-BTL encode different structural biases over the space of partitions. In RC-BTL, the indifference with respect to cluster sizes simplifies posterior sampling, since all partitions with the same number of groups are equally likely, but it also implicitly favors balanced groups: if partitions are drawn uniformly conditional on $R$, the overwhelming majority will allocate players approximately equally across groups~\cite{peixoto2019bayesian, peixoto2017nonparametric}. Depending on the amount of available data and prior knowledge about the items being ranked, this may introduce artificial structure into the inferred rankings. In contrast, the PR formulation explicitly samples the group sizes as part of the prior hierarchy, remaining agnostic about balance until the data are observed. For a more detailed comparison between the PR and RC-BTL models see Appendix~\ref{appendix:appendix_f}.

By minimizing Eq.~\eqref{eq:pr_dl}, we can determine the optimal ranks $\bm{\hat r}$ and the corresponding optimal strengths $\bm{\hat \sigma}=\{\hat\sigma_{\hat r_i}\}_{i=1}^{N}$ that best fit the observed data. It is important to emphasize that, although this work focuses primarily on MAP estimates of the strengths and rankings, Eq.~\eqref{eq:pr_dl} defines a well-posed (negative) log-posterior distribution. This posterior can be explored using sampling techniques such as Markov Chain Monte Carlo (MCMC), enabling estimation of the full posterior distribution and associated uncertainty quantification rather than relying solely on point estimates.

We can see that the first term in Eq.~\eqref{eq:pr_dl} is a constant independent of the underlying node ranking. The second and third terms synergistically penalize having a large number of rankings as they will increase $\mathcal{L}(\bm{r}, \bm{\sigma})$. Indeed, it is easy to prove that the sum of these two terms 
\begin{equation}
    f(R) = \log \binom{N - 1}{R - 1} + \log\binom{N}{n_1 \ldots n_R}
\end{equation}
has a global minimum when $R = 1$, see Appendix~\ref{appendix:objective_contributions}. The final two terms are harder to interpret, as they will depend on the particular realization of the matches. However, empirically, we observe that the fourth term will also penalize having a large number of rankings, while the last term will tend to favor them.

An advantage of working with a Bayesian framework is that it provides us with a principled way to compare results obtained by different models. Suppose we have two different ranking models, which we call $\mathcal{H}_1$ and $\mathcal{H}_2$, that correspond to two different hypotheses for the generating mechanism that produced some observed network of outcomes $\bm{W}$. Let $\bm{r}_1$ and $\bm{r}_2$ be the corresponding most likely rankings obtained by maximizing their respective posterior distributions. A principled approach to decide which of the two rankings/model combinations better represents the data is to compute the posterior odds ratio~\cite{jaynes2003probability}
\begin{equation}
    \frac{P(\bm{r}_1, \mathcal{H}_1 | \bm{W})}{P(\bm{r}_2, \mathcal{H}_2 | \bm{W})} = 
    \frac{P(\bm{W} | \bm{r}_1, \mathcal{H}_1)P(\bm{r}_1 | \mathcal{H}_1)P(\mathcal{H}_1)}{P(\bm{W} | \bm{r}_2, \mathcal{H}_2)P(\bm{r}_2 | \mathcal{H}_2)P(\mathcal{H}_2)},
\end{equation}
where a ratio above (below) $1$ indicates that we should favor model $\mathcal{H}_1$ ($\mathcal{H}_2$) based on posterior probability.
If we assume that the two models are a priori equally likely ahead of observing any data so that $P(\mathcal{H}_1) = P(\mathcal{H}_2)$, we have that the posterior odds ratio can be written as
\begin{equation} \label{eq:POR}
    \frac{P(\bm{r}_1, \mathcal{H}_1 | \bm{W})}{P(\bm{r}_2, \mathcal{H}_2 | \bm{W})} = 
    \frac{P(\bm{W} | \bm{r}_1, \mathcal{H}_1)P(\bm{r}_1 | \mathcal{H}_1)}{P(\bm{W} | \bm{r}_2, \mathcal{H}_2)P(\bm{r}_2 | \mathcal{H}_2)}.
\end{equation}
By taking the logarithm of Eq.~\eqref{eq:POR}, we can rewrite the posterior odds ratio as
\begin{align}
    \log \frac{P(\bm{r}_1, \mathcal{H}_1 | \bm{W})}{P(\bm{r}_2, \mathcal{H}_2 | \bm{W})} &= \log P(\bm{r}_1, \mathcal{H}_1 | \bm{W}) - \log P(\bm{r}_2, \mathcal{H}_2 | \bm{W}) \notag \\
    &= \mathcal{L}_2(\bm{r}_2, \mathcal{H}_2) - \mathcal{L}_1(\bm{r}_1, \mathcal{H}_1).
\end{align}

\subsection{Optimization}
\label{subsec:optimization}

\begin{algorithm}
\caption{Greedy Algorithm for MAP inference of Partial Rankings} \label{alg:greedy}
\begin{algorithmic}[1]
\State Initialize ranking vector $r$ with each node in its own group (i.e. $R = N$)
\State Initialize strength vector $\sigma$ with initial BT estimates
\State $L_{\text{current}} \gets \text{ComputeLoss}(r,\sigma)$
\State $L_{\text{best}} \gets L_{\text{current}}$
\State $r_{\text{best}} \gets r$, $\sigma_{\text{best}} \gets \sigma$
\vspace{0.5em}
\While{$R > 1$} \Comment{Continue merging until a single group remains}
    \State \textbf{/* Evaluate potential merges between adjacent groups */}
    \For{each adjacent pair $(i, i+1)$ in $r$}
        \State $\Delta L[i] \gets \text{ComputeDeltaL}(\text{MergeGroup}(i,i+1), r, \sigma)$
    \EndFor
    \State $(i_{\text{merge}}, j_{\text{merge}}) \gets \arg\min_{(i,i+1)} \Delta L[i]$
    \vspace{0.5em}
    \State \textbf{/* Merge the selected groups */}
    \State $r \gets \text{MergeGroups}(r, i_{\text{merge}}, j_{\text{merge}})$
    \State $\sigma_{\text{merged}} \gets \text{SolveForMergedSigma}(i_{\text{merge}}, j_{\text{merge}}, r, \sigma)$
    \State Update $\sigma$ to incorporate the merged group and adjust the remaining strengths accordingly
    \vspace{0.5em}
    \State $L_{\text{current}} \gets \text{ComputeLoss}(r, \sigma)$
    \If{$L_{\text{current}} < L_{\text{best}}$}
        \State $L_{\text{best}} \gets L_{\text{current}}$
        \State $r_{\text{best}} \gets r$, $\sigma_{\text{best}} \gets \sigma$
    \EndIf
    \State $R \gets$ number of groups in $r$
\EndWhile
\vspace{0.5em}
\State \Return{$r_{\text{best}}, \sigma_{\text{best}}$}
\end{algorithmic}
\end{algorithm}

We propose a fast nonparametric agglomerative algorithm based on an alternating optimization strategy to approximate the MAP estimates $\bm{\hat r}$, $\bm{\hat \sigma}$ for the node rankings and strengths. Starting from an initial condition in which each node belongs to its own group, so that $R = N$, the algorithm alternates between updating the unique strengths $\bm{\sigma}$ and rankings $\bm{r}$ so as to minimize the negative log-posterior $\mathcal{L}$ (Eq.~\eqref{eq:pr_dl}).

Taking the gradient to optimize $\mathcal{L}$ with respect to $\bm{\sigma}$, while keeping the rank vector $\bm{r}$ fixed, yields
\begin{align} \label{eq:saddle_point}
   \sigma_r = \frac{1+\sum_{r'\neq r}\omega_{rr'}\sigma_{r'}/(\sigma_{r}+\sigma_{r'})}{2/(\sigma_r+1)+\sum_{r'\neq r}\omega_{r'r}/(\sigma_{r}+\sigma_{r'})}
\end{align}
Eq.~\eqref{eq:saddle_point} defines a set of $R$ self-consistent equations for the $\sigma_r$'s which can be iteratively solved until convergence. If $E(\bm{r})$ is the number of non-zero entries in the matrix $\bm{\omega}$ when the ranks are equal to $\bm{r}$, then the complexity of this algorithm is $O(E(\bm{r}))$. These iterative updates for the player's strengths are of the same form as those introduced in Eq.~\eqref{eq:MAP_update} and benefit from the same convergence guarantees.

To update $\bm{r}$ given $\bm{\sigma}$, we can greedily identify the pair of consecutive ranks $r<r'$ that produces the largest decrease $\Delta \mathcal{L}(r,r')$ in the negative log-posterior when the two ranks are merged. To do this, it is useful to rewrite the negative log-posterior as follows:
\begin{equation}
    \mathcal{L}(\bm{r}) = C(R) + \sum_{r=1}^R g(r) + \sum_{r, r' = 1}^R f(r, r')
\end{equation}
where
\begin{align}
    &C(R) = \log N + \log \binom{N - 1}{R - 1} + \log N!, \label{eq:C_R} \\
    &g(r) = -\sum_r\log n_r! + \sum_r \log\left[\frac{(\sigma_r + 1)^2}{\sigma_r}\right], \label{eq:g_r} \\
    &f(r, r') = \sum_{r, r'} \omega_{r,r'}\log\left[\frac{\sigma_r + \sigma_{r'}}{\sigma_r}\right] \label{eq:f_rs}
\end{align}
and we have dropped the dependence on the strength parameters as we consider them fixed. $C(R)$ accounts for the contribution to the negative log-posterior given by the number of rankings $R$, $g(r)$ describes contributions from group sizes and strengths, and $f(r,r')$ captures the interaction terms between the rank groups based on the comparison matrix $\bm{\omega}$ among the rank groups.

Let $R$ represent the current number of ranks, and let $\bm{n}$ denote the sizes of the rank groups before the merge. Then the change in the negative log-posterior resulting from merging $r$ and $r'$ is given by:
\begin{align} \label{eq:delta_L}
    &\Delta\mathcal{L}(r, r') = C(R-1) - C(R) + g((r, r')) - g(r) - g(r') \notag \\
    &+ f((r,r'), (r,r')) - f(r,r) - f(r',r') - f(r, r') - f(r', r) \notag \\
    &+ \sum_{r''\neq r,r'} \left[f((r,r'),r'')+f(r'',(r,r'))-f(r,r'')-f(r',r'')\right. \notag \\
    &\left.\hspace{1.5cm}-f(r'',r)-f(r'',r')\right],
\end{align}
where $(r,r')$ represents the new rank group formed by merging groups $r,r'$.

Once the ranks to merge have been identified, the optimal strength $\sigma_{(r,r')}$ for the newly merged rank can be determined by solving the following equation:
\begin{equation}
  \pder{\Delta \mathcal{L}(r,r')}{\sigma_{(r,r')}} = 0,
\end{equation}
which leads to
\begin{align}
  &\sigma_{(r,r')} = \frac{1+\sum_{r''\neq r,r'}\omega_{(r,r')r''}\sigma_{r''}/(\sigma_{(r,r')}+\sigma_{r''})}{2/(\sigma_{(r,r')}+1)+\sum_{r''\neq r,r'}\omega_{r''(r,r')}/(\sigma_{(r,r')}+\sigma_{r''})},
\end{align}
which is the standard iterative update of Eq.~\eqref{eq:saddle_point} and can be solved efficiently with a computational complexity of $O(E(\bm{r})/R)$, where $E(\bm{r})$ is now the number of non-zero entries in the matrix $\bm{\omega}$ for the current ranks $\bm{r}$. When applied across all $R-1$ possible adjacent rank pairs $(r,r')$, the total complexity for this update and merge process is then $O(E(\bm{r}))$.

Having computed the strength for the merged cluster $(r,r')$, we can repeat the process and do another full update for the strengths $\bm{\sigma}$ given the new ranking $\bm{r}$ after the merge. Again, this requires using the previous iterative update with complexity $O(E(\bm{r}))$. We can then continue alternating the updates $\bm{\sigma}\vert \bm{r}$ and $\bm{r}\vert \bm{\sigma}$ until all groups have been merged, at which point we can inspect all the examined rankings and select the ranking $\bm{ \hat r}$ associated with the highest posterior probability (the lowest value of $\mathcal{L}(\bm{r}, \bm{\sigma})$). Pseudocode for the algorithm is shown in Algorithm~\ref{alg:greedy}.

\begin{figure*}[!ht]
    \centering
    \includegraphics[width=0.8\textwidth]{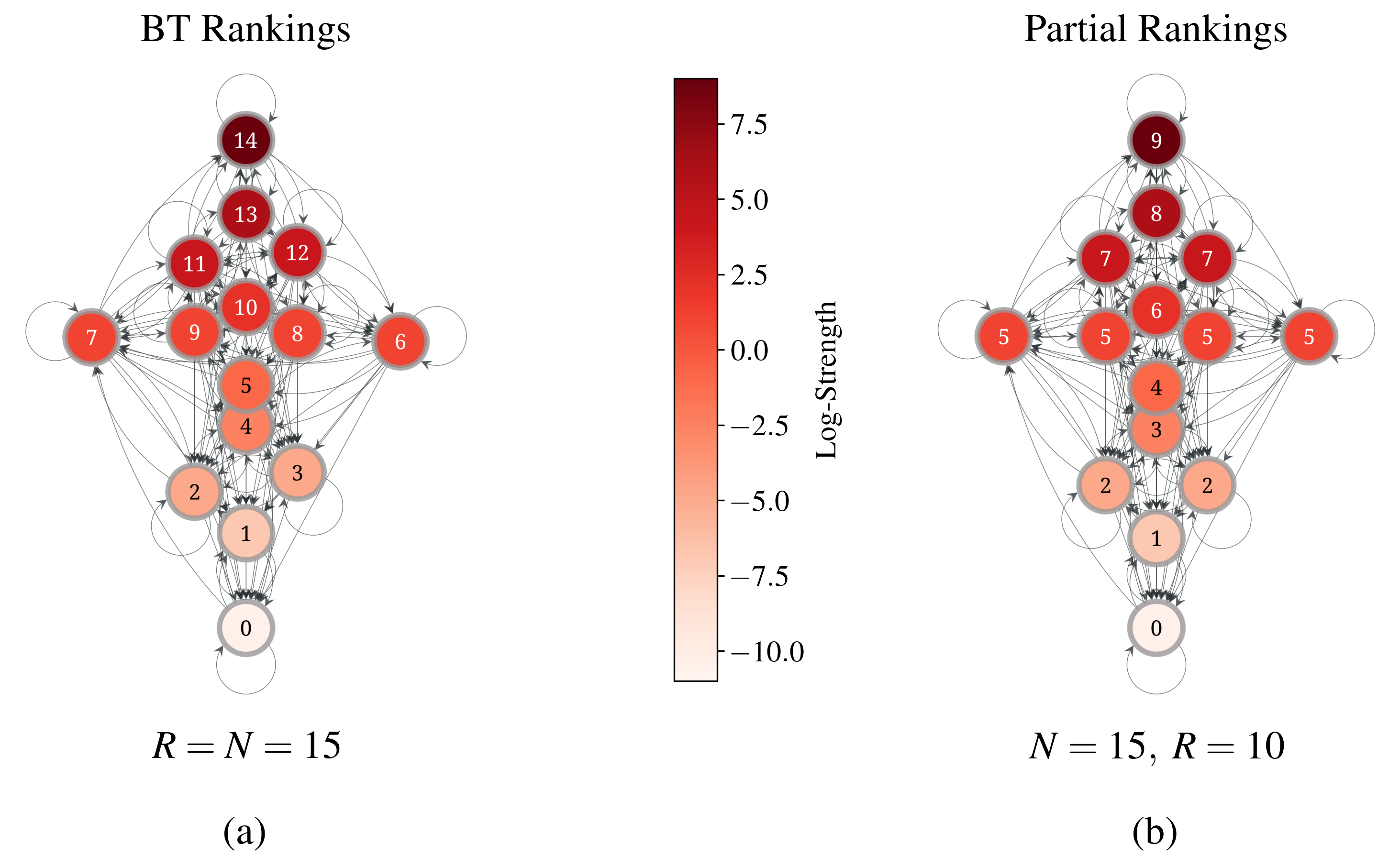}
    \caption{\textbf{Partial rankings in a small example network.} \textbf{(a)} Rankings inferred by the BT model and \textbf{(b)} the PR algorithm, for a dataset capturing hierarchical relationships in a pack of wolves. The nodes are labeled according to their inferred ranking. The distance along the y-axis and the node colors are proportional to the inferred strength $\pi_i$ of each node $i$, with the strongest nodes placed at the top. In this case, there was not enough statistical evidence in the edges to justify separating the node ranks $\{2,3\}$, $\{6,7,8,9\}$, and $\{11,12\}$ on the left hand side, so the partial ranking method grouped these nodes together into the same partial rankings.}
    \label{fig:diagram}
\end{figure*}

The overall complexity of this algorithm is given by:
\begin{align}
  O\left(\sum_{R=N}^{2}E(\bm{r})\right) = O(N^{2+\alpha}),    
\end{align}
where $N$ is the number of nodes or players in the network, $\bm{r}$ is the optimal rank vector for each value $R$ of the number of unique ranks during the merge process, and $\alpha\in [0,1]$ depends on the density of the matrix $\bm{\omega}$ as the ranks are progressively merged. In the worst-case scenario, where the matrix $\bm{\omega}$ is maximally dense ($E(\bm{r})=R^2$ at every step), the complexity would scale as $O(N^3)$. However, such maximal density is highly unlikely for $R \approx N$, especially in real, sparse networks. In practice, we observe a runtime scaling of approximately $O(N^2)$ with the network size $N$; see Fig.~\ref{fig:rw_time_complexity} in Appendix~\ref{appendix:appendix_b}. We note that this algorithm is greedy in nature, and therefore is only guaranteed to identify a local optimum for the MAP estimates $\bm{\hat r}$, $\bm{\hat \sigma}$. However, exact optimization over node groupings is likely NP-hard (as is the case with many other clustering problems \cite{welch1982algorithmic}), making exact inference intractable in all but the smallest of networks. Greedy agglomerative algorithms have been shown to closely approximate the optimal log posterior probabilities obtained through exact enumeration and simulated annealing for other network clustering tasks in which we expect to preserve spatial contiguity \cite{kirkley2022spatial,morel2024bayesian} or an initial temporal ordering \cite{kirkley2024inference}, similar in nature to preserving the initial ordering of the BT scores under the final inferred partial rankings, which is overwhelmingly the case in practice. 

Fig.~\ref{fig:diagram} shows the results of applying both the BT model and our Partial Rankings (PR) algorithm (with the BT likelihood of Eq.~\eqref{eq:bt_likelihood}) to a small dataset of dominance interactions in a pack of wolves~\cite{van198711}. We notice that our model offers a more concise interpretation of the data by grouping individuals with similar BT scores into the same rankings.

Code implementing our partial ranking method can be found in an updated release of the \texttt{PANINIpy} package for nonparametric network inference \cite{kirkley2024paninipy} and at~\url{https://github.com/seb310/partial-rankings}.

\section{Results}
\label{sec:results}

\subsection{Synthetic Match Datasets}
\label{sec:synthetic}


\begin{figure*}
    \centering
    \includegraphics[width=\textwidth]{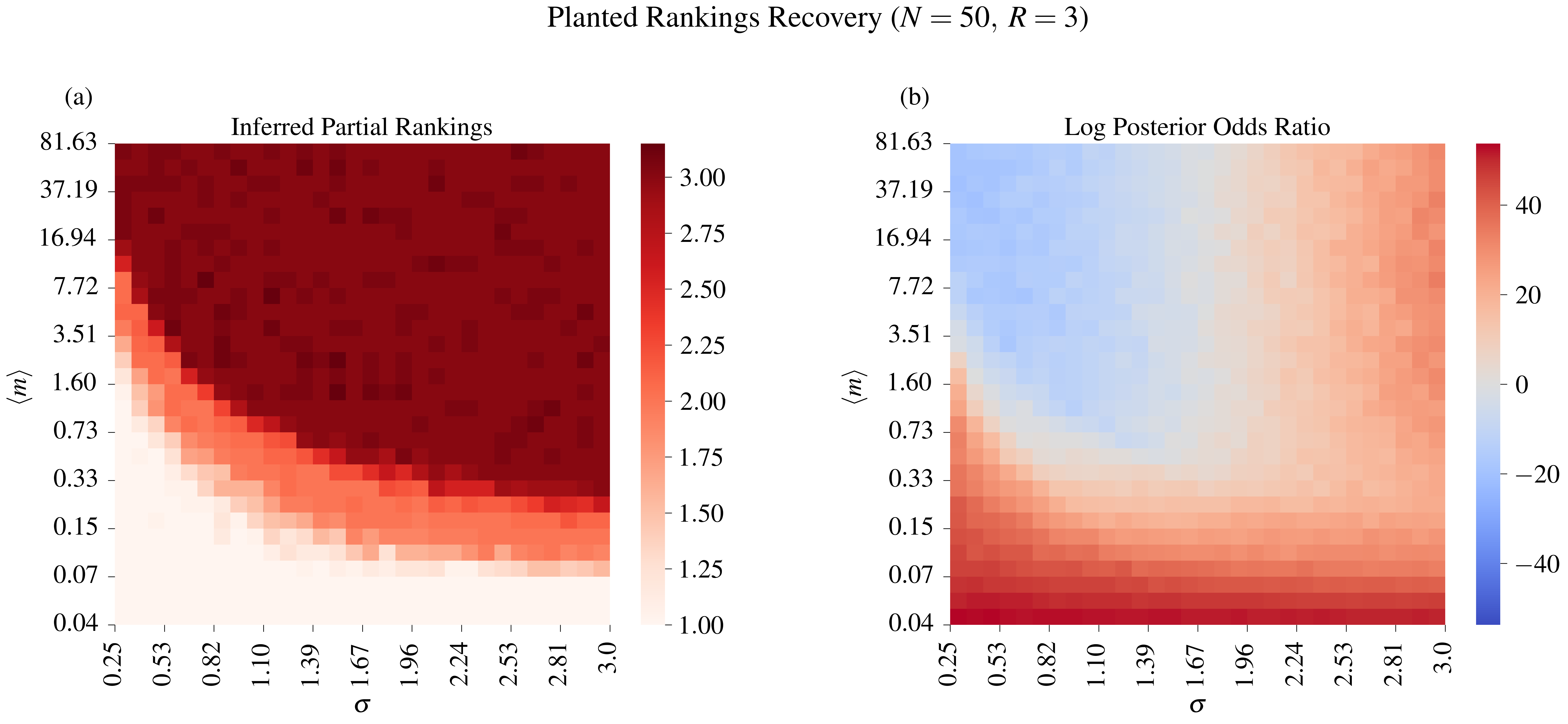}
    \caption{\textbf{(a)} Heatmap of the number of rankings $R$ inferred by the partial rankings model, as a function of $(\sigma, \expec{m})$ for a dataset of $N=50$ nodes planted into $R = 3$ rankings as described by the synthetic model in Sec.~\ref{sec:synthetic}. \textbf{(b)} Heatmap of the log posterior odds ratio (Eq.~\eqref{eq:PORbtpr}) between the BT and the partial rankings model across the simulations. Positive values indicate a preference for the partial rankings model, and negative values a preference for the BT model.}
    \label{fig:2_km_sigma_phase_heatmaps}
\end{figure*}

\begin{figure}
    \centering
    \includegraphics[width=\linewidth]{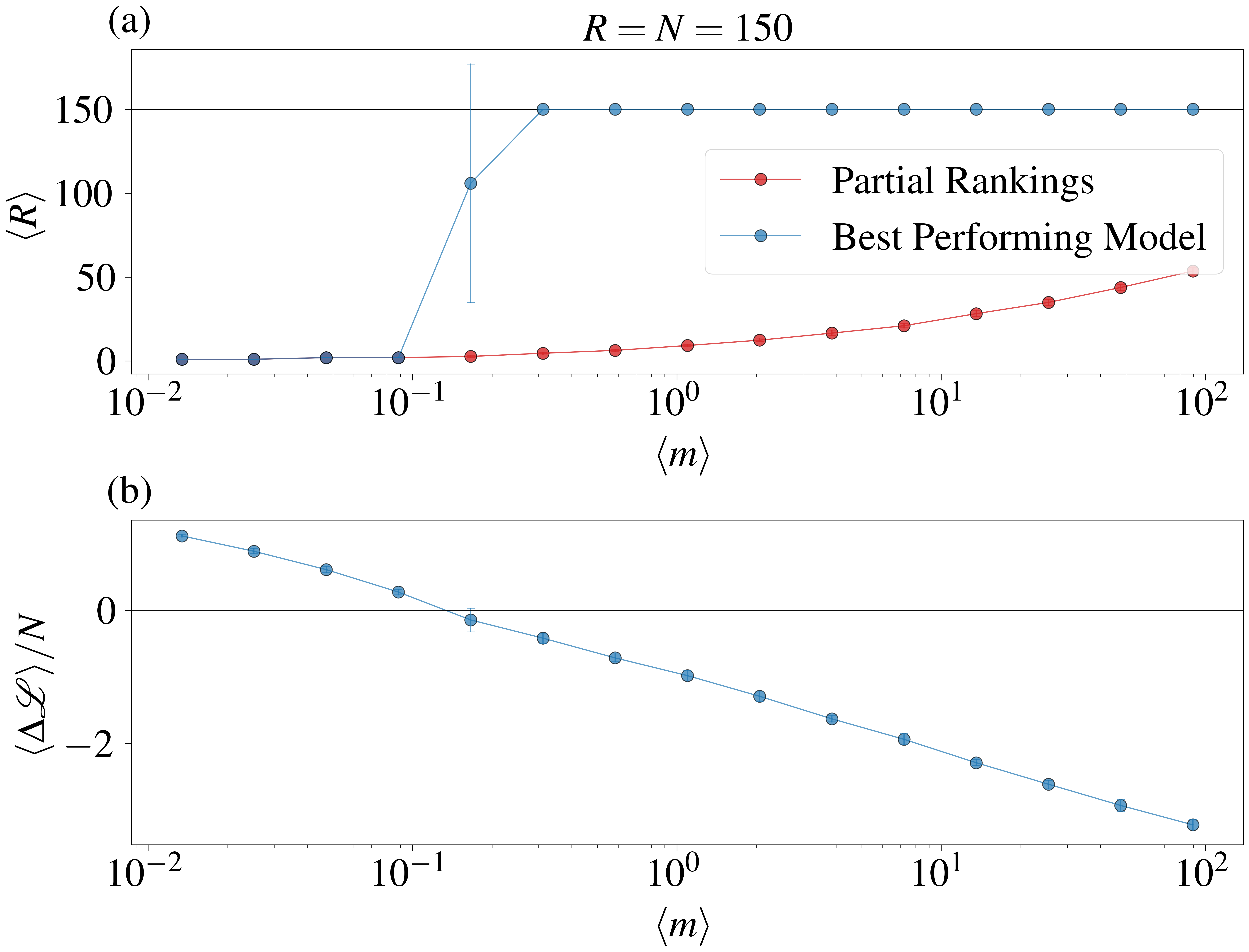}
    \caption{(a) Average number of rankings inferred by the best-performing model (blue) and our partial rankings model (red) as a function of the average number of matches per player pair $\langle m \rangle$ for the case in which no planted partial rankings are present. (b) Log-posterior-odds ratio (normalized per node) between the two models. Negative values of this difference indicate a preference for the BT model and positive values a preference for the partial rankings model. All results were obtained by averaging over $20$ different simulations from the synthetic network model of Sec.~\ref{sec:synthetic}, and error bars indicate $2$ standard errors from the mean. Except near the transition point, error bars are smaller than the marker size.}
    \label{fig:2_km_sigma_phase_N150R150}
\end{figure}

We begin our analysis by evaluating the performance of our algorithm in recovering partial rankings from synthetic data. Specifically, we examine two key sources of uncertainty in ranking recovery: the number of matches played and the separation in player strengths. To achieve this, we design a synthetic model with two tunable parameters, $M$ and $\sigma$, which control the number of matches played and the strength separation between players, respectively. The model requires a set of $N$ players and generates matches as follows:
\begin{enumerate}
    \item For a given value of $\sigma$, assign three scores $[-\sigma, 0, \sigma]$ (representing player strengths $[e^{-\sigma}, 1, e^{\sigma}]$) to define the planted partial rankings.

    \item Randomly assign each of the $N$ players to one of the three rankings, ensuring that each rank includes at least one player.

    \item For a chosen number of matches $M$, randomly select pairs of players (with repetition) and simulate match outcomes using the BT model.
\end{enumerate}
By varying $M$ and $\sigma$, we generate networks of matches with different levels of sparsity and strength separation, allowing us to assess the performance of our algorithm across a wide range of scenarios. While the networks produced by this synthetic model differ from those typically observed in real data---in terms of density, clustering, or other structural features---working with such simplified models allows us to isolate and systematically examine how specific factors, namely the number of matches per pair and the separation of underlying player strengths, influence the algorithm’s performance. To that end, we adopt the simplest possible null model, in which matches are generated as random graphs, leaving all other structural features unspecified.

Results, shown in Fig.~\ref{fig:2_km_sigma_phase_heatmaps}, illustrate the algorithm’s ability to recover the planted rankings as a function of $\sigma$ and the average number of matches played per pair of nodes%
\footnote{%
In a simple graph—i.e., a network with no self-loops or multi-edges—$\langle m \rangle$ corresponds to the \emph{density} of the network, measuring the fraction of realized edges out of all possible edges. 
In our case, however, the network of matches is a \emph{directed multigraph}, where multiple games may occur between the same pair of teams. 
Hence, $\langle m \rangle$ quantifies the expected number of matches played by a pair of nodes chosen uniformly at random from the network. 
This can be seen as a natural generalization of network density to multigraphs, and we adopt it throughout this manuscript.%
}
\begin{equation}
    \langle m \rangle = \frac{M}{\binom{N}{2}},
\end{equation}
for $N=50$ and different total numbers of matches $M$.

For each $(\sigma, \expec{m})$ pair, $20$ different networks were generated, and the results were then averaged. The ranges of values for $\expec{m}$ and $\sigma$ were chosen so as to capture all the relevant behaviors of the model.

As expected, we observe that for low values of $\expec{m}$, there is not enough evidence to recover the planted rankings regardless of the strength separation between players. In such sparse conditions, the algorithm assigns all players to the same rank, as shown in Fig.~\ref{fig:2_km_sigma_phase_heatmaps}(a). This behavior contrasts sharply with the BT model, which still produces a complete ranking of the nodes even in these highly sparse conditions. As the number of matches increases and the networks become denser, the model gains sufficient statistical evidence to distinguish between the rankings and eventually recovers the planted structure. Notably, the number of matches required to correctly recover the rankings decreases rapidly with greater strength separation between players. In these cases, stronger players consistently outperform weaker ones, creating clearer signals that enable the ranking structure to be resolved with less data.

Perhaps more notable is the fact that our model does not consistently outperform the BT model in scenarios where players are nearly equal in strength. Fig.~\ref{fig:2_km_sigma_phase_heatmaps}(b) shows a heatmap of the log posterior odds ratio, defined as
\begin{equation}\label{eq:PORbtpr}
    \Delta\mathcal{L} = \log \frac{P_{PR}(\boldsymbol{\pi | A})}{P_{BT}(\boldsymbol{\pi | A})},
\end{equation}
as a function of $\expec{m}$ and $\sigma$. We observe that, when the separation in player strengths is small, our partial rankings model outperforms BT in sparse networks. In these cases, the PR model effectively leverages its ability to group players into shared ranks, avoiding overfitting. However, as the networks become denser, while our model accurately recovers the planted rankings even in this low-separation regime, it does not offer a more parsimonious description of the data (in terms of the posterior odds ratio of Eq.~\eqref{eq:PORbtpr}) compared to the standard BT model which, in contrast, infers a fully ranked structure for the nodes with $R = N = 50$. This suggests a noisy regime where the evidence supporting partial rankings is relatively weak. The BT likelihood provides a more flexible fit for fine-gained variations in the data in these dense regimes (effectively overfitting the data), even if its prior poorly reflects the underlying data.

In Fig.~\ref{fig:2_km_sigma_phase_N150R150} we analyze the performance of both models in the absence of partial rankings, using networks of $N = 150$ players, each assigned a unique score sampled from a logistic distribution with mean zero and scale one---a setting in which the BT model is expected to accurately capture the data. Averaged over 20 network realizations, the results show the behaviour of our PR model and the \emph{best performing model}, defined as the model (BT or PR) with the smallest negative log-posterior. For small values of $\expec{m}$, our PR model offers a more parsimonious fit, as the limited number of interactions prevents reliable inference of a complete ranking. Consequently, the log-posterior-odds ratio is positive in this regime, and the best performing model coincides with PR. As $\expec{m}$ increases, however, we observe a sharp transition in which the BT model quickly surpasses the PR model in explanatory power---the log-posterior odds ratio becomes negative, and the best performing model switches to BT, which infers $R = N = 150$ distinct rankings. We also observe that the PR model continues to infer more rankings as $\expec{m}$ grows, albeit very slowly, indicating that, in principle, it is capable of recovering the full ranking given sufficient data. We note that care has to be taken with large values of $R$ at high $\expec{m}$, as, in these large data regimes the influence of the prior terms on the log posterior diminishes and so does the regularization they provide. See Appendix~\ref{appendix:appendix_c} for a more detailed analysis of the behavior of the model for large $R$ and $\expec{m})$.

Interestingly, as the inferred number of groups increases with $\expec{m}$, the log posterior odds ratio decreases. This is somewhat counterintuitive, as one might expect the posterior odds ratio to improve as the number of rankings inferred by the PR model approaches the true number of rankings. However, as mentioned in Sec.~\ref{sec:methods}, the PR model tends to heavily penalize the presence of a large number of rankings so that the net effect of increasing the number of rankings is a reduction in the posterior odds ratio, with the PR model becoming a less efficient encoding for the rankings as the number of inferred groups grows. In practice, the best practice is to evaluate both an original ranking method and our corresponding partial ranking method to determine which provides a more parsimonious fit to the data in terms of log-posterior odds. 

To assess the goodness of the recovered partitions, in Appendix~\ref{appendix:appendix_d} we compute the Kendall rank correlation coefficient ($\tau_B$), adjusted for ties, between the ground truth rankings and the rankings inferred by the BT and PR models. This comparison is performed both when partial rankings are present and when they are not. We observe that when partial rankings are present in the network, the PR model can quickly attain perfect recovery as the networks become denser. On the other hand, while the BT model still achieves high $\tau_B$ scores, it never achieves perfect ordinal association and always underperforms relative to the PR model. However, when no partial rankings are present, and each player is assigned a unique score, the roles are reversed, and BT always displays a higher ordinal association with the ground truth rankings. We also compare the player strengths inferred by the PR and BT models to the ground truth, finding that both models recover scores that closely match the planted values. However, the PR model is able to leverage the coarse grained structure of the player strengths to group them into partial rankings.

We also compare our model with the partial rankings derived by applying 1D mean shift (MS) clustering~\cite{comaniciu2002mean} to the rankings inferred by the BT model in Appendix~\ref{appendix:appendix_e}. We observe that our algorithm can always recover the correct number of partial rankings when these are present, while MS fails to do so. More importantly, our algorithm can adjust the number of inferred clusters as more data becomes available, even in cases when no partial rankings exist, resulting in a more accurate description of the rankings. In contrast, mean shift clustering consistently infers roughly the same number of partial rankings regardless of the evidence provided by the data.

\subsection{Real Match Datasets}
\label{sec:real}

\begin{figure*}
    \centering
    \includegraphics[width=\textwidth]{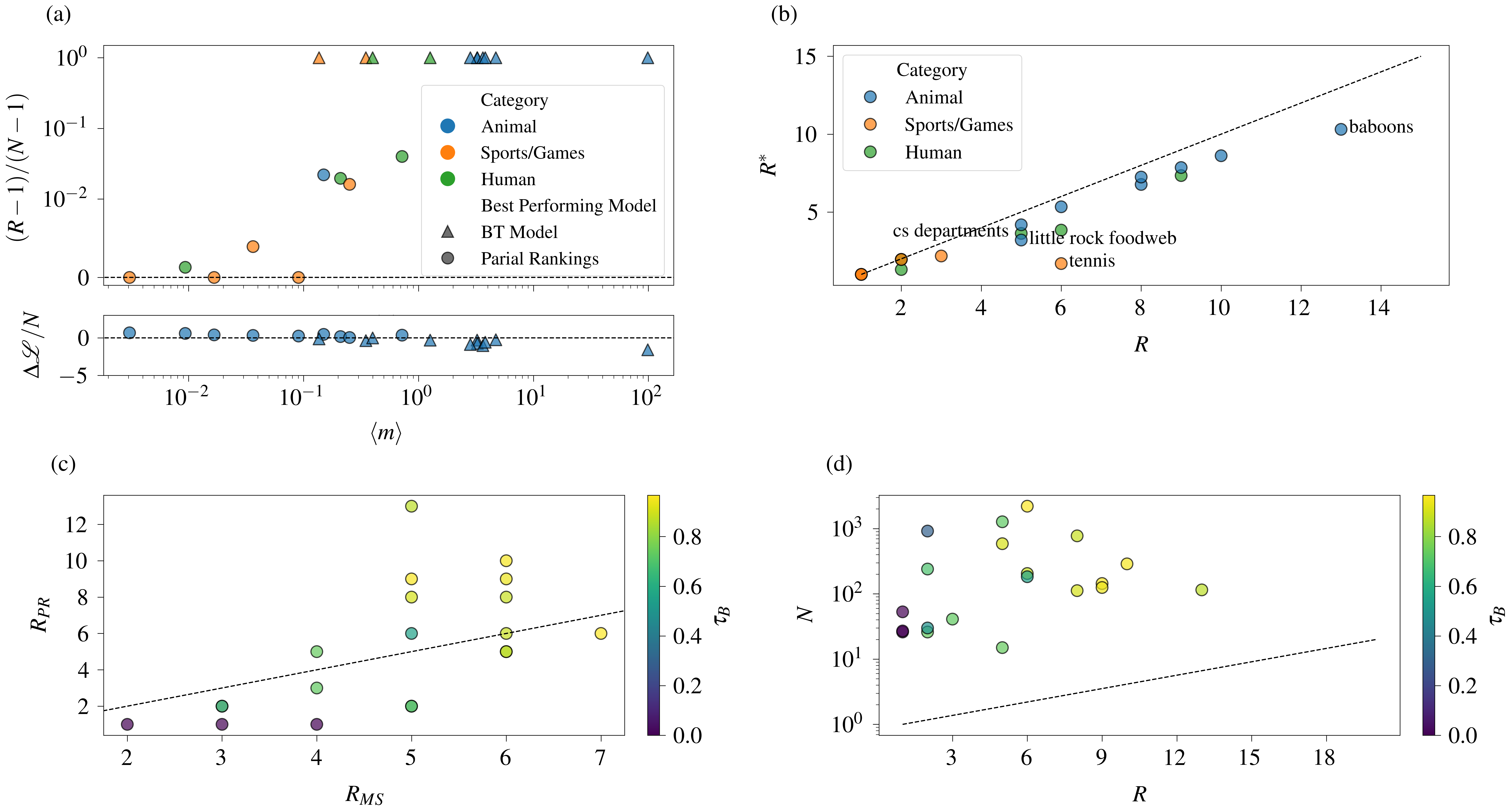}
    \caption{\textbf{Partial rankings in real networks of pairwise comparisons.} \textbf{(a)} (top) Rescaled number of ranks per node inferred by the best-performing model and (bottom) log posterior odds ratio per node, both as a function of $\expec{m}$ for all real-world networks considered in the study (see Table~\ref{tab:datasets}). The colours of the points indicate the different categories to which the datasets belong, while the shape of the markers indicates which model (PR or BT) emerged as the best-performing method (in terms of the posterior odds ratio). Positive values of the log posterior odds ratio indicate evidence in favor of the PR model; negative values indicate evidence in favor of the BT model. \textbf{(b)} Effective number of ranks, $R^*$ (Eq.~\eqref{eq:R_eff}), as a function of the number of unique ranks $R$ inferred by the PR algorithm. The black dashed line represents the line $R^* = R$. \textbf{(c)} Number of rankings inferred via the partial rankings algorithm as a function of the number of rankings inferred via mean shift clustering. Point colors indicate the $\tau_B$ score between the two inferred rankings. The black dashed line represents the line $R_{PR} = R_{MS}$. \textbf{(d)} Number of nodes of each network as a function of the number of rankings $R$ inferred by the partial rankings algorithm. Point colors indicate the $\tau_B$ score between the rankings inferred via the PR algorithm and those inferred by the BT model. The dashed black line indicates the line $R = N$.}
    \label{fig:real_world_data}
\end{figure*}

\begin{table*}[!ht]
\setlength\tabcolsep{2.14pt} 
\scriptsize\centering
\begin{tabular*}{\textwidth}{| l | c | c | c | c | c | c | l | l | c|}
\hline
  Data set  & $N$ & $M$ & $\expec{m}$ & $R$ & $R^*$ & $\Delta \mathcal{L}$ & Category & Description & Ref. \\
\hline
  Soccer& 2204 & 7438 & 0.003  & 1 & 1 & 1469.3 & Sports/games & Men's international association football matches 2010 - 2019 & \cite{soccer} \\
  Friends& 774 & 2799  & 0.009 & 2 & 2.0 & 450.1 & Human & High-school friend nominations & \cite{high_school_friendships} \\
  Tennis& 1272 & 29397 & 0.036 & 6 & 1.7 & 404.9 & Sports/games & Association of Tennis Professionals matches 2010–2019 & \cite{tennis} \\
  Chess & 917 & 7007 & 0.017 & 1 & 1 & 357.5 & Sports/games & Online chess games on lichess.com in 2016 & \cite{chess} \\
  Little Rock food web& 183 & 2494 & 0150 & 5 & 3.2 & 83.7 & Animal & Food web among the species in Little Rock Lake in Wisconsin & \cite{martinez1991artifacts} \\
  CS departments& 205 & 34388 & 0.210 & 5 & 3.6 & 33.4 & Human & PhD graduates of one department hired as faculty in another & \cite{clauset2015systematic} \\
  College football& 115 & 593 & 0.090 & 1 & 1 & 26.8 & Sports/games & NCAA College Football matches 2013-2023 & \cite{collegefootball} \\
  Dutch school friends & 26 & 234 & 0.720 & 2 & 1.3 & 9.2 & Human & Friendships at secondary school in The Netherlands, 2003-2004 & \cite{snijders2010introduction} \\
  Video games& 125 & 1951 & 0.252 & 3 & 2.2 & 4.7 & Sports/games & \textit{Super Smash Bros Melee} tournament matches in 2022 & \cite{videogames} \\
  History departments& 144 & 4112 & 0.400 & 6 & 3.8 & -3.2 & Human &  PhD graduates of one department hired as faculty in another & \cite{clauset2015systematic} \\
  Hyenas & 29 & 1913 & 4.712 & 9 & 7.9 & -7.6 & Animal & Dominance interactions among hyenas in captivity & \cite{strauss2019social} \\
  Sparrows & 26 & 1238 & 3.809 & 8 & 7.3 & -15.4 & Animal & Attacks and avoidances among sparrows in captivity & \cite{watt1986relationship} \\
  Baboons & 53 & 4464 & 3.239 & 13 & 10.3 & -16.3 & Animal & Dominance interactions among baboons in captivity & \cite{franz2015self} \\
  Dogs & 27 & 1143 & 3.256 & 6 & 5.3 & -20.3 & Animal & Aggressive behaviors in a group of domestic dogs & \cite{silk2019elevated}\\
  Wolf & 15 & 10382 & 98.876 & 10 & 8.6 & -24.0 & Animal & Dominance interaction among wolves in captivity & \cite{van198711}\\
  Mice & 30 & 1230 & 2.828 & 5 & 4.2 & -26.8 & Animal & Dominance interactions among mice in captivity & \cite{williamson2016mouse} \\
  Business departments & 112 & 7856 & 1.264 & 9 & 7.3 & -35.2 & Human & PhD graduates of one department hired as faculty in another & \cite{clauset2015systematic} \\
  Vervet monkeys & 41 & 2980 & 3.634 & 8 & 6.8 & -42.7 & Animal & Dominance interactions among a group of wild vervet monkeys & \cite{vilette2020comparing} \\
  Scrabble & 587 & 23477 & 0.137 & 2 & 2.0 & -86.0 & Sports/games & \textit{Scrabble} tournament matches 2004–2008 & \cite{scrabble} \\
  Basketball & 240 & 10002 & 0.349 & 2 & 2.0 & -90.7 & Sports/games & National Basketball Association games 2015–2022 & \cite{basketball} \\
  Soccer (aggregated) & 288 & 7438 & 0.180 & 6 & 4.8 & -131.1 & Sports/games & Soccer dataset above aggregated across 2010 - 2019 & \cite{soccer}\\
\hline
\end{tabular*}
\caption{Datasets analysed using the PR and BT algorithms, in order of decreasing log posterior odds ratio in Eq.~\eqref{eq:PORbtpr} (i.e. increasingly in favor of the original BT model). $N$, $M$, and $\expec{m}$ indicate the number of nodes, edges, and \emph{density} of each network, respectively. $R$ and $R^*$ indicate the number of partial rankings inferred via the PR algorithm and the effective number of partial rankings (Eq.~\eqref{eq:R_eff}). $\mathcal{L}$ indicates the logarithm of the posterior odds ratio between the PR and BT models (Eq.~\eqref{eq:PORbtpr}).} \label{tab:datasets}
\end{table*}

In this section, we focus on how our PR algorithm may be used to derive patterns in real-world data. We consider the datasets compiled in~\cite{jerdee2024luck} and available at~\cite{jerdeegit}. In addition, we also consider a dataset of social hierarchies among a pack of wolves~\cite{van198711, newman2023efficient}, and other sets of directed networks available at~\cite{netzschleuder}. The datasets are summarized in Table~\ref{tab:datasets}.

We apply our PR algorithm and the BT model to all match lists in the dataset and compare the results. In Fig.~\ref{fig:real_world_data}(a), we show the number of inferred ranks per node and the value of the log posterior odds ratio per node as a function of $\expec{m}$ for each network. To standardize the number of ranks per node within the interval $[0, 1]$, we rescale it as $(R-1) / (N-1)$, where $N$ is the number of nodes in the network and $R$ is the total number of ranks inferred by the best-performing model. This quantity will be $1$ for perfect orderings in which $R = N$. It will be $0$ if no rankings are inferred and all nodes are assigned the same strength, and it will take intermediate values when non-trivial partial rankings are present. 
The data reveals three distinct phases based on the density of the underlying network. For low densities, $\expec{m}$, the evidence is insufficient to support any meaningful ranking, leading to all actors being assigned the same strength. As connectivity increases, we observe an intermediate phase, in which partial rankings can often offer a more parsimonious description of the data compared to a perfect ordering of the nodes. In this intermediate regime, non-trivial subdivisions emerge, where actors are grouped by strength: all nodes within a group share the same strength, but strengths vary across groups. Finally, as the networks become sufficiently dense, enough data is available to support a perfect ordering of the nodes, and the BT model provides a more parsimonious description of the data.

To evaluate the homogeneity in the distribution of rank sizes, we adapt the concept of the \emph{effective number of groups} from the community detection literature~\cite{riolo2017efficient} and define the \emph{effective number of rankings}. If $R$ is the number of rankings inferred by the PR algorithm, we can express the effective number of rankings $R^*$ as
\begin{equation} \label{eq:R_eff}
    R^* = \exp\left(-\sum_{r=1}^R\frac{n_r}{N}\log\frac{n_r}{N}\right),
\end{equation}
where $n_r$ is the size of rank $r$ and $N$ is the total number of nodes in the network. This measure reflects the balance in the size distribution of rankings: $R^*$ will achieve a maximum value of $R$ when all the rankings are of equal size. In contrast, it approaches a minimal value close to one when the distribution is highly imbalanced, with most nodes concentrated in a single rank. Fig.~\ref{fig:real_world_data}(b) shows $R^*$ as a function of the inferred partial rankings $R$ for the PR model. While for most datasets $R^* \simeq R$, indicating that in most cases the PR algorithm infers roughly equally sized rankings, there are some notable exceptions, most notably the dataset of ATP tennis matches between 2010 and 2019, which displays a considerably smaller effective number of groups, reflecting a pronounced imbalance in the sizes of the inferred ranks. Specifically, the strongest 15 players are distributed across three partial rankings, while over $84\%$ of all players are grouped into the weakest rank. This skewed decomposition likely stems from the structure of professional tennis tournaments. ATP matches are generally organized in a knock-out format, where losers are progressively eliminated. While this structure is efficient and straightforward, it can result in noisy rankings, as strong players may be eliminated early due to random fluctuations or unfavorable matchups. To mitigate this, the ATP organizes multiple knock-out tournaments throughout the year, ensuring that the strongest players consistently rise to the top despite potential variability in individual tournaments. However, this system disproportionately concentrates matches among the strongest players, who predominantly compete against one another as they progress through the brackets. 
Consequently, the dataset contains ample information to distinguish the strongest players but far less data to differentiate among weaker players, leading to highly unbalanced partial rankings in terms of size.

Again, we compare the results obtained via our PR algorithm with those obtained by applying 1D Mean Shift clustering to the rankings inferred by the BT model. In Fig.~\ref{fig:real_world_data}(c), we plot the number of partial rankings, $R_{PR}$, inferred by the PR algorithm against the number of partial rankings, $R_{MS}$, inferred by Mean Shift. While a general positive correlation is observed between the two quantities, there is significant variability in the number of rankings inferred by the two methods. Moreover, even when the total number of rankings inferred by both algorithms is similar, the rankings themselves may differ substantially, as evidenced by the Kendall rank correlation ($\tau_B$) scores also shown in Fig. \ref{fig:real_world_data}(c). Specifically, $\tau_B$ scores appear to increase with the total number of groups inferred by both algorithms, regardless of how close $R_{PR}$ and $R_{MS}$ are. This suggests that the increase in $\tau_B$ scores is likely an artefact resulting from the fact that the algorithms are likely to align on most pairwise relationships as the number of groups grows rather than reflecting a similarity in the group structures inferred by the two approaches.

Finally, we also compare the results inferred by our algorithm with those obtained by the BT model. Fig.~\ref{fig:real_world_data}(d) shows the number of nodes $N$ in each network as a function of the rankings $R$ inferred by the partial rankings algorithm, where the points are colored according to the $\tau_B$ score between the rankings inferred by the PR and BT models. Again, we observe a general tendency for the $\tau_B$ scores to increase as $R$ increases.

Across real datasets from various domains we have found the emergence of three distinct regimes in rank detectability as a function of the network density. For low densities, there is too little statistical evidence available for distinguishing rankings and we infer a single partial rank, with the partial rankings model being preferred over the standard BT model. In an intermediate density regime of around $\expec{m}\simeq 0.02$ matches per node pair, we find a modest number of partial ranks much less than $N$, and that the partial rankings model is often more parsimonious than BT in terms of posterior odds. Finally, for higher densities, we find that the BT model is most parsimonious, so that partial rankings are no longer needed as there is sufficient evidence in the data from the matches to distinguish the rankings of the nodes. These results are consistent with those in Sec.~\ref{sec:synthetic}. We also find of interest the considerable heterogeneity in inferred cluster sizes, suggesting that the underlying topology of the interaction network can influence the inferred partial rankings. This is of particular interest for ranking models such as BT that do not model edge placement, i.e. how the matches are decided. We additionally identify substantial discrepancies between the results of our pairwise partial ranking and those obtained by simply performing 1D clustering of the ranks, which highlights the importance of properly regularized ranking approaches that infer ties simultaneously while inferring the rankings. These partial rankings are shown to have increased correlation with the standard BT ranks as the number of ties approaches zero.

\subsection{Case Study: CS Faculty Hiring}
\label{sec:casestudy}

\begin{figure*}
    \centering
    \includegraphics[width=0.8\textwidth]{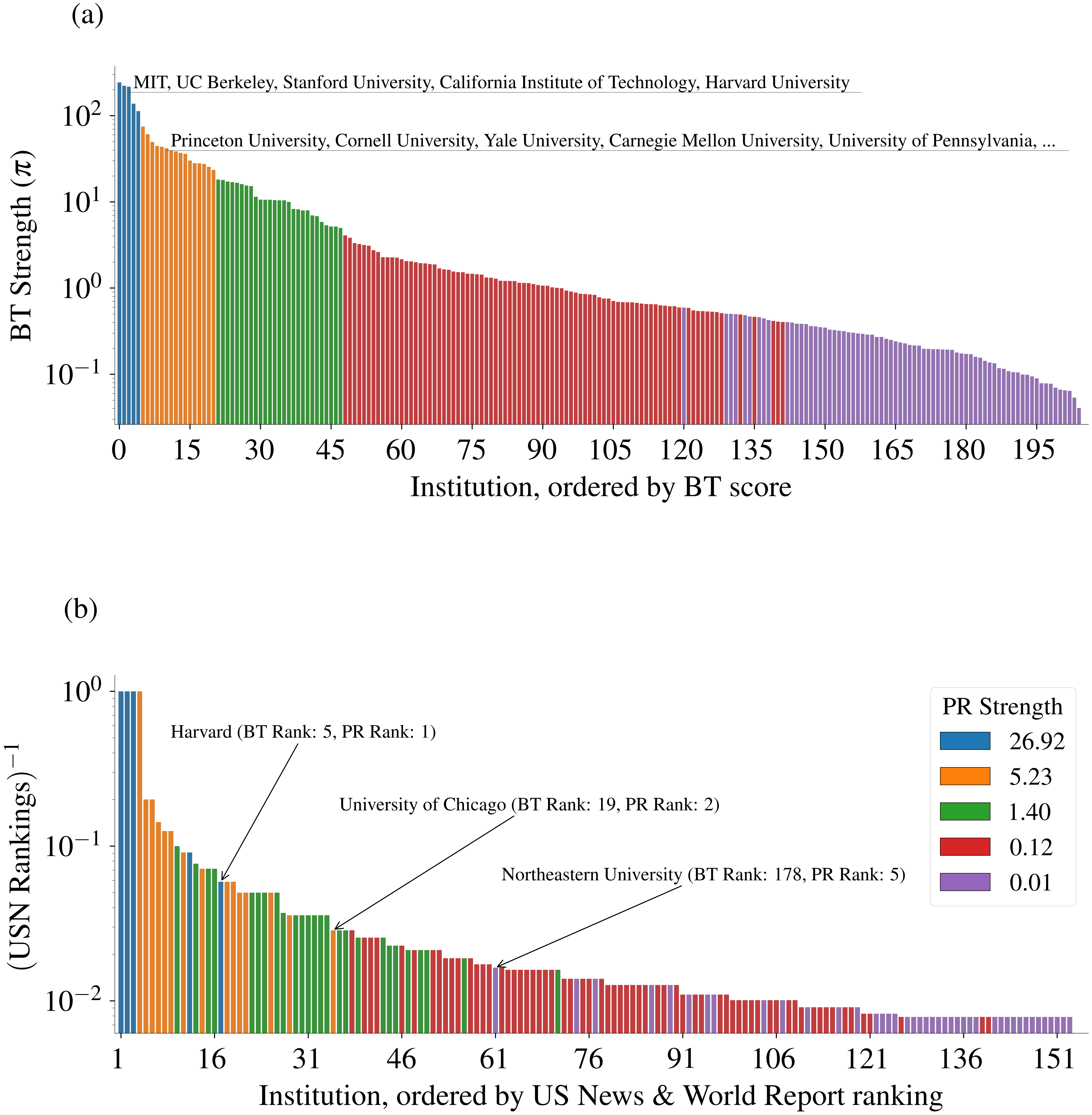}
    \caption{\textbf{Partial rankings of CS departments according to faculty hiring patterns.} \textbf{(a)} Barplot of BT strengths ($y$-axis) for the 205 PhD-granting institutions included in the dataset, ordered according to their inferred BT rankings ($x$-axis). Colors indicate the PR membership of each institution. A legend displaying the numerical values of these PR strengths is shown in panel (b). The names of the five institutions making up the strongest PR cluster and of the first five institutions in the second-strongest PR cluster (ordered in terms of their BT rank) are shown in the figure. \textbf{(b)} Barplot depicting the PR groups (colored as before) with respect to the USN ranking ($x$-axis) instead of BT rank, for the 153 institutions for which USN data was available. Some notable inconsistencies between the USN rankings of some institutions and those inferred by pairwise comparison methods are shown in the figure. The inverse of USN ranking was plotted along the $y$-axis to provide an analogue to the BT score for these pre-determined rankings for easier visualization.}
    \label{fig:cs_depts_barplots}
\end{figure*}

\begin{figure*}
    \includegraphics[width=\textwidth]{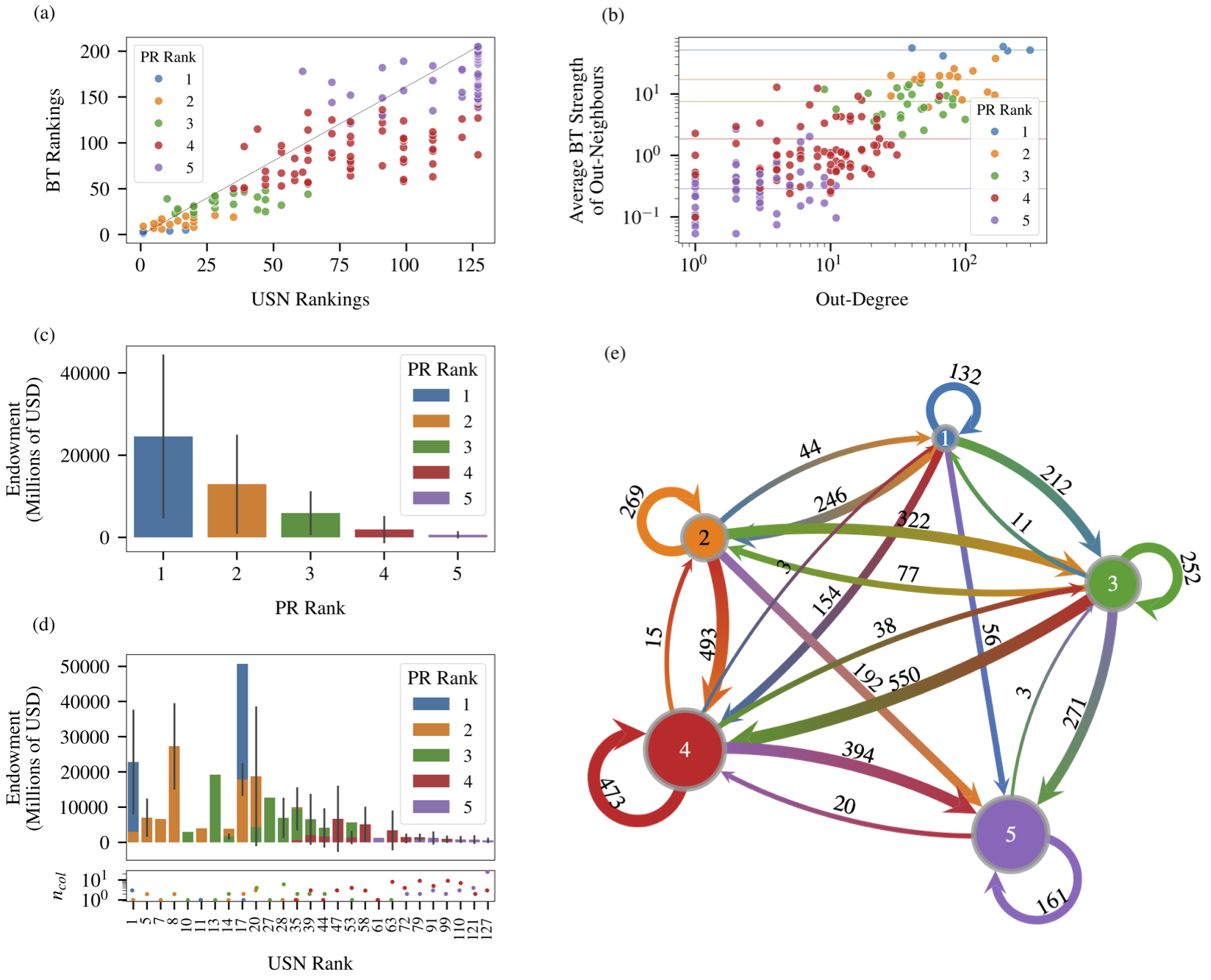}
    \caption{\textbf{Partial rankings of CS departments and their external metadata.} \textbf{(a)} BT rankings as a function of USN rankings for all 153 PhD-granting institutions that had an associated USN ranking. Points are colored according to their PR rank. The black dashed line corresponds to the case $R_{BT} = R_{USN}$. \textbf{(b)} Average BT strength of a node's out-neighbors as a function of the node's out-degree. Points are colored according to their PR rank. The horizontal lines represent the average BT score of the out-neighbors disaggregated by PR group. \textbf{(c)} Average endowment (in millions of USD) for the institutions in each partial rank. Error bars represent the standard deviations within each partial rank group. \textbf{(d)} (Top panel) Average endowment (in millions of USD) as a function of the USN rank, disaggregated by partial rankings. Error bars represent the standard deviations within each group of universities binned by USN rank (including ties). (Bottom panel) Strip plot of the the number of institutions within each (USN rank, PR rank) group, i.e. the number of institutions belonging to each bar. \textbf{(e)} Network representation of faculty hiring flows among the partial rank groups inferred by our algorithm. Each node represents a partial ranking, with labels and colors indicating the corresponding group. Node sizes are proportional to the number of institutions assigned to each partial ranking. Directed edges represent the hiring flows between groups, with edge thickness proportional to the net flow. Edge colors transition from the origin node’s color at the base to the destination node’s color at the arrowhead and the net flow value is annotated on each edge.}
    \label{fig:cs_depts_panel2}
\end{figure*}

As a case study, we apply our algorithm to a dataset of faculty hires in Computer Science (CS) departments at U.S. higher education institutions, originally compiled in \cite{clauset2015systematic} for analyzing hierarchies in academic hiring using a different ranking methodology. Each node in the network represents one of the 205 PhD-granting institutions included in the study, and a directed edge $(i, j)$ is included for each individual who received their doctorate from institution $i$ and held a tenure or tenure-track faculty position at institution $j$ during the data collection period (May 2011 to March 2012). For 153 of these institutions, we have access to their \textit{U.S. News \& World Report} (USN) rankings, which we use as node metadata to compare against the rankings and partial rankings inferred by the two pairwise ranking methods explored in this study. 

The USN rankings are constructed from a broad set of indicators---including reputation surveys, research output, student selectivity, and institutional resources---and therefore reflect a metrics-based assessment of institutional prestige. In contrast, the rankings inferred from hiring data capture a more behavioral notion of prestige, one that emerges from the collective decisions of faculty members and hiring committees. Comparing these two types of rankings allows us to identify not only their areas of agreement but also systematic differences between them, revealing how the hierarchies implied by hiring dynamics may diverge from those defined by traditional metrics. These differences highlight the distinction between perceived prestige, i.e., how academics act upon their professional judgments when hiring or accepting positions, and evaluated prestige, i.e., how institutions are ranked based on quantifiable performance indicators such as publications, grants, and resources. Additionally, we manually collected endowment data for 151 of the 153 institutions with USN rankings to explore potential correlations with inferred prestige.

Fig.~\ref{fig:cs_depts_barplots}(a) shows the results of applying the BT model and our PR algorithm to the entire set of 205 PhD-granting institutions. Each bar represents an institution, ordered by its BT rank, with bar heights representing the BT strength scores. Bars are color-coded according to their PR rankings, with corresponding strengths shown in the legend in Fig.~\ref{fig:cs_depts_barplots}(b). The rankings are dominated by a small group of elite universities displaying significantly higher strengths, followed by a sharp decline in strength values further down the list. The PR and BT rankings show strong ordinal association ($\tau_B = 0.82$; p-value $= 1.2 \times 10^{-53}$) with only a small number of rank violations---instances where $\text{rank}_{BT}(A) > \text{rank}_{BT}(B)$ but $\text{rank}_{PR}(A) < \text{rank}_{PR}(B)$--- at the boundary between the two lowest-scoring PR groups, possibly indicating a lack of information to accurately determine the group boundaries.

In Fig.~\ref{fig:cs_depts_barplots}(b), we compare the partial rankings inferred by our model with the \textit{U.S. News \& World Report} rankings. Each bar in the plot represents a PhD-granting institution, ordered by its USN rank, with bar heights corresponding to the inverse of their USN rank. Again, bars are color-coded according to their partial rankings inferred by the PR model.

One first thing to notice with regards to the USN rankings is that they allow for partial rankings. For example, the four highest-scoring universities according to the USN methodology are all assigned an equal rank of 1. Fig.~\ref{fig:cs_depts_barplots}(b) reveals that institutions with equal USN ranking often belong to the same partial rankings according to the PR model. However, the PR model generally produces broader rankings, grouping institutions that span a wider range of USN ranks, suggesting that hiring-based prestige hierarchies are consistent with, but less granular than, those defined by metrics-based rankings, reflecting a structure of academic reputation that operates at the level of broad institutional tiers rather than fine rank distinctions. Another striking feature is the significantly higher number of rank violations, although the overall ordinal association remains strong ($\tau_B = 0.76$; p-value $= 1.37 \times 10^{-33}$). Unlike in the BT case, where rank violations likely result from limited information, the presence of rank violations across the entire ranking spectrum suggests that they reflect genuine differences between evaluated prestige---as measured by metrics-based rankings---and perceived prestige---as revealed through hiring dynamics. Indeed, some of these rank violations are particularly striking. For example, Harvard University, ranked $5^{th}$ in the BT model and belonging to the top-scoring PR group, is ranked $17^{th}$ according to the USN rankings, behind institutions including the University of Maryland, College Park ($26^{th}$ in the BT model and in the third PR group) that scored significantly lower according to the hiring network structure. Similarly, the University of Chicago is ranked $19^{th}$ by the BT model and belongs to the second-highest-scoring PR group, but is ranked $35^{th}$ in the  USN rankings, suggesting that these factors may not directly align with academic job market preferences. In the case of Harvard University or the University of Chicago, one might surmise that the prestigious reputation of the institutions might be influencing the job market more than their USN rankings. The opposite effect can also be observed. For instance, Northeastern University, ranked $61^{st}$ in the USN rankings, is assigned $178^{th}$ place by the BT model and belongs to the lowest PR group despite its considerably higher USN ranking. This suggests that perhaps this program was undervalued in the job market relative to its academic credentials during the period studied.

To evaluate how the BT rankings compare to the USN rankings, Fig.~\ref{fig:cs_depts_panel2}(a) plots the BT rank of each institution against its corresponding USN rank, with points color-coded based on their PR rank. The highest-ranking institutions, according to both models, tend to concentrate around the $R_{BT} = R_{USN}$ line, where $R_{BT}$ and $R_{USN}$ represent the BT and USN rankings respectively---indicating close agreement between the two ranking systems for top-tier institutions. As we move towards lower-ranked institutions, we observe an increasing divergence between the rankings. While the overall ordinal association remains strong ($\tau_B = 0.73$, p-value $= 1.07 \times 10^{-38}$), the spread in rankings widens, and institutions with a given USN rank are often assigned considerably different BT ranks, with the BT model frequently assigning these institutions higher positions compared to the USN rankings. This discrepancy is particularly pronounced among institutions in the two lowest-scoring PR groups. In general, these results suggest that, outside the top-ranked institutions, subjective perceptions of prestige can significantly shape professional mobility, elevating or diminishing institutions relative to their metric-based ranking. Likewise, the observation that institutions with similar BT ranks span a broad range of USN ranks indicates that perceived prestige, as expressed through hiring and faculty mobility, does not always correspond directly to quantitative measures of institutional performance.

An interesting question about the pairwise ranking methods considered in this work is whether an institution's rank benefits more from ``who you win against'' than from the sheer number of ``wins''. In other words, does an institution's ranking improve more by producing a large number of doctoral graduates who secure tenure-track positions across a wide range of institutions or by producing fewer graduates who secure positions at highly ranked institutions? To address this, in Fig.~\ref{fig:cs_depts_panel2}(b), we plot the average BT strength of an institution's out-neighbors (capturing the quality of an institution's ``wins'') as a function of its out-degree (representing the total number of ``wins'' an institution has). We observe a clear positive correlation (Pearson $p = 0.78$; p-value $= 1.32 \times 10^{-43}$) with the highest ranked institutions having both a large number of ``wins'' (i.e. producing a large number of graduates that go on to secure tenure-track positions at other institutions), as well as out-neighbors with high BT strengths (i.e. their graduates tend to secure positions at high-ranking institutions). Perhaps more revealing is the fact that the PR groups appear to stratify according to the average BT strength of their out-neighbors, suggesting a well-defined hierarchy characterized by a predominance in horizontal mobility as opposed to vertical mobility. 
This observation is reinforced by analyzing the network of mobility flows between the PR groups; see Fig. \ref{fig:cs_depts_panel2}(e). The results reveal a well-defined hierarchy with minimal upward mobility across ranks. Most mobility occurs downward within the hierarchy, a trend partly explained by the relatively small size of the top-ranked groups. Nonetheless, horizontal mobility emerges as a critical driver of hiring dynamics across all levels, particularly within the lower-ranked groups, where it becomes the dominant hiring pattern. Horizontal mobility is also particularly pronounced at the topmost level of the hierarchy, where the top five institutions comprising this elite group hire over twice as many faculty members from within their group as they do from all other PhD-granting institutions combined.

Finally, we note that the ranking patterns possess a moderately strong ordinal association with the endowments of the institutions. Although there is considerable variability, higher-ranked institutions generally tend to have significantly larger endowments, as illustrated in Fig.~\ref{fig:cs_depts_panel2} (c)-(d). The high variability can be partly attributed to differences in the number of institutions contributing to each (USN rank, PR rank) group---i.e., the number of universities represented by each bar. Group sizes vary considerably across the data set, although most bars consist of relatively few institutions (1–25 universities per bar; median = 2, mode = 1). As a result, standard deviations are generally uninformative for most bars, with variability estimates becoming more reliable only for larger groups.

\section{Conclusion}
\label{sec:conclusions}

We have introduced a probabilistic generative model for inferring partial rankings in directed networks and a fast, fully nonparametric, agglomerative algorithm for efficient inference. We have shown that, particularly with limited observations available, our model can provide a more parsimonious description of pairwise comparison data than models that inherently assume complete rankings, such as the BT model. Specifically, in extremely sparse regimes, our model effectively recognizes that insufficient evidence exists to infer any meaningful ranking among the compared entities. As the network's connectivity increases, partial rankings typically emerge as a more compact and accurate description of the data. Finally, as we move towards more dense regimes, we reach a point where sufficient information is available to infer a complete ordinal ranking of all the nodes. When applied to a network of faculty hiring among U.S. computer science departments, our algorithm inferred five distinct partial rankings, which align closely with the ordinal rankings inferred by the BT model and reveal a well-defined hierarchy dominated by a small number of elite universities. The inferred rankings also highlight the limited upward mobility within the hierarchy, with lateral and downward movements being more prevalent.

There are several directions in which our work can be extended. While widely used, the classical BT model represents a relatively simplistic approach to pairwise interactions between entities. Numerous extensions to the model have been proposed, including accounting for ties~\cite{rao1967ties, davidson1970extending}, home-field advantage~\cite{agresti2012categorical}, randomness in match outcomes, and imbalances in strength or skill between the average pair of  players~\cite{jerdee2024luck}. Any of these extensions could, in principle, be incorporated into our partial rankings framework, enabling the development of more expressive models.

Another characteristic of both the BT model and our partial rankings framework is that they do not explicitly account for the placement of edges in the network---that is, the likelihood does not include a term that models the probability of observing a given set of matches in the first place. As we observed in Sec.~\ref{sec:casestudy}, similarly (partially) ranked institutions tend to interact primarily with one another. A similar observation holds for the ATP tennis dataset, where the strongest players, who advance through the tournament, tend to play more matches and preferentially against each other. While other methods explicitly model edge placement \cite{de2018physical, peixoto2022ordered}, this aspect remains absent from the current BT and partial rankings models. Incorporating it into the framework could help to explore how network topology influences ranking recovery.

Our framework could also be extended to consider other cases of interest, such as dynamic rankings, where the ranks of the individual entities can rise or fall over time~\cite{della2024model}, personalised rankings, in which ranking results are tailored to individual users based on their preferences~\cite{rendle2012bpr}, or a combination of both.

Finally, while we have developed a fully Bayesian nonparametric framework to infer partial rankings, we have focused on inferring point estimates of these rankings without considering uncertainty quantification as done in other works~\cite{pearce2024bayesian, peixoto2022ordered}. However, as noted in Sec.~\ref{subsec:model}, the Bayesian nature of our model lends itself naturally to the exploration of the full posterior, which can be explored via the development of suitable inference algorithms---such as Expectation-Maximization (EM) or MCMC methods---to enable posterior sampling or approximation, thereby allowing for principled uncertainty quantification in inferred rankings.


\section*{Acknowledgments}
\vspace{-\baselineskip}
The authors thank Max Jerdee for useful discussions about the datasets, and Michael Pearce and Elena Erosheva for helpful suggestions to improve the initial manuscript. 






\section*{Data accessibility}
\vspace{-\baselineskip}
All data used in the study are open source and freely
available on the Internet. We have provided links and references in the text and electronic supplementary material for all the data used in the study.









\section*{Conflict of interest declaration}
\vspace{-\baselineskip}
We declare we have no competing interests.



\section*{Funding}
\vspace{-\baselineskip}
The authors acknowledge funding support from the HKU-100 Start Up Fund and an HKU Urban Systems Institute Fellowship Grant.




\appendix

\section{\label{appendix:objective_contributions} Minimization of $f(R)$}

We prove that the function
\begin{equation} \label{eq:f_R}
    f(R) = \log \binom{N - 1}{R - 1} + \log\binom{N}{n_1 \ldots n_R}
\end{equation}
has a global minimum at $R = 1$.

We begin proving that the second term in Eq.~\eqref{eq:f_R} is maximized when all groups are of equal size such that $n_r = n = N/R~~\forall r$.
\begin{proposition} \label{proposition:multinomial}
    Let $N \in \mathbb{Z}^+$ and let $\{n_r\} \in \mathbb{Z}^+$ be a set of positive integers such that $\sum_{r = 1}^R n_r = N$. Then the multinomial coefficient $\binom{N}{n_1 \ldots n_R}$ is maximized when $n_r = n = N/R~~\forall r$.
\end{proposition}
\begin{proof}
    By definition
    \begin{equation} \label{eq:multinomial_coeff}
        \binom{N}{n_1 \ldots n_R} = \frac{N!}{n_1! \ldots n_R!}.
    \end{equation}
    Using Stirling's approximation of the factorial,
    \begin{equation}
        \log k! \approx k \log k - k + \frac{1}{2}\log(2\pi k),
    \end{equation}
    we have that, to first order
    \begin{align}
        \binom{N}{n_1 \ldots n_R} &= \exp\left\{N \log N - N - \sum_{r=1}^R (n_r\log n_r - n_r)\right\} \\
        &= \exp\left\{N \log N - \sum_{r=1}^R n_r \log n_r\right\},
    \end{align}
    where we have used the fact that $\sum_r n_r = N$.

    To maximize Eq.~\eqref{eq:multinomial_coeff}, we then have to solve the following constrained optimization problem
    \begin{align}
        &\max_{\{n_r\}} \left(N\log N - \sum_{r=1}^R n_r \log n_r\right) \\
        &\textrm{s.t.} \quad \sum_{r=1}^R n_r = N,
    \end{align}
    which corresponds to extremizing the following Lagrangian
    \begin{equation}
        \mathcal{L}(\{n_r\}, \lambda) = -\sum_{r=1}^R n_r\log n_r + \lambda\left(\sum_{r=1}^R n_r - N\right),
    \end{equation}
    where we have ignored constant terms. Taking the partial derivatives with respect to $n_r$ and equating to zero, we have
    \begin{equation}
        \frac{\partial \mathcal{L}}{\partial n_r} = -(\log n_r + 1) + \lambda = 0 \implies n_r = e^{\lambda - 1} \equiv C.
    \end{equation}
    From the sum constraint, we then have
    \begin{equation}
        \sum_{r=1}^R n_r = RC = N \implies C = n_r = \frac{N}{R}~~\forall r.
    \end{equation}
\end{proof}
Given Proposition~\ref{proposition:multinomial} and in view of the monotonicity of the logarithm, it follows that the second term on the right-hand side of Eq.~\eqref{eq:f_R} is maximized when all groups are of equal size.

Given this result, we can now prove our original statement.
\begin{proposition}
    The function
    \begin{equation}
        f(R) = \log \binom{N - 1}{R - 1} + \log\binom{N}{n_1 \ldots n_R} \notag
    \end{equation}
    has a global minimum at $R=1$ with $f(R = 1) = 0$.
\end{proposition}
\begin{proof}
    From Proposition~\ref{proposition:multinomial} and by virtue of the monotonicity of the logarithm, it follows that Eq.~\eqref{eq:f_R} is upper bounded by
    \begin{equation} \label{eq:bound_1}
         f(R) \leq \log\binom{N-1}{R-1} + \log\frac{N!}{n!^R},
    \end{equation}
    with $n = N/R$. Eq.~\eqref{eq:bound_1} can be written as
    \begin{align}
        f(R) &\leq \log\binom{N-1}{R-1} + \log N! - R\log n! \\
        &\approx \log\binom{N-1}{R-1} + N\log N - N - R\left(\frac{N}{R}\log\frac{N}{R} - \frac{N}{R}\right) \\
        &=\log\binom{N-1}{R-1} + N\log R, \label{eq:bound_2}
    \end{align}
    where we have once more made use of Stirling's approximation. The first term on the right-hand side of Eq.~\eqref{eq:bound_2} is a concave function of $R$, reaching its minimum value of zero at both $R=1$ and $R=N$. The second term, however,  increases monotonically with $R$, starting from a minimum of zero when $R=1$. Since these two terms combine additively in Eq.~\eqref{eq:bound_2}, their sum must attain its global minimum of zero at $R = 1$, where both terms simultaneously achieve their lowest values. Finally, since the binomial and multinomial coefficients  are always greater than or equal to one, it follows that Eq~\eqref{eq:f_R} is also lower bounded by zero, meaning it must attain a global minimum of zero at $R = 1$.
\end{proof}

\section{\label{appendix:appendix_b} Time Complexity}

\begin{figure*}
    \centering
    \includegraphics[width=\textwidth]{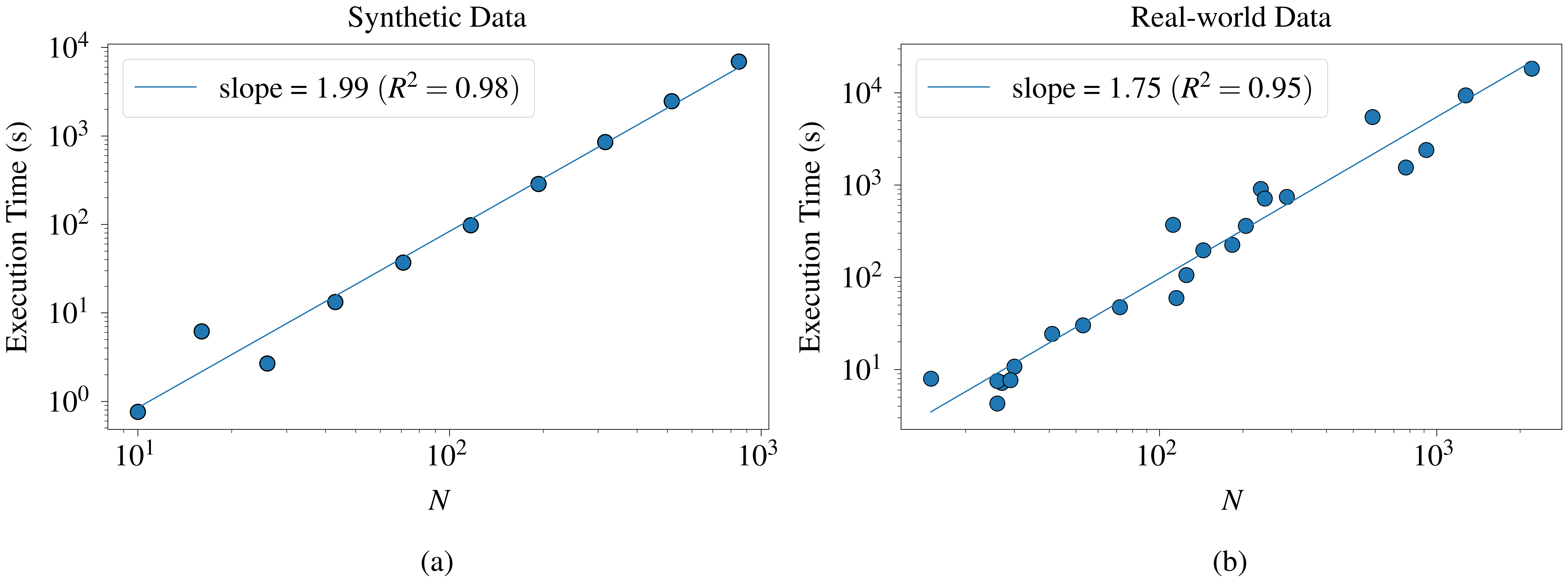}
    \caption{\textbf{Runtime scaling on real and synthetic networks.}
    \textbf{(a)} Runtimes of the partial rankings algorithm across a set of synthetically generated networks with average degree $\langle k \rangle = 28.5$, showing a quadratic scaling with the network size. \textbf{(b)} Runtimes of the partial rankings algorithm across all empirical networks considered in this study, displaying a scaling of $\sim O(N^{1.75}) < O(N^2)$, with variability attributable to heterogeneity in the edge densities of the real datasets.
    }
    \label{fig:rw_time_complexity}
\end{figure*}

In Fig.~\ref{fig:rw_time_complexity}, we plot the runtime of our algorithm as a function of the number of nodes $N$ in the network for both synthetically generated data (Fig.~\ref{fig:rw_time_complexity}(a)) and empirical networks (Fig.~\ref{fig:rw_time_complexity}(b)). The results display an approximately quadratic scaling of the runtime with $N$ (see Sec.~\ref{sec:methods}), with some variation in the real datasets due to varying densities of edges present.

\section{\label{appendix:appendix_c} Increasing $R$}

\begin{figure*}
    \centering
    \includegraphics[width=\textwidth]{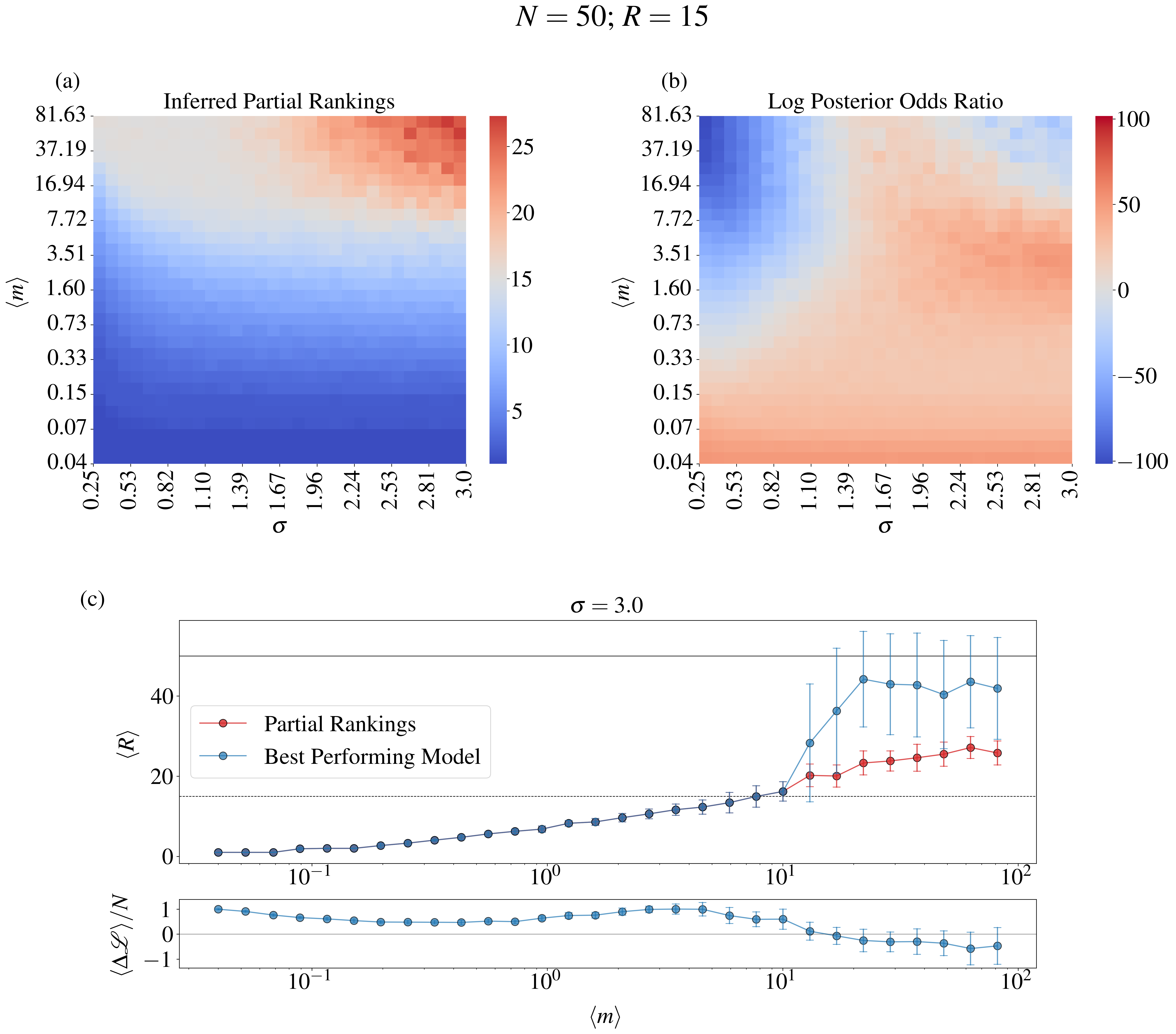}
    \caption{\textbf{Partial rankings in synthetic datasets with $N = 50$ and $R=15$.} \textbf{(a)} Heatmap of the number of rankings $R$ inferred by the partial rankings model, as a function of the parameters $(\sigma, \expec{m})$ of the synthetic model in Sec.~\ref{sec:synthetic} when $N = 50$ and $R = 15$. \textbf{(b)} Heatmap of the log posterior odds ratio (Eq.~\eqref{eq:PORbtpr}) between the BT and the partial rankings model across the simulations. Positive values indicate a preference for the partial rankings model, and negative values a preference for the BT model. \textbf{(c)} The top panel displays the average number of rankings inferred by the best-performing model (blue) and our partial rankings model (red) as a function of the network density $\expec{m}$ of the network of matches for the slice at $\sigma = 3.0$. The bottom panel displays the log posterior odds ratio (normalized per node) between the two models. Negative values of this difference indicate a preference for the BT model and positive values a preference for the partial rankings model. All results were obtained by averaging over $20$ different simulations from the synthetic network model of Sec.~\ref{appendix:appendix_c} with $N = 50$ and $R = 15$, and error bars indicate $2$ standard errors in the mean.}
    \label{fig:k_sigma_phase_R15}
\end{figure*}


\begin{figure*}
    \centering\includegraphics[width=\textwidth]{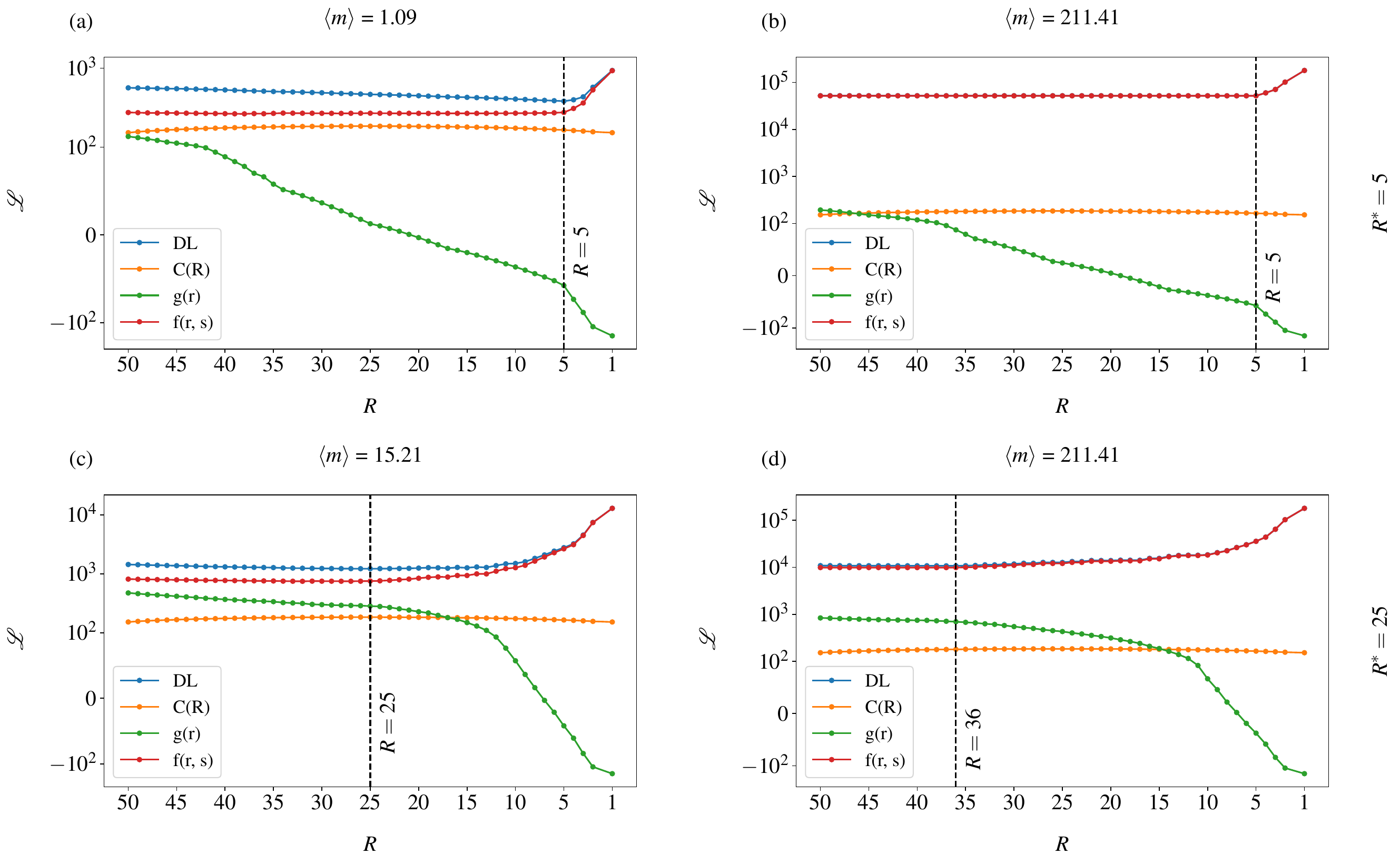}
    \caption{\textbf{Contributions to the inference objective}. Behavior of the log posterior $\mathcal{L}$ and the various contributions to $\mathcal{L}$---$C(R)$, $g(r)$, and $f(r,s)$--- as a function of the inferred number of rankings $R$ for different values of $\expec{m}$ and planted rankings $R^*$. The black vertical dashed lines indicate the global minimum of $\mathcal{L}$.}
    \label{fig:dl_decomp}
\end{figure*}

We analyze the recovery performance of our algorithm as the number of partial rankings increases. Fig.~\ref{fig:k_sigma_phase_R15} shows the behavior of the model for a network with $N = 50$ nodes and $R = 15$ planted rankings for different values of $\expec{m}$ and $\sigma$

By increasing the number of planted rankings $R$, we observe that the number of matches required to successfully recover the correct ranking structure also increases, especially at low noise levels ($\sigma$). This is expected, as distinguishing a larger number of groups demands more evidence in the form of pairwise comparisons.

At low $\sigma$, the algorithm continues to accurately recover the planted rankings across a wide range of conditions. However, the model still yields a less parsimonious description of the data compared to the BT model as observed in Fig.~\ref{fig:2_km_sigma_phase_heatmaps}. This is due to the inconsistency between the extensive number of rank groups and the prior, and can in principle be addressed with a weaker prior, e.g. one that is uniform over rank partitions rather than the number of ranks.

Increasing the score separation at high $R$ values pushes the algorithm to find a larger number of ranks than expected. This arises from the behavior of the log-likelihood in the high-$\expec{m}$ regime as the number of planted rankings $R$ increases. Figure~\ref{fig:dl_decomp} illustrates the log posterior $\mathcal{L}$, along with the contributions of its individual components---namely $C(R)$, $g(r)$, and $f(r,s)$---as functions of the number of inferred rankings $R$, for varying values of $\expec{m}$ and planted rankings $R^*$. Figures~\ref{fig:dl_decomp}(a) and (b) show the behavior of the algorithm at low and high values of $\expec{m}$, respectively, for $R^* = 5$ planted rankings with equally spaced scores.

At low $\expec{m}$, the three terms contribute to produce a global minimum of the log posterior at $R = R^* = 5$, thereby correctly identifying the true number of rankings. In contrast, in the high-$\expec{m}$ regime, the contributions from $C(R)$ and $g(r)$ diminish, and the log-posterior becomes dominated by the likelihood term $f(r, s)$. This marks the transition to a high-data regime where the influence of the prior effectively ``washes out'',  at which point our PR model behaves similarly to the BT model---which is prone to overfitting. Specifically, the likelihood term $f(r, s)$ decreases monotonically with increasing $R$, favoring more complex models.

However, as seen in Fig.~\ref{fig:dl_decomp}(b), this decrease is sufficiently slow that the prior still regularizes the model adequately, and the global minimum remains at $R = 5$. But this balance deteriorates as $R^*$ increases. While the behavior at low $\expec{m}$ remains qualitatively unchanged (Fig.~\ref{fig:dl_decomp}(c)), in the high-$\expec{m}$ regime the monotonic decrease of $f(r, s)$ with $R$ becomes steeper. As a result, the regularizing influence of the prior weakens further, reducing its ability to constrain model complexity and allowing the algorithm to overfit.

\section{\label{appendix:appendix_d} Tau Recovery}

\begin{figure*}
    \centering
    \includegraphics[width=\textwidth]{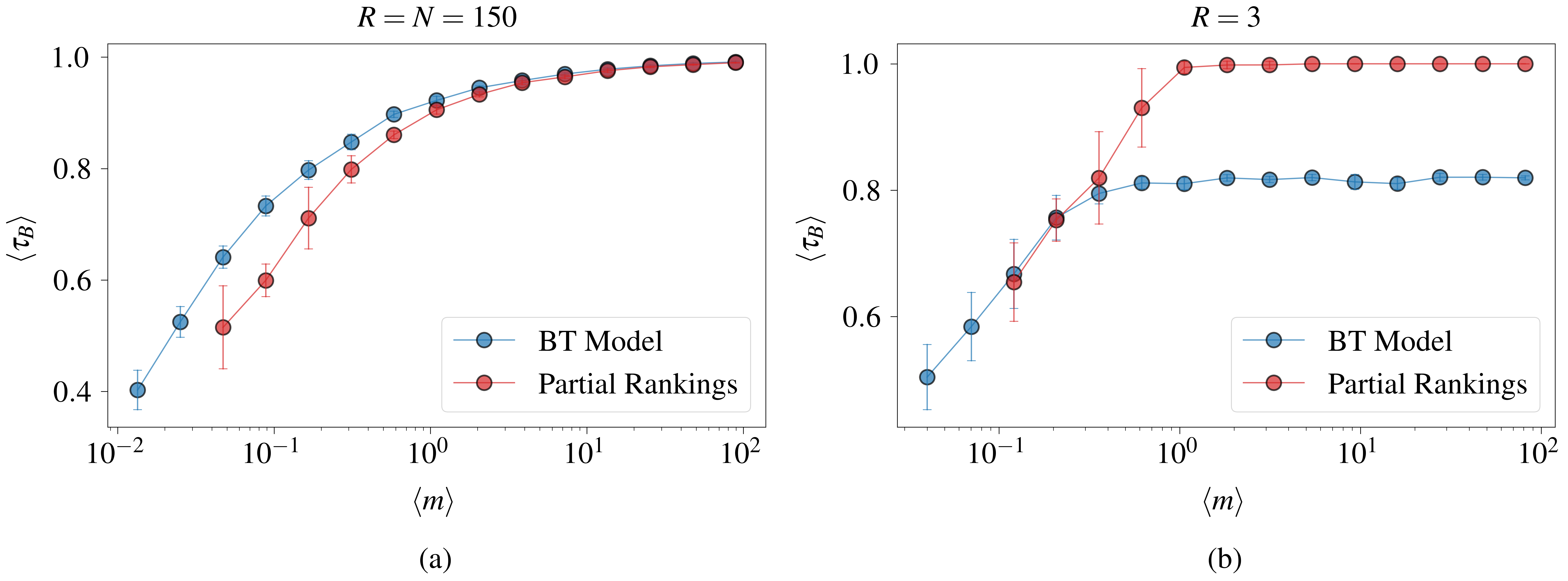}
    \caption{\textbf{Correlations of inferred rankings with planted rankings.} \textbf{(a)} $\tau_B$ scores between the rankings inferred by the BT and PR models with respect to the ground truth ranking for the case in which there are no partial rankings ($R = N = 150$). \textbf{(b)} $\tau_B$ scores between the rankings inferred by the BT and PR models with respect to the ground truth ranking for the case in which there are three planted partial rankings ($R = 3$). All results are averaged over 10 different simulations and error bars represent 2 standard errors from the mean.}
    \label{fig:app_tau_recovery}
\end{figure*}

\begin{figure*}
    \centering
    \includegraphics[width=\textwidth]{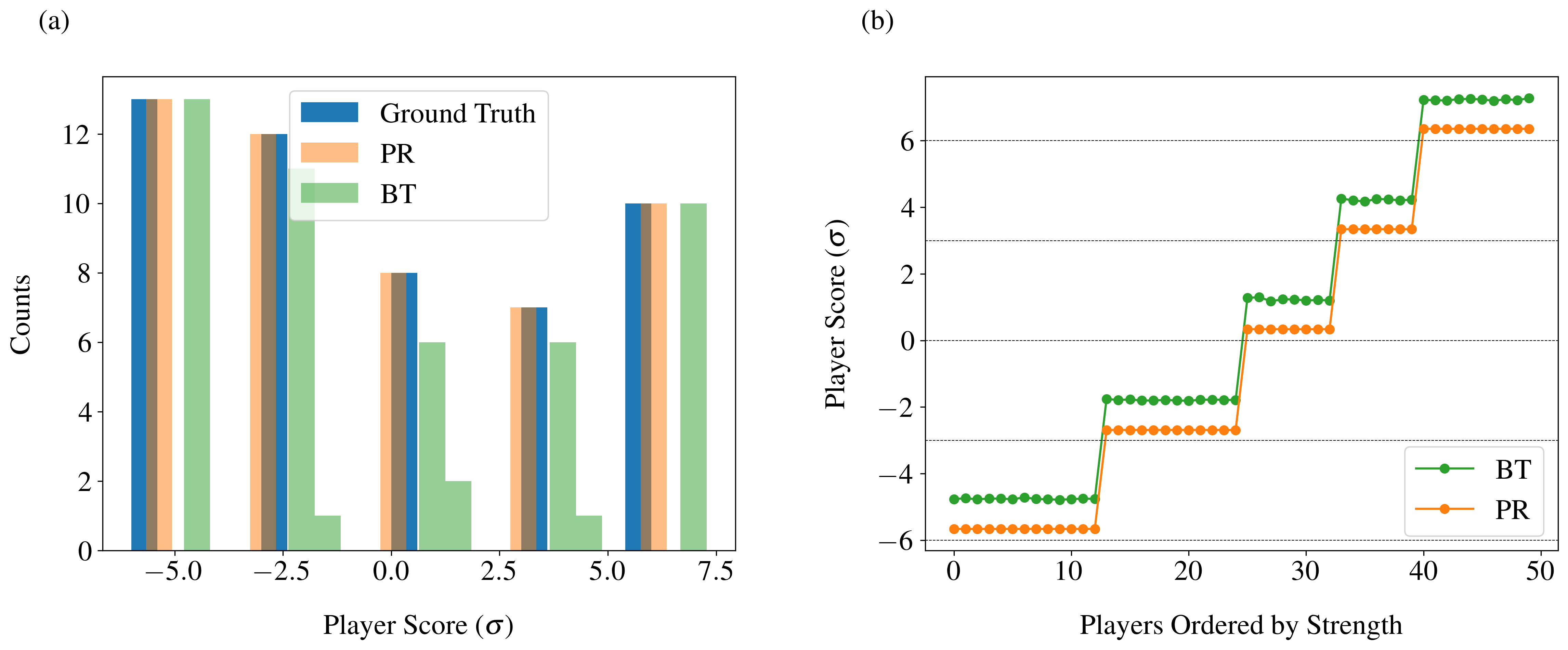}
    \caption{\textbf{Inference of planted scores.} (a) Histograms of planted ground truth scores, BT inferred scores, and PR inferred scores. (b) BT and PR inferred scores, ordered by increasing player strength. The horizontal black lines correspond to the ground-truth planted scores. All results were obtained for $N=50$ nodes and $R = 5$ planted partial rankings}
    \label{fig:app_score_recovery}
\end{figure*}

In Fig.~\ref{fig:app_tau_recovery}, we show Kendall's $\tau_B$ rank correlation coefficient as a function of $\expec{m}$ for synthetically generated networks consisting of $N = 150$ nodes. Panel (a) corresponds to the case $R = N = 150$, where each node is assigned a unique ranking, implying no partial rankings are present. Panel (b) illustrates the scenario where $R = 3$ unique partial rankings are imposed, with the nodes distributed at random among these three ranks.

In the case where no partial rankings are present (panel (a)), we observe that the BT model consistently outperforms the PR model in terms of rank correlation with the ground truth rankings, with the PR model obtaining similar $\expec{\tau_B}$ scores only at very high values of $\expec{m}$. Note that, as shown in Fig.~\ref{fig:2_km_sigma_phase_N150R150}, the PR model never recovers the correct number of rankings, even in dense regimes. However, the PR model is able to increase the inferred number of partial rankings as more data becomes available, which improves its rank correlation with the ground truth. By the time the PR model infers approximately $R = 50$ rankings, it achieves $\tau_B$ scores of roughly 0.99. 

When partial rankings are present (panel (b)), the $\expec{\tau_B}$ value for the BT model plateaus around 0.8. This behaviour likely stems from the BT model's inherent assumption of a complete ranking, which limits its ability to adapt once it has achieved the best possible alignment it can with the ground truth rankings. On the other hand, the PR model is able to obtain perfect correlation already at relatively low values of $\expec{m}$, thanks to its ability to account for and adapt to the presence of partial rankings in the data.

Fig.~\ref{fig:app_score_recovery} shows a comparison of the player scores inferred by both the BT and PR models with respect to the ground-truth planted scores for the case $N=50$, $R=5$. We observe that both the BT and PR models are able to recover scores which are very close to the planted ones. However, the BT model is unable to appropriately account for small fluctuations in match outcomes and assigns each node a unique score, even when these are very close. On the other hand, the PR algorithm is able to leverage the coarse-grained structure of the scores to group players into the correct number of partial rankings.

\section{\label{appendix:appendix_e} Comparison with 1D Clustering}

\begin{figure*}
    \includegraphics[width=\textwidth]{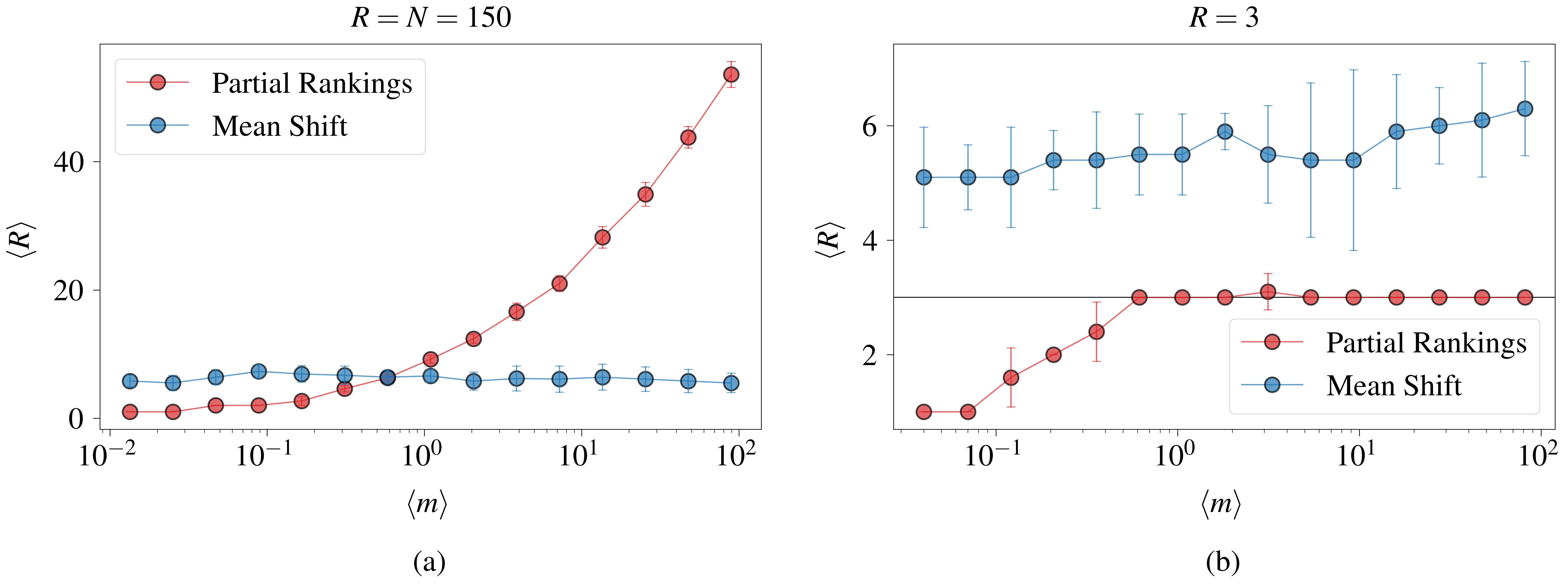}
    \caption{\textbf{Comparison of PR with 1D clustering of BT scores.} \textbf{(a)} Number of ranks inferred by the partial rankings algorithm and mean shift clustering as a function of $\expec{m}$ for the case where there are no partial rankings ($R = N = 150$). \textbf{(b)} Number of ranks inferred by the partial rankings algorithm and mean shift clustering as a function of $\expec{m}$ for the case in which there are three planted rankings ($R = 3$, indicated by the solid black line). Results are averaged over $10$ different network realizations for each $\expec{m}$ value, and error bars indicate two standard errors from the mean.}
    \label{fig:app_ms_comparison}
\end{figure*}

In Fig.~\ref{fig:app_ms_comparison}, we show the number of partial ranks inferred by using our partial rankings algorithm and mean shift clustering~\cite{comaniciu2002mean} for (a) the case in which no partial rankings are present ($R = N = 150$) and (b) the case with $R= 3$ planted partial rankings to which the nodes are assigned. When partial rankings are present (panel (b)), we observe that our partial rankings algorithm is able to reliably recover the correct number of rankings without the need to impose any kind of parameter (such as the bandwidth parameter in mean shift clustering). On the other hand, mean shift clustering is unable to recover the correct number of rankings. When partial rankings are not present (panel (a)), we observe that neither method is able to recover the correct number of ranks. Crucially, however, our partial rankings algorithm is able to adapt the number of inferred rankings as more evidence becomes available, progressively identifying additional ranks. On the other hand, mean shift clustering remains insensitive to the amount of statistical evidence, consistently inferring approximately the same number of clusters regardless of the number of matches played.

\clearpage
\section{\label{appendix:appendix_f} Comparison with the RC-BTL model}

The \emph{Rank-Clustered Bradley-Terry-Luce} model (RC-BTL), introduced by Pearce and Erosheva~\cite{pearce2024bayesian}, is a Bayesian method to infer rank-clusters which employs a spike-and-slab prior to encourage parameter fusion. Specifically, the generating process behind the RC-BTL is as follows:
~
\begin{enumerate}
    \item Sample the total number of unique ranks (rank-clusters) from a Poisson distribution with parameter $\lambda$
    \begin{equation}
        P(R|\lambda) = \frac{e^{-\lambda} \lambda^R}{R!}.
    \end{equation}
    This controls the ``spikiness'' of the distribution. Lower values of $\lambda$ favor a smaller number of clusters, encouraging parameter fusion. On the other hand, larger values of $\lambda$ promote a larger number of clusters, decreasing the probability of ties in the observed rankings.

    \item Draw the unique strengths for each cluster from a Gamma distribution with parameters $a_\gamma$ and $b_\gamma$
    \begin{align}
        P(\boldsymbol{\sigma} | R, a_\gamma, b_\gamma) &= \prod_{r=1}^R \text{Gamma}(\sigma_r | a_\gamma, b_\gamma) \notag \\
        &= \prod_{r=1}^R\frac{b_{\gamma}^{a_\gamma}}{\Gamma(a_\gamma)} \sigma_r^{a_\gamma - 1} e^{-b_\gamma \sigma_r}.
    \end{align}
    The parameters $a_\gamma$ and $b_\gamma$ control the distribution of the unique player scores. However, as pointed out by the authors in~\cite{pearce2024bayesian}, the player strengths are invariant to multiplicative transformations, so that the choice of $a_\gamma$ and $b_\gamma$ is generally non-influential. The primary advantage of this prior is that it allows for a closed-form Gibbs sampling of the posterior via data augmentation, resulting in an efficient MCMC sampler~\cite{pearce2024bayesian, caron2012efficient}.

    \item Draw the individual strengths $\boldsymbol{\pi}$ using the deterministic prior
    \begin{equation}
        P(\boldsymbol{\pi} | \boldsymbol{\sigma)} = \prod_{i=1}^N\delta(\pi_i, \sigma_{r_i}).
    \end{equation}

    \item Finally, the likelihood of observing a certain matrix $\bm{W}$ of match outcomes is given by the BT model
    \footnote{
         Since we are dealing with pairwise comparisons, the BT model is the most appropriate. However, the RC-BTL model can accommodate any of the distributions in the Bradley-Terry-Luce (BTL) family of models as a likelihood.
    }
    \begin{equation}
        P(\bm{W} | \boldsymbol{\pi}) = \prod_{ij}\left( \frac{\pi_i}{\pi_i + \pi_j} \right)^{w_{ij}}.
    \end{equation}
    
\end{enumerate}
Putting it all together, the RC-BTL is described by the following generative model:
\begin{align} \label{eq:rc_btl_posterior}
    &P(\boldsymbol{\pi} | \bm{W}, a_\gamma,b_\gamma,\lambda) \propto P(\bm{W} | \boldsymbol{\pi})P(\boldsymbol{\pi} | \boldsymbol{\sigma)}P(\boldsymbol{\sigma} | R, a_\gamma, b_\gamma)P(R | \lambda) \notag \\
    &= \prod_{ij}\left( \frac{\sigma_i}{\sigma_i + \sigma_j} \right)^{w_{ij}} \cdot \prod_{r=1}^R\frac{b_{\gamma}^{a_\gamma}}{\Gamma(a_\gamma)} \sigma_r^{a_\gamma - 1} e^{-b_\gamma \sigma_r} \cdot \frac{e^{-\lambda} \lambda^R}{R!},
\end{align}
and the standard BT model can be recovered, up to $O(\log N)$, by setting $R = \lambda = N$.

The primary differences between RC-BTL and our PR model lie in the formulation of the partition prior. In RC-BTL, the probability of observing a specific partition into $R$ clusters is, by design, independent of the cluster sizes, so that any partition into $R$ groups is equally probable under the model\footnote{As pointed out in~\cite{pearce2024bayesian}, there still is a certain degree of implicit dependence of the cluster sizes on the number of groups. For example, if $R=N$, then all clusters must be singletons, such that $n_r = 1~\forall r$.}. As discussed in Section~\ref{sec:methods}, this indifference with respect to group sizes implicitly biases the model toward balanced partitions, since the vast majority of partitions divide nodes into approximately equal-sized groups purely on combinatorial grounds. See Appendix~\ref{appendix:uniform_sampling}. By contrast, the PR model first samples the distribution of group sizes from a uniform prior over all possible histograms of group sizes, conditioned on $R$. Under this construction, clusters of any size are equally plausible a priori, so the model remains agnostic about balance until information from the data is incorporated. Another key difference is that RC-BTL requires us to fix three hyperparameters, $a_\gamma$, $b_\gamma$, and $\lambda$. As previously mentioned, the effect of $a_\gamma$ and $b_\gamma$ on the inference results is limited. $\lambda$, on the other hand, controls the average number of unique rankings we expect to see and can considerably alter the inferred partitions. On the one hand, this is an advantage, as $\lambda$ can be tuned to encourage or discourage rank-clustering depending on how one sees fit. On the other hand, choosing appropriate values for the hyperparameters can be nontrivial and poor choices could potentially bias the results. In~\cite{pearce2024bayesian}, the authors suggest setting $a_\gamma = 5$, $b_\gamma = 3$, and $\lambda = 1$ in a pairwise comparison setting so as to encourage rank clustering, given that pairwise comparisons provide limited ordinal information about the items being ranked. In what follows, we use the same hyperparameter settings for all the considered datasets.

\begin{figure*}[!t]
    \centering
    \includegraphics[width=\textwidth]{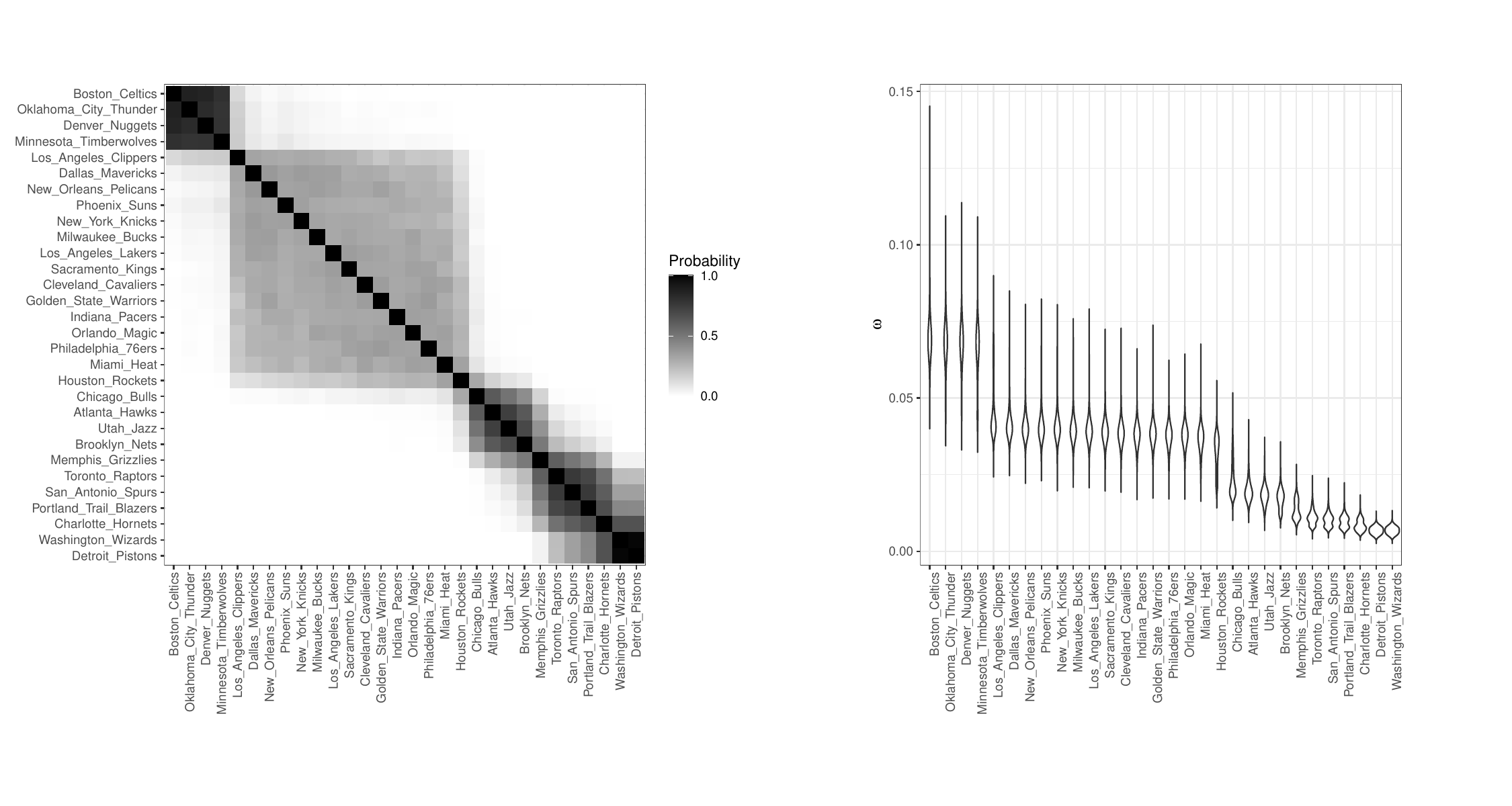}
    \caption{\emph{Left:} Posterior rank-clustering probabilities according to the RC-BTL model for the 2023-24 NBA data. \emph{Right:} Posterior distributions of team strengths as computed by the RC-BTL model. Plots are generated via the \texttt{rankclust} R package~\cite{rankclust}}
    \label{fig:PE_heatmap}
\end{figure*}

The second crucial difference between the two models is that RC-BTL samples the entire posterior via MCMC, while, in its current implementation, the PR model provides only MAP estimates via the agglomerative greedy algorithm described in Section~\ref{sec:methods}. RC-BTL thus has the advantage of yielding a full posterior over the rankings, allowing the user to perform uncertainty quantification. However, this uncertainty quantification comes at the cost of a substantially slower inference process involving MCMC sampling. By contrast, the PR objective can be more easily redefined for different priors and likelihoods, including that of the RC-BTL model as shown below, and provides fast MAP estimates using a greedy agglomerative approach. In principle, different priors could also be used within the MCMC sampling for RC-BTL, but the specific prior structure in RC-BTL is crucial for sampler efficiency as it enables closed-form Gibbs updates of the cluster strengths.

\begin{figure*}[]
    \centering
    \includegraphics[width=\textwidth]{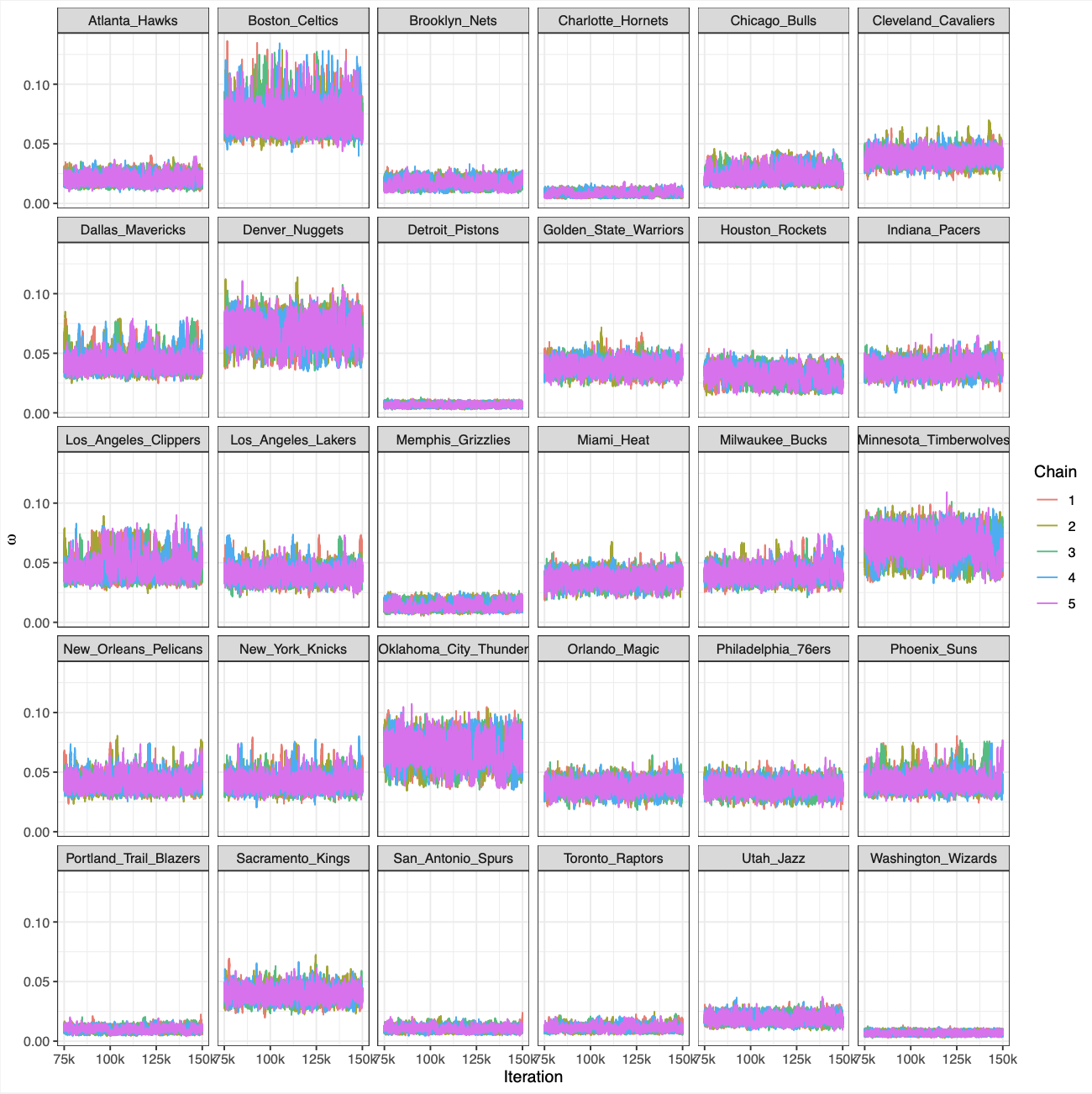}
    \caption{Trace plots of team strengths for the 2023-24 NBA data set. Plots are generated via the \texttt{rankclust} R package~\cite{rankclust}.}
    \label{fig:PE_trace_plots_NBA}
\end{figure*}

In order to compare our PR model results with those inferred by the RC-BTL, we consider the same dataset used by Pearce and Erosheva in~\cite{pearce2024bayesian}, containing all matches from the 2023-24 season of the US National Basketball Association (NBA). The dataset consists of $N=30$ teams, each of which plays $M=82$ games throughout the season. Following~\cite{pearce2024bayesian}, we set $a_\gamma = 5$, $b_\gamma = 3$, and $\lambda=1$, and fit five MCMC chains of $150,000$ iterations each using the \texttt{rankclust} R package provided by the authors~\cite{rankclust}. Results are shown in Fig.~\ref{fig:PE_heatmap}. Fig.~\ref{fig:PE_heatmap}(a) shows the posterior rank-clustering probabilities that two teams belong to the same cluster, while Fig.~\ref{fig:PE_heatmap}(b) shows the posterior distributions of each team's strength. The corresponding trace plots for the strength of each individual team are shown in Fig.~\ref{fig:PE_trace_plots_NBA}.

The results suggest that, with high probability, four of the teams (Boston Celtics, Oklahoma City Thunder, Denver Nuggets, and Minnesota Timberwolves) can be clustered within a single rank representing the strongest teams. Similarly, six of the teams (Toronto Raptors, San Antonio Spurs, Portland Trail Blazers, Charlotte Hornets, Washington Wizards, and Detroit Pistons) show considerable probability of being rank-clustered in last place.

\begin{figure}[!ht]
    \centering
    \includegraphics[width=0.75\linewidth]{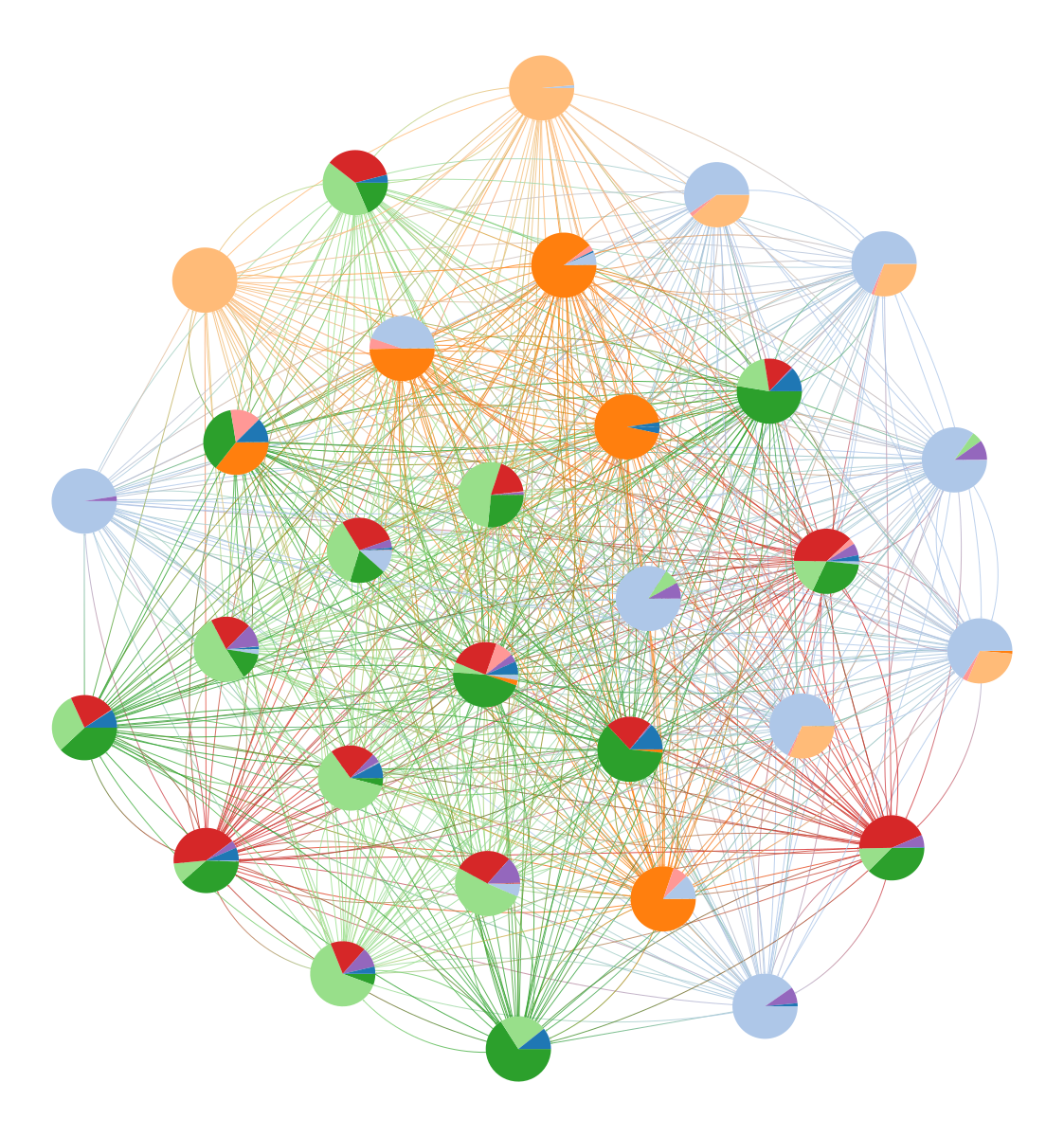}
    \caption{Marginal posterior group membership distribution of NBA teams obtained from relabeled partitions of $25,000$ MCMC samples of the RC-BTL posterior distribution.}
    \label{fig:nba_marginal}
\end{figure}

\begin{figure}[!t]
    \centering
    \includegraphics[width=\linewidth]{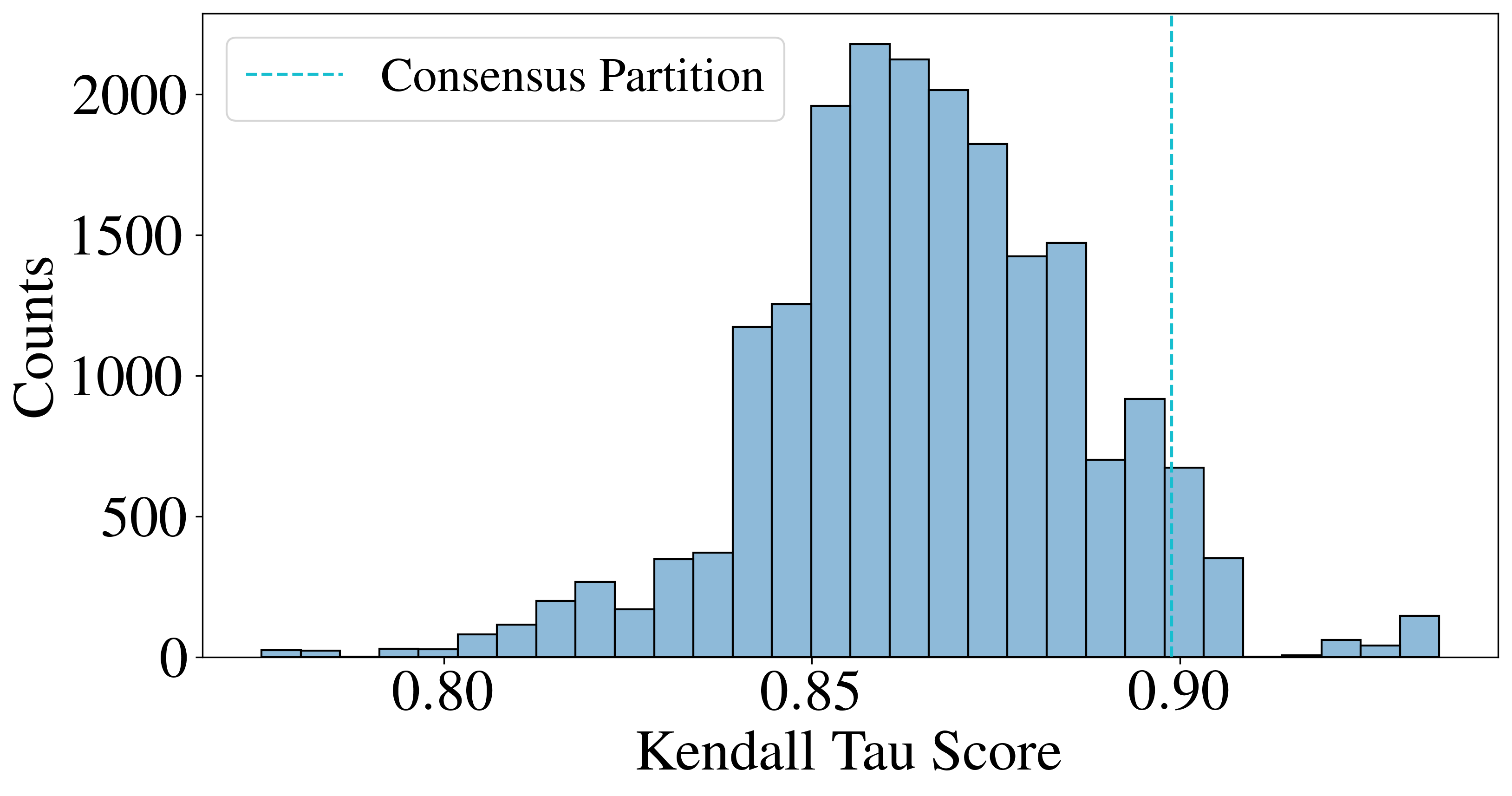}
    \caption{Distribution of Kendall $\tau_B$ scores between the PR partition and the RC-BTL MCMC samples. The dashed cyan line represents the $\tau_B$ score with the consensus partition.}
    \label{fig:2_kendall_tau_histogram_nba}
\end{figure}

To facilitate comparison between the RC-BTL results and the partition inferred via the PR algorithm (which, as noted above, provides only a MAP estimate rather than a full posterior), we summarize the partitions sampled from the RC-BTL posterior into a single consensus representation. Specifically, we take the last $5,000$ MCMC samples of each chain (giving us $25,000$ post-burn-in samples) and use Peixoto’s Random Label Model~\cite{peixoto2021revealing} to resolve label switching and align the sampled partitions to a common group labelling. From these aligned samples, we compute the marginal posterior group membership distribution for each team. Fig.~\ref{fig:nba_marginal} shows the network representation of matches for the 2023–24 season, where each team is represented by a node and colored according to its marginal posterior group membership distribution. The consensus partition is then obtained by assigning each team to the group with the highest marginal posterior probability. For most teams in Fig.~\ref{fig:nba_marginal}, one group clearly dominates the membership distribution, suggesting that the consensus partition provides an accurate summary of the MCMC results.
Table~\ref{table:pr_vs_consensus_partitions} displays the partition inferred by our PR algorithm alongside the RC-BTL consensus partition, with clusters displayed in descending order of (normalized) strength. The two partitions are in excellent accordance ($\tau_B \simeq 0.879$, p-value $\sim O(10^{-9})$), the main difference being that the RC-BTL consensus parition distinguishes three clusters very close in strength (Cluster 1, Cluster 2, and Cluster 3) which are instead merged into a single cluster in the PR partition (Cluster 1). In general, the PR partition displays high ordinal association with all the MCMC samples as shown in Fig.~\ref{fig:2_kendall_tau_histogram_nba}.

These results suggest that our PR algorithm could serve as a useful complement to RC-BTL, providing practitioners with a fast method to infer partial rankings before carrying out a more detailed posterior analysis with RC-BTL. For instance, under the parametrization described above, running a single RC-BTL MCMC chain required roughly 18 hours on a MacBook Pro M4 Max with 64GB of RAM, whereas the PR algorithm required only 0.02 seconds on the same machine to infer the partial rankings. While direct runtime comparisons across implementations in different programming languages (Python for PR vs R for RC-BTL) should be interpreted with caution, it is expected that a full MCMC exploration of the posterior will generally entail a substantially higher computational cost.

\subsection{Assessing the Quality of the Partitions}

Given any partial ranking sampled from the RC-BTL posterior, we can once more attempt to assess which of the two ranking/model combinations better represents the data by computing the log-posterior odds ratio
\begin{equation}
    \Delta\mathcal{L} = \frac{\log P_{PR}(\boldsymbol{\pi}|\bm{A})}{\log P_{RC-BTL}(\boldsymbol{\pi}|\bm{A},a_\gamma, b_\gamma, \lambda)}.
\end{equation}
However, comparing across model families is perhaps not that useful, as the constant terms and priors differ fundamentally between the models. However, we can still compare the partitions obtained by both models by evaluating their negative log-posterior within each model class.

For each RC-BTL sample, we can compute its negative log-posterior according to the PR model by plugging in the corresponding values into Eq.~\eqref{eq:pr_dl}, where the strengths $\sigma_r$ can be computed via Eq.~\eqref{eq:saddle_point}. \\
To compute the negative log-posterior of the PR partition under the RC-BTL model, we first take the negative logarithm of Eq.~\eqref{eq:rc_btl_posterior}, obtaining
\begin{align} \label{eq:pe_dl}
    &\mathcal{L}_{RC-BTL}(\boldsymbol{\sigma}, a_\gamma, b_\gamma ,\lambda) = -\sum_{r, r'}\omega_{rr'}\log \left( \frac{\sigma_r}{\sigma_r\sigma_{r'}} \right) \notag \\
    &- \sum_r \left[ a_\gamma \log b_\gamma - \log\Gamma(a_\gamma)+(a_\gamma - 1)\log \sigma_r - b_\gamma \sigma_r \right] \notag \\
    &+ \lambda - R\log\lambda + \log R!,
\end{align}
where $\omega_{rr'}$ denotes the number of times that teams in rank $r$ beat teams in rank $r'$. Note that $\mathcal{L}_{RC-BTL}$ depends explicitly on the model hyperparameters. In principle, these terms are constants for fixed hyperparameters, but they become relevant when comparing partitions across different hyperparameter settings or across models, so we retain them here.
To evaluate the PR partition under RC-BTL, we must determine the associated strength parameters $\boldsymbol{\sigma}_{RC-BTL}$. In RC-BTL, these parameters are sampled from a Gamma prior during MCMC. However, since the PR algorithm targets a MAP estimate, we can compute the strength parameters by minimizing Eq.~\eqref{eq:pe_dl} with respect to $\sigma_r$ (equivalently, maximizing the posterior) holding the partition fixed. Taking the gradient with respect to $\sigma_r$ of Eq.~\eqref{eq:pe_dl} yields
~
\begin{equation}
    -\frac{a_\gamma - 1}{\sigma_r} + b_\gamma - \sum_{r' \neq r}\frac{w_{rr'}\sigma_{r'}}{\sigma_r(\sigma_r + \sigma_{r'})} + \sum_{r' \neq r}\frac{w_{r'r}}{\sigma_r + \sigma_{r'}}.
\end{equation}
\begin{figure*}[!ht]
    \includegraphics[width=\textwidth]{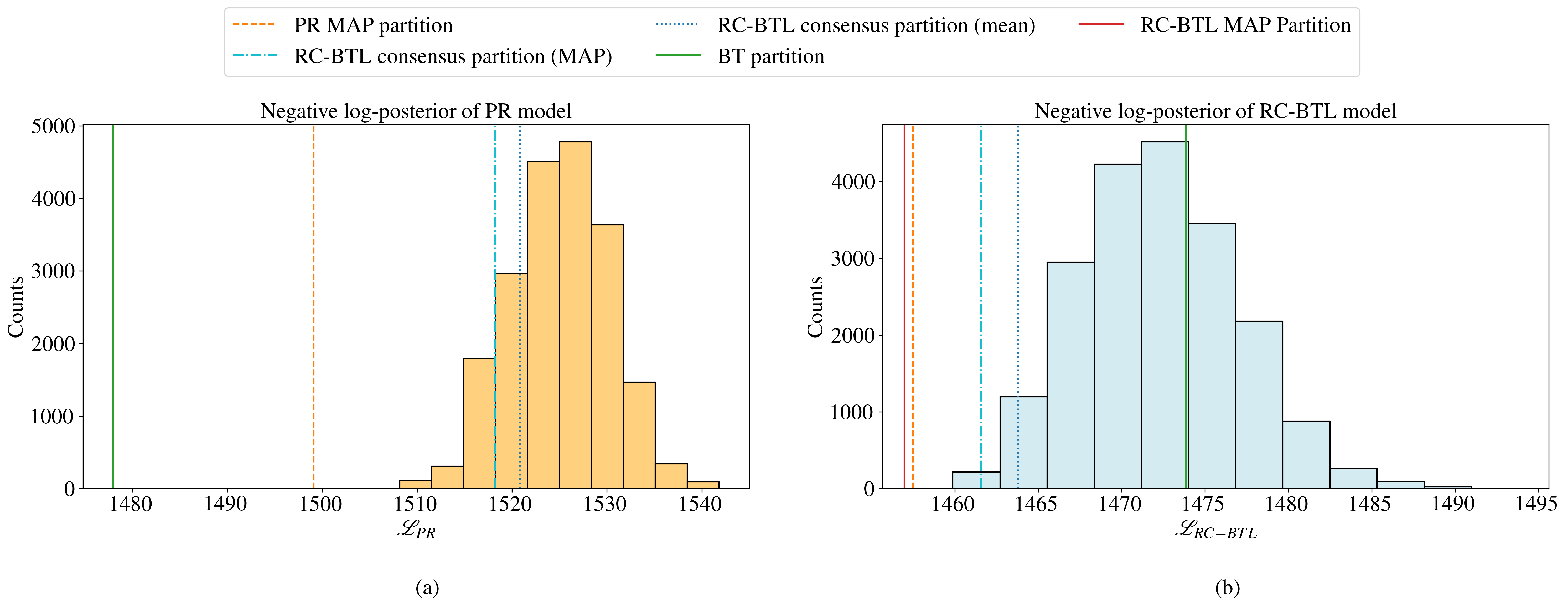}
    \caption{Histograms of negative log-posterior values according to the PR model (panel a) and the RC-BTL model (panel b). The dashed vertical orange lines indicate the negative log-posterior values of the PR MAP partition according to both models, while the vertical dash-dotted cyan lines indicate the negative log-posterior values of the RC-BTL consensus partition with strengths evaluated by minimizing Eq.~\eqref{eq:pe_dl}. The dotted blue lines indicate the negative log-posterior of the RC-BTL consensus partition when the group strengths are evaluated by averaging the strengths of the individual teams making up each cluster. The solid red line corresponds to the MAP estimate of the RC-BTL model computed via our PR algorithm, while the solid green line represents the BT partition where $N = R = 30$.}
    \label{fig:pr_pe_dl_comparison_NBA}
\end{figure*}
Equating to zero and rearranging the terms, we obtain the following self-consistent equation for $\sigma_r$, which can be iterated until convergence
~
\begin{equation} \label{eq:pe_saddle_point}
    \sigma_r = \frac{a_\gamma - 1 + \sum_{r' \neq r}\frac{w_{rr'}\sigma_{r'}}{\sigma_r(\sigma_r + \sigma_{r'})}}{b_\gamma + \sum_{r' \neq r}\frac{w_{r'r}}{\sigma_r + \sigma_{r'}}}.
\end{equation}
Eq.~\eqref{eq:pe_saddle_point} is the equivalent of Eq.~\eqref{eq:saddle_point} for the RC-BTL model. In practice, we observe small differences in the solutions inferred by Eq.~\eqref{eq:pe_saddle_point} and Eq.~\eqref{eq:saddle_point} under the above parametrization.

A similar consideration must be made for the strength parameters of the RC-BTL consensus partition. One option is to assign each cluster in the consensus partition a $\hat\sigma_r$ value equal to the average strength of all the teams in each cluster, where the strengths of each team can be computed as the median strengths along all the MCMC samples. Another option is to proceed as above and compute the MAP $\boldsymbol{\sigma}$ scores of each cluster in the consensus partition, i.e. the values of $\boldsymbol{\sigma}$ that maximize the posterior given the consensus partition. In what follows we adopt both methods. Finally, we can also use the PR algorithm to compute a MAP estimate of the RC-BTL model by substituting the objective of Eq.~\eqref{eq:pr_map_objective} with Eq.~\eqref{eq:pe_dl} and updating the strength parameters via Eq.~\eqref{eq:pe_saddle_point}.

    Fig.~\ref{fig:pr_pe_dl_comparison_NBA} shows a histogram of negative log-posterior values according to the PR model (Fig.~\ref{fig:pr_pe_dl_comparison_NBA}(a)) and the RC-BTL model (Fig.~\ref{fig:pr_pe_dl_comparison_NBA}(b)), with the values for the PR MAP partition, the consensus partition, and the RC-BTL MAP partition, indicated by vertical lines. In both cases, we observe that the ranking inferred by the PR algorithm is associated with a higher posterior probability than the consensus partition according to both models. Furthermore, the partition inferred by the PR model is very close (in terms of negative log-posterior) to the RC-BTL MAP estimate and in good ordinal accordance ($\tau_B \simeq 0.908$, p-value $\sim O(10^{-9})$) suggesting that the PR model returns partitions similar to the RC-BTL model. Interestingly, we observe that under the PR encoding, the BT model achieves a lower negative log-posterior than any other partition, whereas under the RC-BTL encoding it yields a higher negative log-posterior than most posterior samples, as well as than the consensus and MAP partitions.

\begin{figure*}[!ht]
    \includegraphics[width=\textwidth]{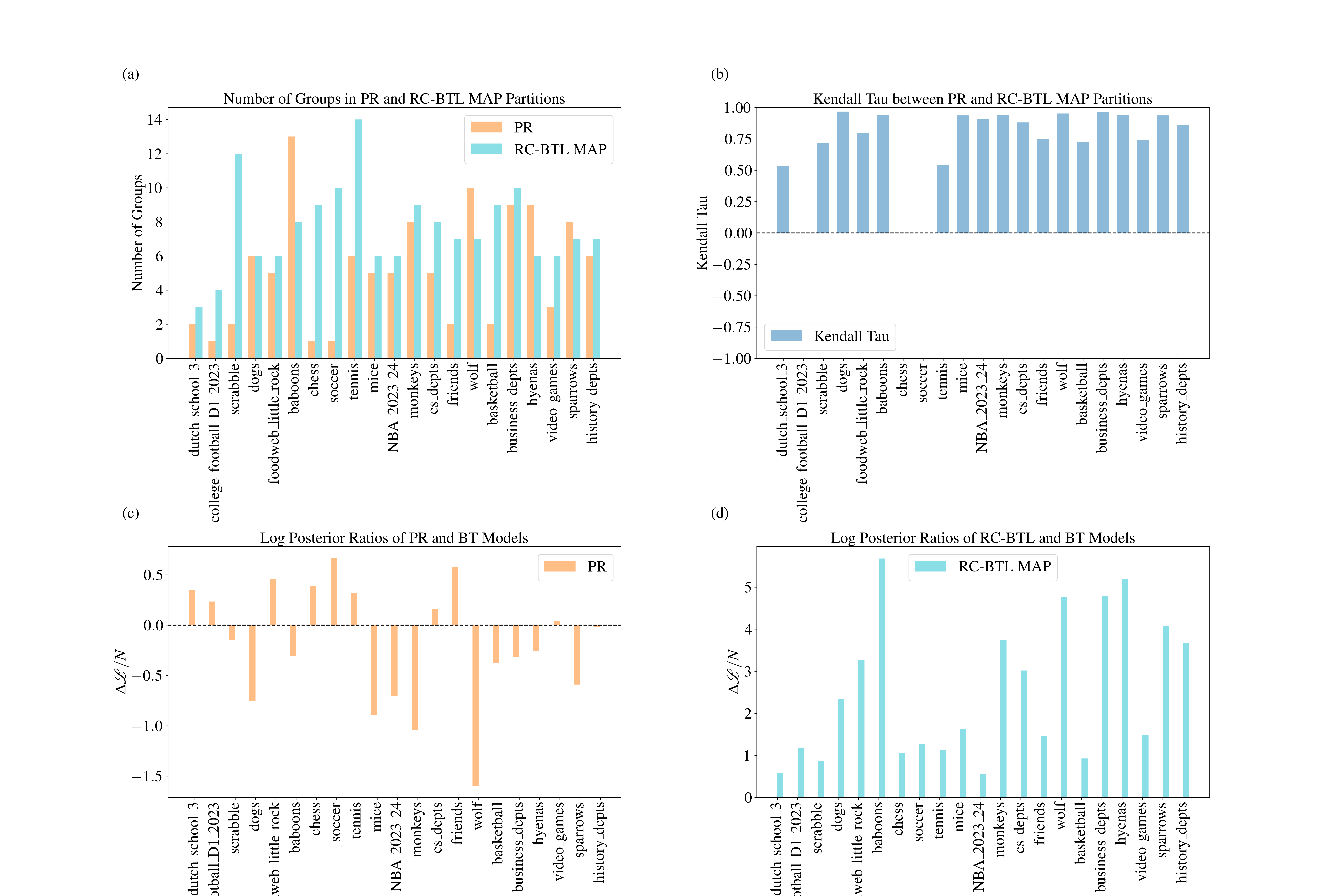}
    \caption{(a) Number of groups inferred by the PR model (orange bars) and the MAP partition of the RC-BTL model (cyan bars). (b) Kendall's rank correlation coefficient $\tau_B$ between the rankings inferred by the PR model and the MAP partition of the RC-BTL. Missing values correspond to cases in which the PR model merged all elements into a single rank. (c) log-posterior-odds ratio between the rankings inferred by the PR model and the BT model according to the encoding of the PR model. Positive values indicate a preference for the PR model, negative values favor the BT model. (d) log-posterior-odds ratio between the rankings inferred by the MAP partition of the RC-BTL model and the BT model according to the RC-BTL encoding. Again, positive values indicate that the RC-BTL model provides a better description of the data, negative ones favor the BT model.}
    \label{fig:pr_vs_rcbtl_rw_comparison}
\end{figure*}
~
Comparing the PR partition with the MAP partition of the RC-BTL model across all the real-world datasets studied in this paper, we observe the same general results, see Fig.~\ref{fig:pr_vs_rcbtl_rw_comparison}. The RC-BTL model, which applies a weaker penalty on the number of groups, tends to infer more clusters than the PR model, although the resulting partitions are generally in strong ordinal agreement. Two notable exceptions are the ``Dutch school'' and ``tennis'' datasets, where we observe moderate values of Kendall’s $\tau_B \simeq 0.5$. These reduced associations can be attributed to strong heterogeneity in the group sizes inferred by the two models, leading to many ties, which $\tau_B$ partially discounts. For instance, in the Dutch school dataset, the PR model infers two groups---one of just two students and a second containing the remaining 24---whereas RC-BTL infers three clusters with sizes $n_r =[2,17,7]$.
More notably, we observe that, in the PR model, the inferred partitions occasionally yield lower negative log-posteriors than BT and occasionally don't. On the other hand, the MAP partitions for the RC-BTL model always provide a better explanation of the data (with respect to BT) under the RC-BTL model class.

\subsection{Recovery of Synthetic Datasets}

In the above section, we observed that in the RC-BTL model, most MCMC samples will provide a better explanation of the data (in terms of the log-posterior) than the BT model in all of the datasets analyzed. We explore this further by considering three synthetically generated datasets. One dataset consists of $N=30$ players, each with its own unique strength $\sigma_i$ generated from the BT model. The second and third datasets both consist of three planted rankings, with each team/player assigned to one of these rankings. These two datasets differ only in the heterogeneity of their planted partitions: one has clusters of comparable size, while the other exhibits strong size imbalance.

\subsubsection{Recovery of planted BT rankings}

\begin{figure*}[!ht]
    \centering
    \includegraphics[width=\textwidth]{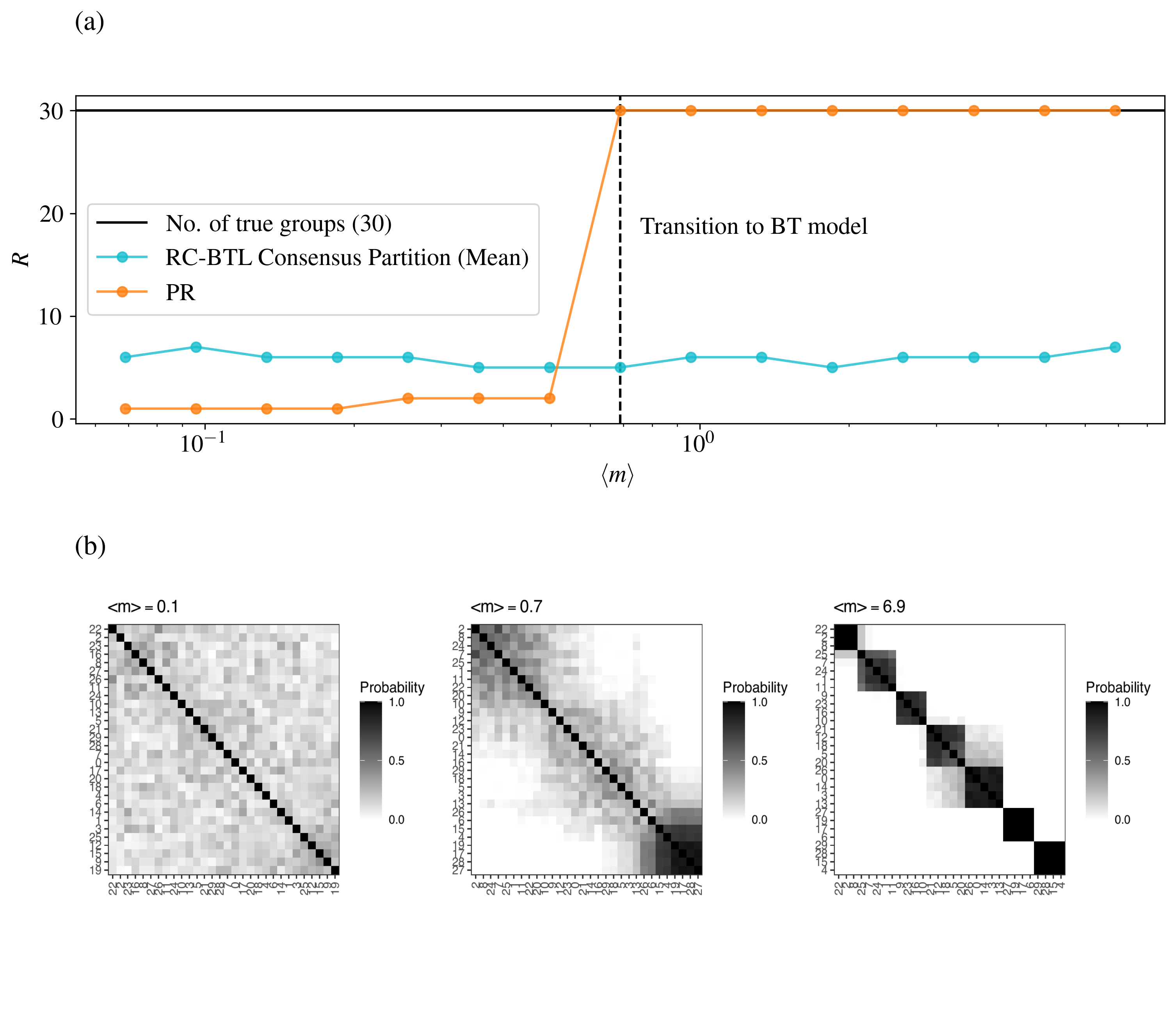}
    \caption{(a) Number of rankings inferred by the PR model (orange line) and the consensus partition of the RC-BTL model with cluster strengths calculated as the  strengths over all elements of the cluster (cyan line). The solid horizontal line indicates the true number of rankings, 30, while the dashed vertical line indicates the point at which the PR model is no longer able to provide a more parsimonious description of the data with respect to the BT model, and the orange line jumps to $R = N = 30$. On the other hand, the RC-BTL consensus partition is always able to give a more parsimonious description of the data with respect to the BT model and no transition is observed. All instances of the RC-BTL have been performed setting the parameters to $a_\gamma = 5$, $b_\gamma = 3$, and $\lambda = 1$. (b) From left to right: posterior rank-clustering probabilities of the RC-BTL model for increasing values of $\expec{m}$ in the case $R = N = 30$. As $\expec{m}$ increases, the RC-BTL begins to infer an increasing number of well-defined clusters.}
    \label{fig:2_bt_recovery_combined}
\end{figure*}

We begin with $N=30$ players, each assigned a unique strength drawn from a logistic distribution with mean zero and scale one under the BT model.\footnote{We repeated the procedure with strengths sampled from a Gamma prior with $a_\gamma = 5$ and $b_\gamma = 3$, and the results presented in the remainder of this section were unchanged.} Given these strengths, we generated datasets by randomly selecting pairs of players and determining match outcomes according to the BT likelihood. We constructed $15$ such datasets, each with an increasing number of matches, thereby raising the density of the corresponding directed network.
Fig.~\ref{fig:2_bt_recovery_combined}(a) shows the number of groups inferred by the PR model and the consensus partition of the RC-BTL model (with cluster strengths computed as the average strength across all members of each cluster) as a function of the density $\langle m \rangle$ of the match network. For small values of $\langle m \rangle$, both models return a more parsimonious fit of the data relative to the BT model, as the limited number of matches does not provide enough information to infer complete rankings. As $\langle m \rangle$ increases, we observe a sharp transition at $\langle m \rangle = 0.7$ in which the PR model is no longer able to provide more parsimonious descriptions of the data and we observe an abrupt jump to the correct value of $R = N = 30$ inferred by the BT model. By contrast, RC-BTL consistently provides a more parsimonious description of the data with respect to the BT model, and as a result it fails to recover the planted number of groups. These results suggest that, while the RC-BTL model offers a powerful framework for clustering ranked data, it may be less suitable for assessing whether partial rankings provide the most appropriate description of a dataset to begin with.
\begin{figure*}[!ht]
    \centering
    \includegraphics[width=\textwidth]{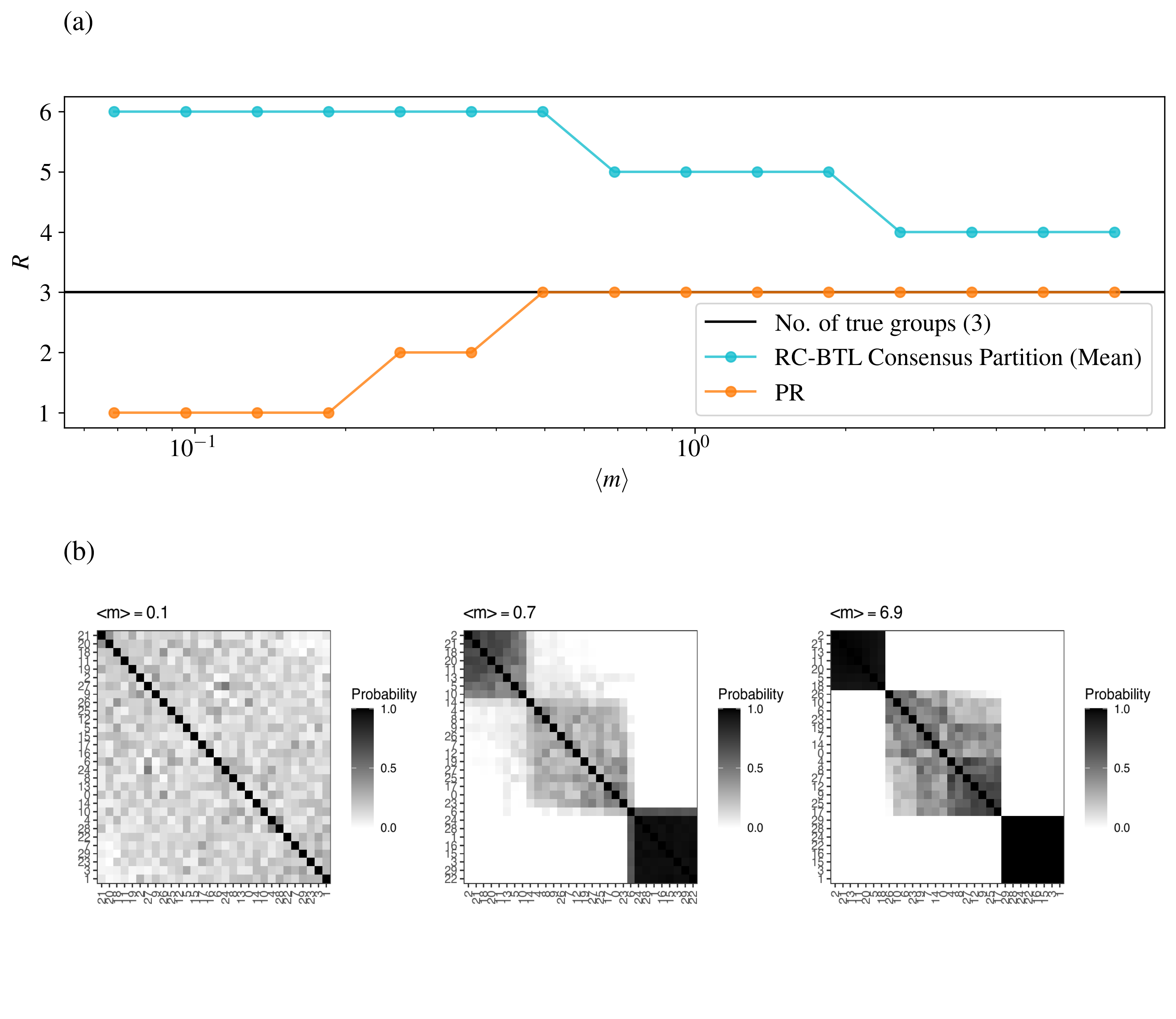}
    \caption{(a) Number of rankings inferred by the PR model (orange line) and the consensus partition of the RC-BTL model with cluster strengths calculated as the average strengths over all elements of the cluster (cyan line). The solid horizontal line indicates the true number of rankings, 3. All instances of the RC-BTL have been performed setting the parameters to $a_\gamma = 5$, $b_\gamma = 3$, and $\lambda = 1$. (b) From left to right: posterior rank-clustering probabilities of the RC-BTL model for increasing values of $\expec{m}$ in the case $R = 3$, $N = 30$. As $\expec{m}$ increases, the RC-BTL begins to cluster around the planted partitions.}
    \label{fig:pr_recovery_combined}
\end{figure*}

Fig.~\ref{fig:2_bt_recovery_combined}(b) shows examples of how the rank-clustering probability changes as a function of the density $\expec{m}$. The general behavior is similar to the PR model. For small $\expec{m}$, the RC-BTL distributes the rank-clustering probability roughly uniformly among all node pairs, as there is not enough evidence to clearly distinguish rank clusters. As $\expec{m}$ increases, RC-BTL begins to infer node clusters, which increase in number and become more defined as the density of the network of matches increases. We should then expect the RC-BTL to eventually recover the ground-truth partition given enough data.

\subsubsection{Recovery of planted PR rankings}

\begin{figure*}[!ht]
    \centering
    \includegraphics[width=\textwidth]{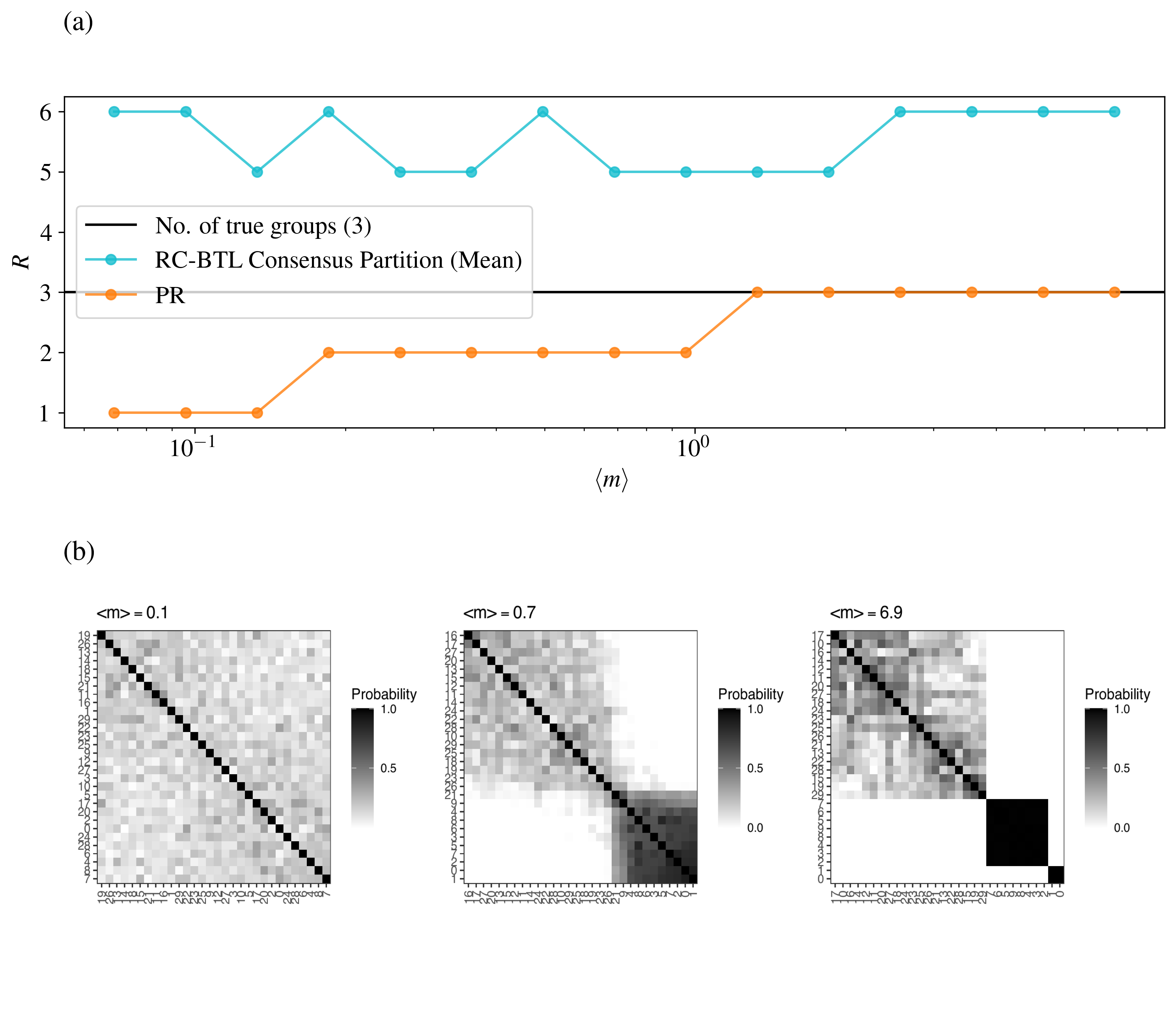}
    \caption{(a) Number of rankings inferred by the PR model (orange line) and the consensus partition of the RC-BTL model with cluster strengths calculated as the average strengths over all elements of the cluster (cyan line) for planted partitions of sizes $n_r = [20, 8, 3]$. The solid horizontal line indicates the true number of rankings, $3$. All instances of the RC-BTL model have been performed setting the parameters to $a_\gamma = 5$, $b_\gamma = 3$, and $\lambda = 1$. (b) From left to right: posterior rank-clustering probabilities of the RC-BTL model for increasing values of $\expec{m}$ in the case $R = 3$, $N = 30$ with cluster sizes $n_r=[20, 8, 2]$. As $\expec{m}$ increases, the RC-BTL begins to cluster around the planted partitions.}
    \label{fig:2_pr_heterogeneous_recovery_combined}
\end{figure*}

Fig.~\ref{fig:pr_recovery_combined}(a) shows the number of inferred rankings by the PR and RC-BTL (consensus) models as a function of $\expec{m}$ for $N=30$ nodes randomly assigned to one of three groups with strengths $e^s$, $s \in [-3, 0, 3]$. In this scenario, both models provide a more parsimonious description of the data compared to the BT model. However, the less constraining prior of RC-BTL leads it to infer more groups than the PR model. In this scenario, where $R = 3$, the PR model quickly identifies the correct number of rankings, whereas RC-BTL requires more matches before the consensus partition reliably begins to reflect the underlying group structure. This relationship may reverse as the number of planted partitions increases: the more restrictive prior of the PR model would then require more data to justify additional inferred rankings.

Importantly, RC-BTL infers the full posterior, so analysis is not limited to the consensus partition. Fig.~\ref{fig:pr_recovery_combined}(b) shows the posterior rank clustering probabilities as the density of the match network increases. Even at relatively low $\expec{m}$, there is strong evidence for a three-group structure, despite the consensus partition being unable to recover the correct rankings. This type of uncertainty quantification is currently absent in the PR model and could substantially enhance its interpretability and robustness.

\begin{figure}[!ht]
    \centering
    \includegraphics[width=\linewidth]{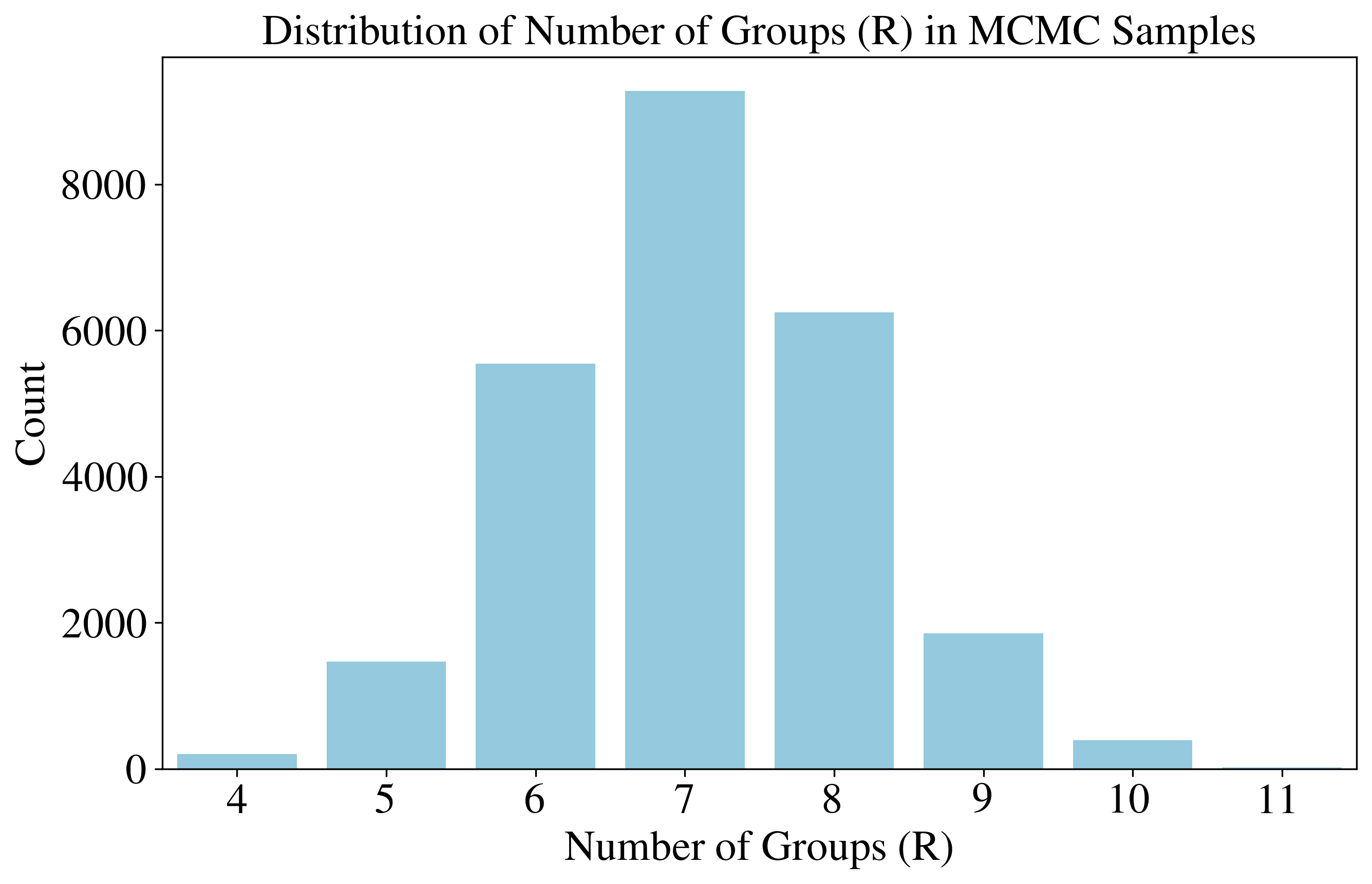}
    \caption{Countplot of the number of inferred rankings by the RC-BTL MCMC samples for a synthetic dataset with three planted rankings of sizes $n_r = [20, 8, 2]$ and strengths $\boldsymbol{\sigma} = [e^3, e^0, e^{-3}]$. We observe that none of the samples is ever able to recover the planted number of rankings.}
    \label{fig:2_rc_btl_mcmc_sample_sizes_countplot}
\end{figure}

A slightly different picture emerges when the cluster sizes become more heterogeneous. In the previous dataset, nodes were randomly assigned to clusters, which produced two clusters of roughly equal size and a larger cluster approximately twice as large as the other two (in decreasing order of strength: $n_r = [7, 15, 8]$). We can consider a more heterogeneous setting in which we explicitly assign nodes to clusters of sizes $n_r = [20, 8, 2]$, again sorted in decreasing order of strength. Fig.~\ref{fig:2_pr_heterogeneous_recovery_combined}(a) shows the number of clusters inferred by the PR algorithm and the consensus partition of the RC-BTL model as a function of the density $\expec{m}$ of the match network. We find that the PR algorithm eventually recovers both the correct number and the correct composition of rankings. In contrast, the RC-BTL consensus partition never achieves exact recovery. Furthermore, the previously observed gradual reduction in the number of inferred groups with increasing evidence is absent, and the consensus partition consistently identifies between five and six groups. In fact, none of the individual MCMC samples recovers the true number of partitions (see Fig.~\ref{fig:2_rc_btl_mcmc_sample_sizes_countplot}). Nevertheless, examining the posterior rank-clustering probabilities of the RC-BTL model reveals that, although a larger number of matches is required compared to the previous setting in order to detect evidence of a three-cluster structure, very strong evidence does emerge when $\expec{m}$ is sufficiently large. See Fig.~\ref{fig:2_pr_heterogeneous_recovery_combined}(b).

How, then, can information about the three-cluster structure emerge in the rank-clustering probabilities when it is seemingly absent from every individual MCMC sample? To address this, we examine the partitions inferred by the RC-BTL model at high $\expec{m}$. We find that the overwhelming majority of sampled partitions correctly identify the two smaller groups. However, inferring a third group of size $n_r = 20$ would yield partitions with extreme heterogeneity in cluster sizes. To avoid this, the RC-BTL model instead tends to subdivide the largest group into multiple subgroups, thereby reducing size imbalance in accordance with its implicit bias toward balanced partitions. Crucially, these subdivisions are not supported by the underlying data; they arise solely from the model’s preference for homogeneity. As a consequence, each sample partitions the largest group differently, with subdivisions amounting to random reshufflings of the same set of nodes (with minor fluctuations in subgroup size and number). When aggregating across the full posterior, this artificial variability cancels out, and nodes in the largest group appear with roughly equal probability of belonging to the same cluster\footnote{This implies that the RC-BTL results must be interpreted at the level of the full posterior: although the marginal distributions clearly indicate three distinct rank clusters, no single sampled partition will provide a representative realization of the underlying structure.}. This effect is also evident in Fig.~\ref{fig:pr_recovery_combined}(b), where MCMC samples frequently split the largest group into two nearly equal subgroups in order to enforce cluster size homogeneity. By contrast, the hierarchy of priors in PR is deliberately designed to be as agnostic as possible with respect to the data, enabling recovery of both the correct number and composition of clusters in its MAP partition. At the same time, this result once again underscores the benefits of sampling from the full posterior rather than relying solely on a single MAP estimate, since posterior sampling can mitigate biases inherent in the underlying model. Consequently, we view the development of an MCMC sampler for the PR model as a promising direction for future work. Such an extension would, however, entail a trade-off, as it would likely reduce computational efficiency, which currently makes PR an attractive alternative to fully Bayesian methods that rely on posterior sampling.

\section{Partition sizes under uniform sampling \label{appendix:uniform_sampling}}

Given a fixed number of groups $R$, we want to evaluate the probability of observing specific distributions of group sizes when sampling partitions of $N$ objects uniformly at random. If the groups are treated as labeled, the problem reduces to counting how many assignments of objects give rise to a particular size profile $\{n_r\}_{r=1}^R$, where $\sum_{r=1}^R n_r = N$. The number of assignments consistent with this profile is given by the multinomial coefficient
\[
    \Omega(n_1, \dots, n_R) = \frac{N!}{n_1!\cdots n_R!}.
\]
In contrast, we are interested in unlabeled partitions, where only the multiset of group sizes matters. For example, the configurations $\{1, 2, 3\}$ and $\{ 2, 3, 1\}$ are indistinguishable. To correct for this, we must account for the number of distinct permutations of the size vector. This equals the total number of permutations of $R$ terms divided by the permutations that leave the vector unchanged due to repeated sizes. Thus,
\begin{equation} \label{eq:multiset_ordered}
\Omega_{\mathrm{lab}}(n_1,\dots,n_R)
\;=\;
\frac{N!}{\prod_{r=1}^R n_r!} \cdot \frac{R!}{\prod_u m_u!},
\end{equation}
where $m_u$ denotes the multiplicity of groups of size $u$. However, Eq. \eqref{eq:multiset_ordered} still counts labeled partitions. To obtain the number of unlabeled partitions with the given size profile, we must divide by the number of ways to relabel the $R$ groups, i.e. by $R!$. This yields
~
\begin{equation}\label{eq:Omega_unlabeled}
\Omega_{\mathrm{unlab}}(n_1,\dots,n_R)
\;=\;
\frac{N!}{\prod_{r=1}^R n_r!\;\prod_u m_u!}.
\end{equation}
~
The total probability of observing a partition with a particular size profile is then given by
\begin{equation} \label{eq:prob_rcbtl_partition}
    p(\{n_r\}) = \frac{\Omega_{\mathrm{unlab}}}{S(N, R)},
\end{equation}
where $S(N, R)$ are the Stirling numbers of the second kind, which count the number of ways to partition a set of size $N$ into $R$ indistinguishable and nonempty groups.

Eq.~\eqref{eq:Omega_unlabeled} shows that the count decomposes into two competing contributions: the multinomial factor $N! / \prod_r n_r!$, which, as shown in Appendix~\ref{appendix:objective_contributions}, is maximized when all groups are as balanced as possible ($n_r \approx N / R$), and the multiplicity factor $\prod_u m_u!$, which penalizes size profiles in which multiple groups share the same size, since each repetition reduces the number of distinct unlabeled partitions. The observed distribution of group sizes is therefore determined by a tradeoff: the multinomial term favors near-equal block sizes, while the multiplicity term discourages ties between group sizes.

We can investigate the behvaior of Eq.~\eqref{eq:prob_rcbtl_partition} in the limit of large $N$. Taking the logarithm of Eq.~\eqref{eq:Omega_unlabeled} we have that
\begin{equation}
\begin{aligned}
    \log \Omega_{\mathrm{unlab}}(\{n_r\}) &= \log \frac{N!}{\prod_r n_r! \prod_u m_u!} \\
    &\approx N \left( -\sum_{r=1}^R q_r \log q_r \right) - \sum_u \log m_u!,
\end{aligned}
\end{equation}
where $q_r = n_r / N$ are the relative sizes of each group and $H(q) = -\sum_r q_r \log q_r$ is the entropy of the relative-size vector. Using the classical asymptotic approximation $S(N, R) \sim R^N / R!$ for the Stirling numbers of the second kind, we then have
\begin{equation} \label{eq:log_rcbtl_partition_prob}
   \log p(\{n_r\}) \approx N(H(q) -\log R) - \sum_u \log m_u! + \log R!.
\end{equation}
Since there are at most $R$ terms in the sum over multiplicities and each term in the sum is at most $\log R!$ (because $m_u \leq R~\forall u)$, the last two terms in Eq.~\ref{eq:log_rcbtl_partition_prob} scale as $O(R\log R)$ and are negligible compared with the leading $O(N)$ term for large $N$. On the other hand, $H(q) \leq \log R$ with equality only in the perfectly balanced case $n_r = N/R~\forall r$. This implies that, in the limit $N\to\infty$, any profile whose entropy is noticably below $\log R$ is exponentially supressed in $N$, so that the multinomial (entropy) term overwhelmingly favors near-uniform partitions.

~
\begin{table*}[!ht]
  \centering
  \renewcommand{\arraystretch}{1.2}
  \setlength{\tabcolsep}{6pt}

  \begin{tabular}{|l|l|}
    \hline
    \textbf{PR Partition} & \textbf{RC-BTL Consensus Partitions} \\
    \hline
    \cellcolor{lightgray!75}\textbf{Cluster 0 ($\bm{\hat\sigma = 0.5007}$)} &
    \cellcolor{lightgray!75}\textbf{Cluster 0 ($\bm{\hat\sigma = 0.3167}$)} \\
    \hline
    \parbox[t]{0.47\textwidth}{\raggedright%
    Minnesota Timberwolves\\Boston Celtics\\Denver Nuggets\\Oklahoma City
    Thunder
    } &
    \parbox[t]{0.47\textwidth}{\raggedright%
    Minnesota Timberwolves\\Boston Celtics\\Denver Nuggets\\Oklahoma City
    Thunder
    } \\
    \hline

    \cellcolor{lightgray!75}\textbf{Cluster 1 ($\bm{\hat\sigma = 0.2654}$)} &
    \cellcolor{lightgray!75}\textbf{Cluster 1 ($\bm{\hat\sigma = 0.1885}$)} \\
    \hline
    \parbox[t]{0.47\textwidth}{\raggedright%
    Los Angeles Lakers\\
    Orlando Magic \\ 
    Sacramento Kings \\ 
    Golden State Warriors \\ 
    Philadelphia 76ers \\ 
    New York Knicks \\ 
    Phoenix Suns \\ 
    Miami Heat \\ 
    Milwaukee Bucks \\ 
    Indiana Pacers \\ 
    Dallas Mavericks \\ 
    Los Angeles Clippers \\ 
    Cleveland Cavaliers \\ 
    Houston Rockets \\ 
    New Orleans Pelicans
    } &
    \parbox[t]{0.47\textwidth}{\raggedright%
    Sacramento Kings \\ 
    New Orleans Pelicans \\ 
    Milwaukee Bucks \\ 
    Phoenix Suns \\ 
    Dallas Mavericks \\ 
    Los Angeles Clippers \\ 
    New York Knicks
    } \\
    \hline

    \cellcolor{lightgray!75}\textbf{Cluster 2 ($\bm{\hat\sigma = 0.1298}$)} &
    \cellcolor{lightgray!75}\textbf{Cluster 2 ($\bm{\hat\sigma = 0.1815}$)} \\
    \hline
    \parbox[t]{0.47\textwidth}{\raggedright%
    Utah Jazz \\ 
    Chicago Bulls \\ 
    Brooklyn Nets \\ 
    Atlanta Hawks
    } &
    \parbox[t]{0.47\textwidth}{\raggedright%
    Los Angeles Lakers \\ 
    Golden State Warriors \\ 
    Cleveland Cavaliers
    } \\
    \hline

    \cellcolor{lightgray!75}\textbf{Cluster 3 ($\bm{\hat\sigma = 0.0696}$)} &
    \cellcolor{lightgray!75}\textbf{Cluster 3 ($\bm{\hat\sigma = 0.1591}$)} \\
    \hline
    \parbox[t]{0.47\textwidth}{\raggedright%
    San Antonio Spurs \\ 
    Charlotte Hornets \\ 
    Toronto Raptors \\ 
    Portland Trail Blazers \\ 
    Memphis Grizzlies
    } &
    \parbox[t]{0.47\textwidth}{\raggedright%
    Philadelphia 76ers \\ 
    Miami Heat \\ 
    Indiana Pacers \\ 
    Chicago Bulls \\ 
    Orlando Magic \\ 
    Houston Rockets
    } \\
    \hline

    \cellcolor{lightgray!75}\textbf{Cluster 4 ($\bm{\hat\sigma = 0.0345}$)} &
    \cellcolor{lightgray!75}\textbf{Cluster 4 ($\bm{\hat\sigma = 0.0780}$)} \\
    \hline
    \parbox[t]{0.47\textwidth}{\raggedright%
    Washington Wizards \\ 
    Detroit Pistons
    } &
    \parbox[t]{0.47\textwidth}{\raggedright%
    Utah Jazz \\ 
    Memphis Grizzlies \\ 
    Brooklyn Nets \\ 
    Atlanta Hawks
    } \\
    \hline

    \multirow{2}{*}{\parbox[t]{0.47\textwidth}{\raggedright~}} &
    \cellcolor{lightgray!75}\textbf{Cluster 5 ($\bm{\hat\sigma = 0.0452}$)} \\
    \cline{2-2}
     & \parbox[t]{0.47\textwidth}{\raggedright
     Charlotte Hornets\\San Antonio Spurs\\Portland Trail Blazers\\Toronto
     Raptors
     } \\
     \hline

    \multirow{2}{*}{\parbox[t]{0.47\textwidth}{\raggedright~}} &
    \cellcolor{lightgray!75}\textbf{Cluster 6 ($\bm{\hat\sigma = 0.0311}$)} \\
    \cline{2-2}
     & \parbox[t]{0.47\textwidth}{\raggedright
    Washington Wizards \\ 
    Detroit Pistons
     } \\
     \hline
  \end{tabular}

  \caption{Comparison of the partition inferred by the PR algorithm (PR
  Partition)
  and the RC-BTL Consensus Partitions for the NBA 2023–24 dataset. Groups are
  sorted in decreasing order of their normalized strengths
  $\boldsymbol{\hat\sigma}$.}
  \label{table:pr_vs_consensus_partitions}
\end{table*}


\end{document}